\pdfoutput=1
\documentclass[aip,jcp,floatfix,reprint]{revtex4-1}

\usepackage{graphicx}
\usepackage{dcolumn}
\usepackage{bm}
\usepackage{color}

\usepackage{graphicx}
\usepackage{subfig}

\begin{document}

\title{The robustness of proofreading  to crowding-induced pseudo-processivity in the MAPK pathway}

\author{Thomas E. Ouldridge,$^\ast$ and Pieter Rein ten Wolde$^\dagger$}

\address{$^\ast$Rudolf Peierls Centre for Theoretical Physics,  University of Oxford, 1 Keble Road, Oxford, OX1 3NP, UK; 
$^\dagger$FOM Institute AMOLF, 1098 XG, Amsterdam, The Netherlands.}

\begin{abstract}
{Double phosphorylation of protein kinases is a common feature of signalling cascades. This motif may  reduce cross-talk between signalling pathways, as the second phosphorylation site allows for proofreading, especially when phosphorylation is distributive rather than processive. 
Recent studies suggest  that phosphorylation can be `pseudo-processive'  in the crowded cellular environment, as rebinding after the first phosphorylation is enhanced by slow diffusion. Here, we use a 
simple model with unsaturated reactants to show that specificity for one substrate over another drops as rebinding increases and pseudo-processive behavior becomes possible. However, this loss of specificity with increased rebinding is typically also observed if two distinct enzyme species are required for phosphorylation, {\it i.e.} when the system is necessarily distributive. Thus the loss of specificity is due to an intrinsic reduction in selectivity with increased rebinding, which benefits inefficient reactions, rather than pseudo-processivity itself.   We also show that proofreading can remain effective when the intended signalling pathway exhibits high levels of rebinding-induced 
pseudo-processivity, unlike other proposed advantages of the dual phosphorylation motif.}\\
{*Correspondence: t.ouldridge@imperial.ac.uk\\
Address reprint requests to Thomas E. Ouldridge, Mathematics Department, Huxley Building, Queen's Gate, London SW7 2AZ, U.K.\\
Editor: XXXX.}
\end{abstract}

\maketitle

\markboth{Ouldridge and Ten Wolde}{Robustness of MAPK proofreading to pseudo-processivity}

\section*{INTRODUCTION}
Cells must sense and respond to their environment, and external signals must be transmitted from cell-surface receptors the
 interior. 
Eukaryotic signal transmission often involves phosphorylation cascades of mitogen-activated protein (MAP) kinases \cite{Chang2001,Gustin1998,Qi2005}. Phosphorylation, the addition of a phosphate group 
to a residue (typically serine, threonine or tyrosine), is a common 
post-transcriptional protein modification. 
{\it Kinases} catalyze phosphorylation, and a kinase cascade involves the successive phosphorylation of downstream kinases by upstream counterparts, with each kinase becoming enzymatically active after
phosphorylation. {\it Phosphatases} catalyze the release of inorganic phosphate  and enzymatic deactivation \cite{Huang1996, Ferrell1997, Swain2002, Kocieniewski2012}. The result is a characteristic {\it push-pull} 
motif in which competition between phosphatases and upstream kinases sets the activation level of a downstream kinase, the first kinase having been activated directly or indirectly by the receptor.

MAP kinases typically require phosphorylation at two residues for activation \cite{Aoki2011, Huang1996, Ferrell1997, Swain2002, Kocieniewski2012}. 
Each stage necessitates the breakdown of an ATP molecule, the cell's fuel source. The need for two phosphorylation events  is thus potentially costly and time consuming, and it is reasonable to assume that such a motif would only survive by conferring a biological advantage. Several possible uses of dual phosphorylation have been proposed.
Firstly, kinases that require double phosphorylation can respond more sensitively, {\it i.e.,} ultrasensitively, to changes in  phosphatase and upstream kinase concentrations \cite{Huang1996}. When the upstream enzymes are saturated, it is even possible to achieve bistability
 \cite{Salazar2006,Ortega2006}. Secondly, dual phosphorylation allows for more discrimination between substrates \cite{Swain2002}. All signalling pathways will experience some degree of cross-reactivity, and the need
 to perform two phosphorylations rather than one  allows for an extra stage of discrimination (or {\it proofreading}). Finally, some kinase cascades involve scaffolding proteins that bind to upstream and downstream kinases simultaneously \cite{Dhanasekaran2007,Engstrom2010}. Such a motif could enhance 
 signalling and improve insulation of pathways \cite{Dhanasekaran2007,Engstrom2010}. It has been claimed \cite{Kocieniewski2012} that this enhancement is only effective when coupled with dual phosphorylation, as the scaffold allows for a single upstream kinase to perform both modifications, rather than requiring two  separate  interactions in the cytosol.

The effectiveness of these motifs depends on whether phosphorylation in the cytosol is naturally  {\it processive} (a single enzymatic molecule can perform both phosphorylations during one interaction) or  {\it distributive} (two separate interactions are required).
The third motif mentioned above obviously requires phosphorylation to be naturally distributive in the cytosol. A reduction
in ultrasensitivity with processivity has also been demonstrated elsewhere \cite{Salazar2009,Takahashi2010}. In their original article on proofreading, Swain and Siggia \cite{Swain2002} considered partially processive kinase operation, in which a certain fraction of phosphorylation events lead directly to the doubly phosphorylated state and the rest cause single phosphorylation via a ``discard pathway". In the limit that they considered, Swain and Siggia showed that specificity is compromised by increasing processivity, and that this decrease is due to a drop in the discrimination at the second stage of phosphorylation.

\begin{figure}[!]
\centerline{\includegraphics[width=18pc]{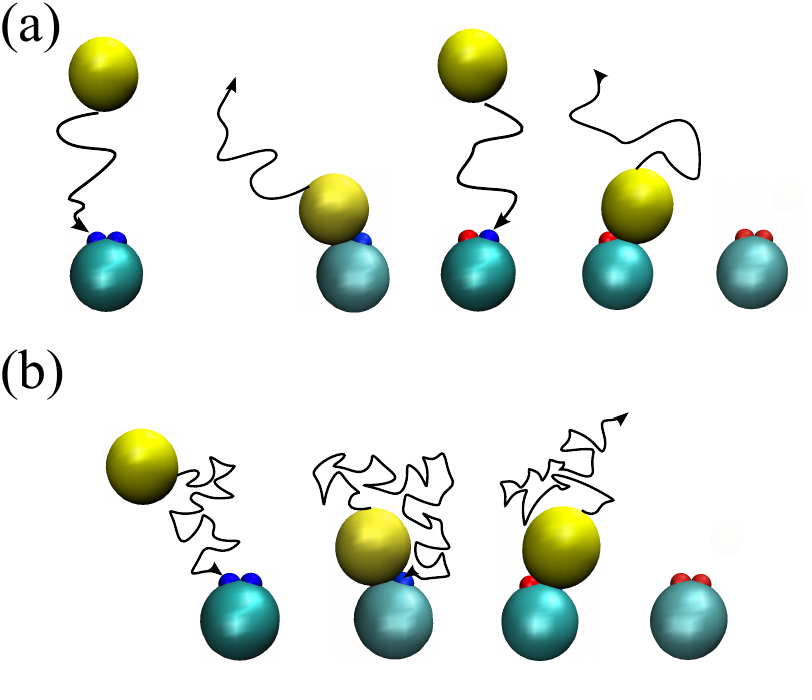}}
\caption{Diffusion induced pseudo-processivity. (a) Conventional distributive phosphorylation of two residues by two distinct kinase molecules. (b) When diffusion is slow
compared to intrinsic reaction rates, the same kinase molecule can rebind and modify the second site, resulting in a pseudo-processive scheme.
\label{pseudo-proc}}
\end{figure}

\begin{figure*}[!]
\centerline{\includegraphics[width=38pc]{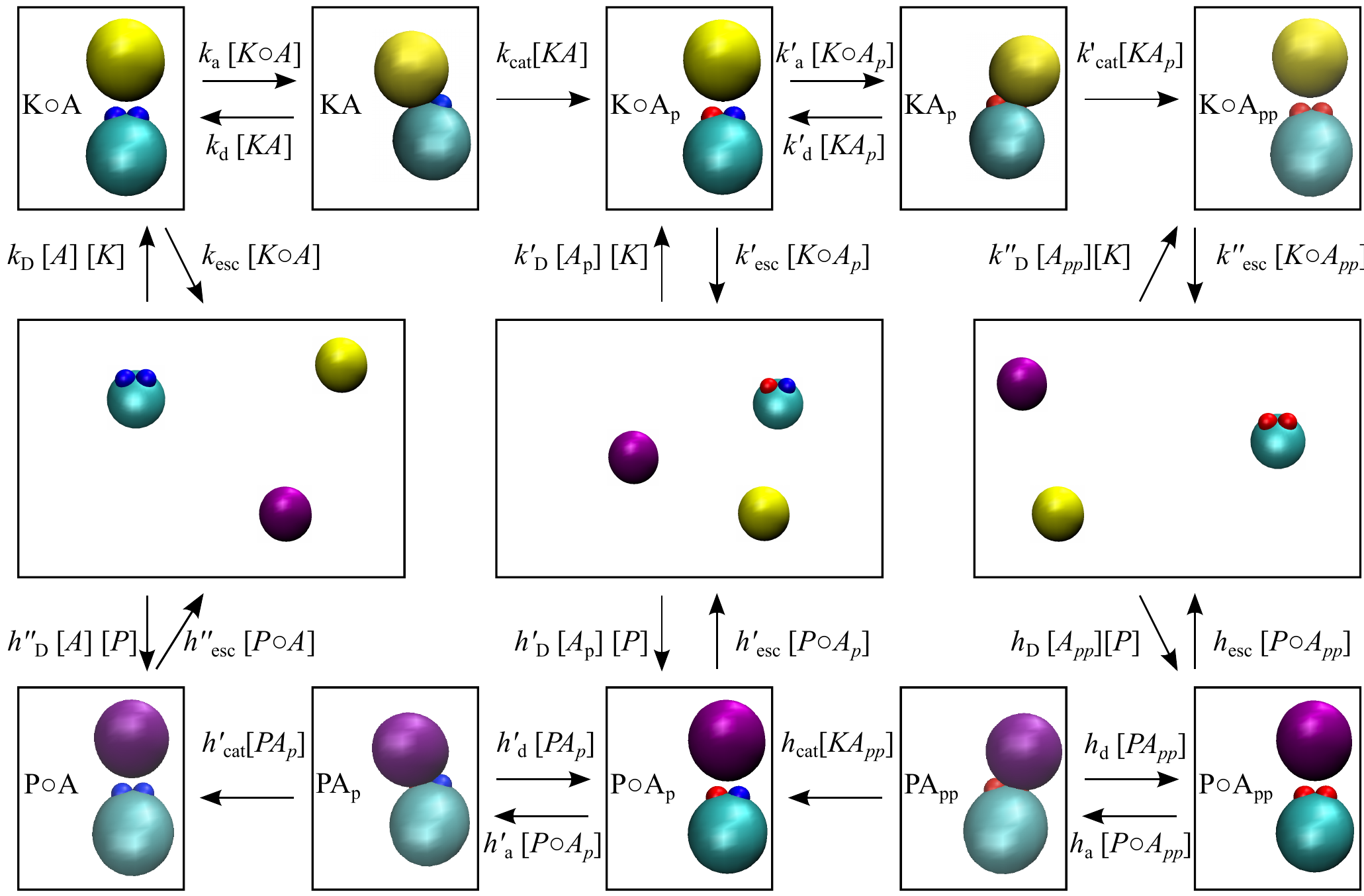}}
\caption{A simple model for pseudo-procesisve phosphorylation. $A$ is phosphorylated in two stages by a kinase (yellow). Firstly, $K$ and $A$ diffuse into
close proximity, a state labelled by $K \circ A$. The two can then bind ($KA$), at which point phosphorylation and release can occur, leaving the kinase and substrate in close proximity
but with the substrate singly phosphorylated $K \circ A_p$. From here, the two can diffuse apart (escape), leaving an isolated $A_p$. Alternatively, the kinase can rebind and perform a second 
phosphorylation. The reverse process can be observed for the phosphatase (purple). Reaction arrows are labelled with rates per unit volume at which reactions occur.
\label{model}}
\end{figure*}

Reactants
that physically separate after
 phosphorylation may nonetheless show {\it pseudo-processive} behavior due to finite rates of diffusion \cite{Takahashi2010,Gopich2014}, as shown in 
Figure \ref{pseudo-proc}. If diffusion is slow enough compared to the intrinsic binding rate, two protein molecules can rebind after the first phosphorylation, allowing effectively processive phosphorylation if the kinase can also catalyze the second step.
Recent experiments \cite{Aoki2011,Aoki2013} and theory \cite{Hellmann2012,Aoki2013} suggest that  molecular crowding  (which slows diffusion relative to intrinsic reaction rates) can cause pseudo-processivity in conditions similar to those found in the cell. Rebinding due to slow diffusion is also relevant in a wide range of biophysical
systems; examples include T cell fate decisions \cite{Govern2010}, signalling involving membrane-bound clusters \cite{Mugler2012}, the accuracy with which surface receptors can sense ligand concentrations \cite{Kaizu2014}, and the dynamics with which 
transcription factors search DNA for their binding sites \cite{Brackley2013}.

Given these insights, characterizing the robustness of dual phosphorylation-based motifs to rebinding-driven processivity is essential. We study a simple model of  pseudo-processivity in the limit of unsaturated reactants. We analyze the consequences of rebinding and pseudo-processivity for the selective phosphorylation of one substrate over another.  
Our results are consistent with the simpler model of Swain and Siggia for parameters that allow comparison \cite{Swain2002}, but our approach reveals key features that arise when rebinding drives pseudo-processivity. High binding probabilities when in close proximity rather than pseudo-processivity {\it per se} are generally responsible for low specificity, and specificity is lost at both stages of phosphorylation. Further, the relative increase in discrimination from adding a second phosphorylation site can remain appreciable with significant pseudo-processivity.  
Finally, we argue that pseudo-processivity does not limit proofreading as it does other uses of dual phosphorylation, which can also
be understood through the same simple model.

\section*{MODEL AND METHODS}

Our model of diffusion and catalysis is based on that of Dushek {\it et al.} \cite{Dushek2011}. We model the system at the level of molecular concentrations. Upstream kinases can bind to  and unbind from substrates, with catalysis and rapid release possible when bound. Substrates can also be dephosphorylated by a phosphatase. Importantly, the model includes states
representing configurations in which two proteins are in close proximity, but unbound \cite{Dushek2011}. These states permit rapid rebinding of molecules, as reactants remain in close proximity for some time after separating. Rebinding either occurs rapidly or 
the reactants diffuse apart and all memory is lost -- such a picture is consistent with theoretical analyses of rebinding in dilute solution \cite{vanZon2006, Kaizu2014}. Transitions between states are quantified by rate constants.

The primary system studied in this paper is illustrated in Figure \ref{model}, which also defines rate constants. Here, the substrate $A$ exists in unphosphorylated ($A$), singly phosphorylated ($A_p$) and doubly phosphorylated ($A_{pp}$) states, and a single kinase $K$ and single phosphatase $P$ can catalyse reactions for both phosphorylation sites. We use $\circ$ to indicate close proximity. This system allows for pseudo-processivity as rebinding and a second catalysis event can occur immediately after the first. In the language of Swain and Siggia \cite{Swain2002}, reactants that diffuse apart after the first phosphorylation follow a ``discard pathway".
We will later introduce a substrate $B$ with different underlying rate constants, and consider the specificity with which $A$ is 
activated over $B$. We will also apply the model to alternative systems in which substrates have only one phosphorylation site, or enzymes can only act on one phosporylation site.

The ``close proximity'' state is assumed to be equally close to both phosphorylation sites, so proteins have no memory of previous binding in that state. This is  reasonable if the phosphorylation sites are close to each other, as is typical \cite{Payne1991, Alessi1994, Cargnello2011}, and pseudo-processivity  is due to reattachment following failure to escape the local environment. A second assumption
is that our model has only one singly phosphorylated state, rather than explicitly considering phosphorylation on either residue. Technically, this assumes an ordered, or sequential, phosphorylation of the sites. This simplification is common in the literature \cite{Takahashi2010,Gunawardena2005,Swain2002,Huang1996}. 
To check that our results are not overly sensitive to this assumption, we consider 
independent phosphorylation sites in Section S11 of the Supporting Material.

For simplicity we assume that reactants are unsaturated; {\it i.e.,} most molecules of each species are not in complexes at any time. States such as $KA$ and $K \circ A$ must therefore be short-lived compared to the time taken for a given reactant to come into close proximity with a reactant partner. For the first stage of phosphorylation, this limit is obtained when
\begin{equation}
\frac{1}{k_{\rm D} [A_0]} ,\frac{1} {k_{\rm D} [K]} \gg \frac{k_{\rm d} + k_{\rm cat} + k_{\rm a}} {k_{\rm esc}k_{\rm d} + k_{\rm esc}k_{\rm cat} +k_{\rm a}k_{\rm cat} },
\label{limit criterion}
\end{equation}
in which $[A_0] $ is the total concentration of substrate $A$.
Similar inequalities must hold for all reactions. The right hand side of Equation (\ref{limit criterion}) is the average time taken for either escape or catalysis to occur once the reactants are in close proximity. It is derived in Section S1  of the Supporting Material, where we also show that the right hand side of Equation (\ref{limit criterion}) is $\leq \rm{max}(1/k_{\rm cat},1/k_{\rm esc})$. Thus fast catalysis {\it and} escape compared to diffusive encounter is a sufficient (but not necessary) condition to ensure that our approximation holds. Note that the saturation of {\it reactants}, which we preclude, should not be confused with mechanisms by which the {\it yield of  product} can become saturated. For example, the yield of $A_{pp}$ can become saturated when $[A_{pp}] \approx [A_0]$.

\begin{figure}[!]
\centerline{\includegraphics[width=18pc]{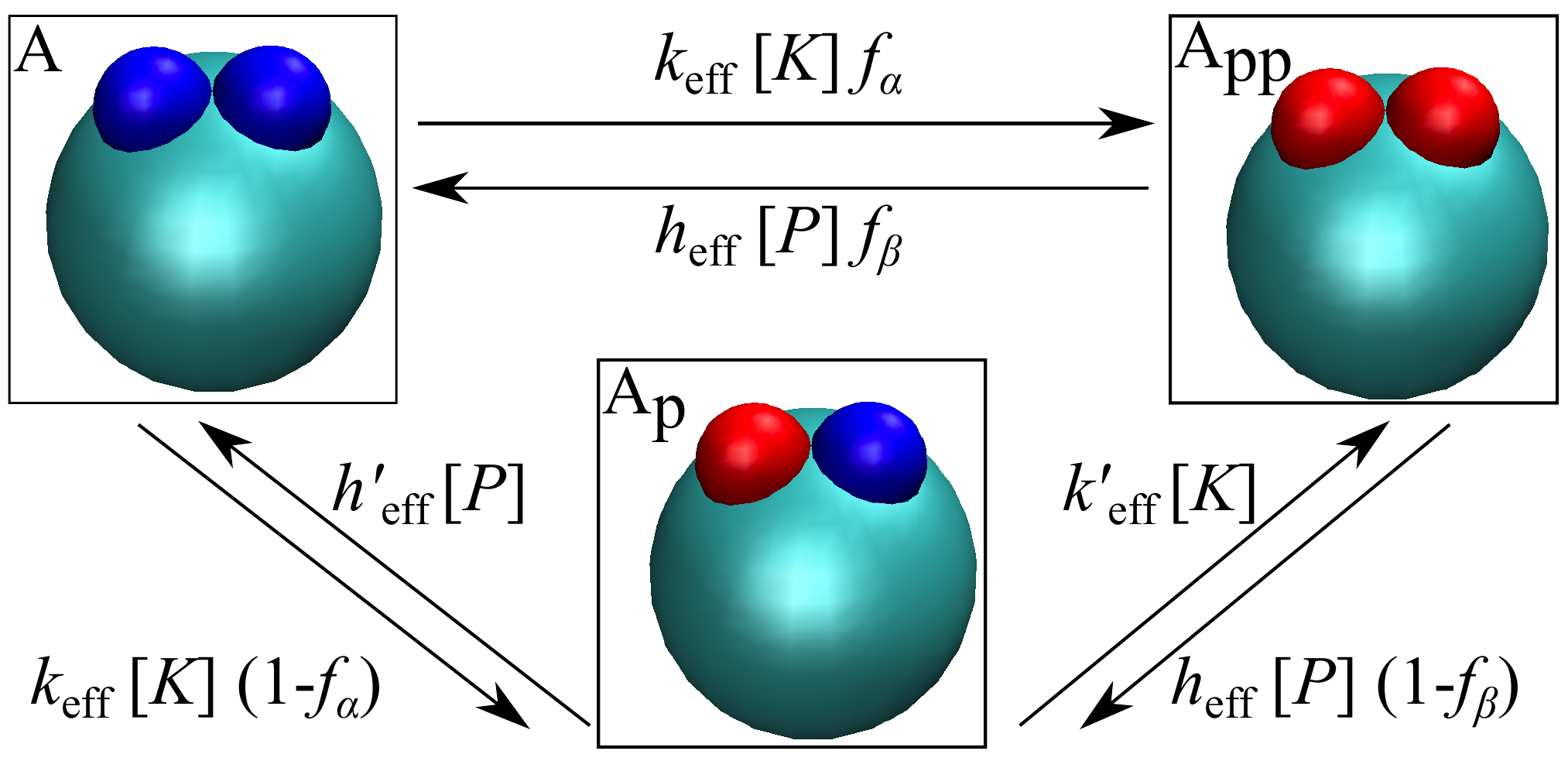}}
\caption{Effective first-order description of the conversion of $A$  between its phosphorylation states that results from the assumption of unsaturated kinetics. Arrows are labelled with effective rate constants.
\label{second-order}}
\end{figure}

In the unsaturated (low concentration) limit, and assuming fixed total concentrations, the model reduces to an effective first-order  interconversion of substrates between phosphorylation states (Figure \ref{second-order}). The rate constants defined in Figure \ref{model} determine the probabilities of various reaction outcomes; the key collective variables that emerge are the effective rate constants ({\it e.g.} $k_{\rm eff}$), and $f_\alpha$ ($f_\beta$), which is the probability that phosphorylation of $A$ (dephosphorylation of $A_{pp}$) leads to modification of both sites rather than just one.
Large $f_\alpha$ and $f_\beta$ indicate substantial  pseudo-processivity.

The effective rate constants, $f_\alpha$, and $f_\beta$, can be expressed via the probabilities that reactants bind given close proximity, and that catalysis occurs given binding. The relevant probabilities are 
\begin{equation}
 \begin{array}{c}
   P_{\rm cat} = \frac{k_{\rm cat}}{k_{\rm cat} +k_{\rm d} }, \hspace{1mm} P^\prime_{\rm cat} = \frac{k^\prime_{\rm cat}}{k^\prime_{\rm cat} +k^\prime_{\rm d} },  \hspace{1mm} 
 P_{\rm on} = \frac{k_{\rm a}}{k_{\rm esc} +k_{\rm a} },\\
  P^\prime_{\rm on} = \frac{k^\prime_{\rm a}}{k^\prime_{\rm a} +k^\prime_{\rm esc} }, \hspace{1mm}
    Q_{\rm cat} = \frac{h_{\rm cat}}{h_{\rm cat} +h_{\rm d} }, \hspace{1mm} Q^\prime_{\rm cat} = \frac{h^\prime_{\rm cat}}{h^\prime_{\rm cat} +h^\prime_{\rm d} }, \\
 Q_{\rm on} = \frac{h_{\rm a}}{h_{\rm esc} +h_{\rm a} }, \hspace{1mm} Q^\prime_{\rm on} = \frac{h^\prime_{\rm a}}{h^\prime_{\rm a} +h^\prime_{\rm esc} }.
 \label{Ps and Qs}
  \end{array}
\end{equation}
Primed probabilities relate to the second stage of phosphorylation (or dephosphoylation), and unprimed probabilities to the first, as in Figure \ref{model}.

The rate constant at which $K$ phosphorylates $A$  is given by the rate constant for $K$ and $A$ coming into close proximity multiplied by the probability that a successful reaction occurs after $n$ binding events, summed over $n$:
$k_{\rm eff} = k_{\rm D} \sum_{n\geq1} P_{\rm on}^n (1-P_{\rm cat})^{n-1} P_{\rm cat} =k_{D} P_{\rm react}  $. $P_{\rm react} $ is the probability that phosphorylation of the first site occurs given that an $A$ molecule is in close proximity to a kinase capable of catalysing the $A\rightarrow A_p$ transition. 
This sum is a simple geometric progression,
\begin{equation}
k_{\rm eff} =  k_{D} P_{\rm react} = k_{D} \frac{P_{\rm cat} P_{\rm on}}{1 - P_{\rm on}(1-P_{\rm cat})}.
\label{k_eff eqn}
\end{equation} 
Similar quantities can be calculated for other reactions, 
\begin{equation}
 \begin{array}{c}
     k^\prime_{\rm eff} = k^\prime_{D} P^\prime_{\rm react} = \frac{k^\prime_{D}  P^\prime_{\rm cat} P^\prime_{\rm on}}{1 - P^\prime_{\rm on}(1-P^\prime_{\rm cat})},\vspace{2mm}\\
    h_{\rm eff} = h_D Q_{\rm react} = \frac{ h_{D}  Q_{\rm cat} Q_{\rm on}}{1 - Q_{\rm on}(1-Q_{\rm cat})},\vspace{2mm}\\
   h^\prime_{\rm eff} = h_D Q^\prime_{\rm react} = \frac{h^\prime_{D}  Q^\prime_{\rm cat} Q^\prime_{\rm on}}{1 - Q^\prime_{\rm on}(1-Q^\prime_{\rm cat})}.\\
  \end{array}
  \label{effective-rates}
\end{equation} 
$P_{\rm react}^\prime$ is the probability that the second site will be modified given that an appropriate kinase is close to an $A_p$ molecule. A kinase that has just modified the first site will be in close proximity to the substrate $A_p$. If this kinase can  also catalyze the phosphorylation of the next site, {\it i.e.} if both sites are modified by the same kinase species (as assumed hitherto), then the fraction of  pseudo-processive modifications  $f_\alpha = P_{\rm react}^\prime$. Similarly, $f_\beta = Q_{\rm react}^\prime$.  We will later consider a system in which two distinct kinases and phosphatases are needed, in which case this identification is inappropriate; we thus retain distinct variables.

Neglecting noise, Figure \ref{second-order} 
implies
differential equations for the concentrations of $[A]$, $[A_p]$ and $[A_{pp}]$. The steady-state solution is simple as the equations are linear.
The results are easiest to express in terms of the ratios $\theta = h_{\rm eff} / h^\prime_{\rm eff}$, $\phi = k_{\rm eff} / k^\prime_{\rm eff}$,
$\psi = k_{\rm eff} / h_{\rm eff}$ and $Y=[K]/[P]$. Low $\phi$ would imply that the second stage of phosphorylation is faster than the first. $\theta$ has
the same meaning for dephosphorylation,  and $\psi$ and $Y$ simply quantify the relative activity and concentrations of kinases and phosphatases. In terms of these variables,
\begin{equation}
 \begin{array}{c}
  \frac{[A]}{[A_0]} = \frac{\phi + (Y \psi) \theta f_\beta}{\phi + (Y \psi) (\theta f_\beta + \phi f_\alpha + \theta \phi (1-f_{\alpha} f_{\beta})) + (Y \psi)^2 \theta},\vspace{1mm}\\
    \frac{[A_p]}{[A_0]} = \frac{(Y \psi) \theta \phi (1-f_{\alpha}  f_{\beta})}{\phi + (Y \psi) (\theta f_\beta + \phi f_\alpha + \theta \phi (1-f_{\alpha} f_{\beta})) + (Y \psi)^2 \theta},\vspace{1mm}\\
    \frac{[A_{pp}]}{[A_0]} = \frac{(Y \psi) \phi f_\alpha + (Y \psi)^2 \theta}{\phi + (Y \psi) (\theta f_\beta + \phi f_\alpha + \theta \phi (1-f_{\alpha}f_\beta)) + (Y \psi)^2 \theta}.\\
  \end{array}
  \label{solution}
  \end{equation}
Although the model is simple, it  maps to the results of a previous analysis of rebinding based on continuum diffusion \cite{Gopich2014}. When the standard diffusion equation is a good description of particle motion, and neglecting behavior on short timescales \cite{Gopich2014}, phosphorylation can be treated as a second-order reaction involving a diffusion-influenced rate constant with a finite probability that more than one phosphorylation event occurs during an encounter. In Section S2 of the Supporting Material, we show that our model is consistent with this result and reproduces the rates at which different products form. In this analogy, $k_{\rm D}$ is the diffusion-limited rate constant and  $k_{\rm a} P_{\rm cat} k_{\rm D}/k_{\rm esc}$ the rate constant in the limit of infinitely fast diffusion (the reaction-limited rate constant). To understand this assignation, note that $k_{\rm D}/k_{\rm esc}$ quantifies the probability that enzyme and substrate are in close proximity, and $k_{\rm a} P_{\rm cat}$ is a reaction rate given close proximity. We emphasize, however, that our model does not rest upon a particular description of diffusion. In the cell, crowding molecules mean that reactants do not diffuse as they would in a simple solution, tending to show sub-diffusive behavior on short timescales. In Section S3 of the Supporting Material, we show that a lattice model also produces results that support our simple finite-state analysis. Dushek {\it et al.} also verified that explicit lattice simulations reproduced results obtained with a similar model \cite{Dushek2011}.

\section*{RESULTS}
To explore specificity, we introduce a substrate $B$ that is less efficiently phosphorylated by the kinase but obeys similar differential equations to $A$. Our model has many parameters;  
we wish to explore system behavior  
as they are varied, but there are too many to do this exhaustively. We therefore
assume that all diffusion rates are identical; encounter rates are described by a single $k_D$,
and escape rates by a single $k_{\rm esc}$. 
As in Ref. \cite{Swain2002}, we shall assume that differential catalytic activity is entirely due to 
variations in unbinding rates $k_d$. We consider alternatives in Section S10 of the Supporting Material. 
Finally, we shall assume that the phosphatases do not discriminate between substrates. The reduction in free parameters is summarized below.
\begin{equation}
 \begin{array}{c}
  k^{A,B}_D, k^{A \prime,B \prime}_D, k^{A\prime \prime,B \prime \prime}_D, h^{A,B}_D, h^{A \prime,B \prime}_D, h^{A\prime \prime,B\prime \prime}_D = k_D, \vspace{1mm}\\
  k^{A,B}_{\rm esc}, k^{A \prime,B \prime}_{\rm esc}, k_{\rm esc}^{A \prime \prime,B\prime \prime}, h^{A,B}_{\rm esc}, h^{A \prime,B \prime}_{\rm esc}, h_{\rm esc}^{A \prime \prime,B\prime \prime} = k_{\rm esc}, \vspace{1mm}\\
  k^{A,B}_{a}, k^{A \prime,B \prime}_{a}, h^{A,B}_{a}, h^{A \prime ,B \prime}_a = k_a, \vspace{1mm}\\
   k^{A,B}_{\rm cat}, k^{A \prime,B \prime}_{\rm cat}, h^{A,B}_{\rm cat}, h^{A \prime,B \prime}_{\rm cat}  = k_{\rm cat},\vspace{1mm}\\
  h^{A,B}_{d} = h_d \hspace{3mm} {\rm and} \hspace{3mm} h^{A \prime,B \prime}_d = h^\prime_d.\\ 
  \end{array}
  \label{reduction}
  \end{equation}
  As a result of this simplification, $P^{A,B}_{\rm on} = P^{A,B \prime}_{\rm on} = Q^{A,B}_{\rm on} = Q^{A,B \prime}_{\rm on} = P_{\rm on}$. $P_{\rm on}$ is the probability of binding given close proximity, and hence the probability of rebinding after dissociation. ``High $P_{\rm on}$" and ``frequent rebinding" are used synonymously in this work.  
  
$A$ and $B$ then differ only  in their binding free energies with $K$:
    $ \Delta \Delta G = kT \ln ( {k^A_d }/{k^B_d})$, $\Delta \Delta G^\prime = kT \ln ( {{k^A_d}^\prime}/{ {k^B_d}^\prime})$. The maximum possible discrimination factor is $\exp(-(\Delta \Delta G + \Delta \Delta G^\prime) /kT)$. 
This discrimination is not necessarily manifested, however; we can define kinetic selectivity factors $S$ and $S^\prime$
\begin{equation}
 \begin{array}{c}
      S = \frac{k^A_{\rm eff}}{k^B_{\rm eff}}= \left(\frac{P^A_{\rm cat}}{P^B_{\rm cat}} \right) \left( \frac{1- P_{\rm on}(1- P^B_{\rm cat})}{1- P_{\rm on}(1- P^A_{\rm cat})}\right), \vspace{2mm}\\
      S^\prime = \frac{k^{A \prime}_{\rm eff}}{k^{B \prime}_{\rm eff}}  =
      \left( \frac{{P^{A \prime}_{\rm cat}}} {{P^{B \prime}_{\rm cat}}}\right) \left(\frac{1- {P_{\rm on}}(1- {P^{B \prime}_{\rm cat}})}{1- {P_{\rm on}}(1- {P^{A \prime}_{\rm cat}})}\right).
      \end{array}
      \label{S-equation}
    \end{equation}
$S$ is the ratio (see Equation (\ref{k_eff eqn})) of rates for going from $K \circ A \rightarrow K \circ A_p$ and  $K \circ B \rightarrow K \circ B_p$ (regardless of whether another phosphorylation occurs immediately). $S^\prime$ is the equivalent for the second
step, and $S,S^\prime \geq 1$ as $A$ is the intended substrate.  $S \leq  \exp(-\Delta \Delta G/ kT)$; selectivity is reduced when $P_{\rm on}$ and $P_{\rm cat}^{A}$ are large.
Note
\begin{equation}
 \begin{array}{c}
      S = S_0 (1 - P_{\rm react}^A) + P_{\rm react}^A, \vspace{2mm}\\
      S^\prime = S_0^\prime (1 - P_{\rm react}^{A \prime}) + P_{\rm react}^{A \prime}. 
      \end{array}
      \label{S-equation2}
    \end{equation}
    Here, $S_0  = P^A_{\rm cat}/P^B_{\rm cat}$ and $S_0^\prime  = P^{A\prime}_{\rm cat}/P^{B\prime}_{\rm cat}$ are the selectivities in the limit of no rebinding. We also define a metric
for the overall specificity 
$X = \lg ([A_{pp}]/[A_0]) -\lg ([B_{pp}]/[B_0])$ (here $\lg$ stands for $\log_{10}$).  
Using Equation (\ref{solution}), and  $f_\alpha^B = f_\alpha^A/S^\prime$, $ f_\beta^B = f_\beta^A$, $\theta^B = \theta^A$, $\phi^B = (S/S^\prime) \phi^A$ and $\psi^B = \psi^A/S$, 
      \begin{equation}
      \begin{array}{c}
       X = \lg \left(S S^\prime \right) + \vspace{1mm}\\
      \lg \left( \frac{
      \phi^A 
      +  \frac{Y \psi^A}{S^\prime} 
      \left(\frac{f^A_\alpha \phi^A}{S} + f^A_\beta \theta^A 
      + \frac{S^\prime-f^A_\alpha f^A_\beta}{S} {\theta^A \phi^A}\right)
      + \frac{\left(Y \psi^A \right)^2}{ S S^\prime} {\theta^A}
      }
      {\phi^A +  Y \psi^A  \left(f^A_\alpha \phi^A + f^A_\beta \theta^A + \left(1-f^A_\alpha f^A_\beta \right) \theta^A \phi^A \right)  + \left(Y \psi^A \right)^2 \theta^A }\right). 
      \end{array}
      \label{X-full}
      \end{equation}
      The two terms in $X$ describe separate contributions. The first represents  the difference in effective phosphorylation rates of $A$ and $B$  (see Supplementary Section S7). The second determines whether that difference in rates is  manifest in the overall yield of $A_{pp}$ and $B_{pp}$.
      
      \subsection*{The low kinase activity limit}
      To understand   Equation (\ref{X-full}), we first consider the limit in which phosphatases dominate over kinases ($Y \psi^A = {[K] k^A_{\rm eff}}/{[P] h^A_{\rm eff}} \rightarrow 0$).  The second term of Equation (\ref{X-full}) then tends to zero; using Equation (\ref{S-equation2}), the first term is
      \begin{equation}
      \begin{array}{c}
       X=
        \lg \left(S_0(1-P_{\rm react}^A) + P_{\rm react}^A\right) 
        \vspace{1mm} \\
        + \lg \left(S_0^\prime (1 - P_{\rm react}^{A \prime}) + P_{\rm react}^{A \prime}  \right).
      \end{array}
      \label{X-low-psiY}
      \end{equation}
      In the previous section, we argued that $f^A_\alpha = P_{\rm react}^{A \prime}$ when a single kinase catalyzes both phosphorylation steps. Thus specificity drops as $f_\alpha^A\rightarrow 1$; one might na\"{i}vely say that proofreading is compromised by pseudo-processivity  (although it is independent of $f^A_\beta$). This argument, however, is misleading in two ways. 
      
      Firstly, low specificity is {\it correlated} with pseudo-processivity, but not {\it caused by it} (increased pseudo-processivity does not lead mechanistically to a decrease in specificity). Frequent rebinding (due to high $P_{\rm on}$) is itself responsible.  To understand the distinction, note that rebinding only causes pseudo-processivity if a kinase is physically capable of catalyzing phosphorylation at both sites, as we have assumed hitherto.
Instead, we could consider a system with two chemically distinct kinase species (of equal concentration) and two chemically distinct phosphatase species (of equal concentration) that each can only interact with one of the two residues in question. Here, pseudo-processivity is impossible; $A \rightarrow A_{pp}$ requires the action of two distinct kinases. 
The new system is still governed by the differential equations implied by Figure \ref{second-order}, but primed rate constants (and underlying reaction probabilities) now refer to the action of the {\it second} enzyme, and $f_{\alpha}, f_{\beta}  = 0$ in this necessarily distributive system. Equation (\ref{solution}), with $f^A_{\alpha},f^A_{\beta}=0$, solves this system. Equation (\ref{X-low-psiY}) still holds, but now $P^{A \prime}_{\rm react} \neq f^A_{\alpha}  = 0$, as $P^{A \prime}_{\rm react}$ is a property of the second kinase and $f^A_{\alpha}$ is a property of the first. If  
the parameters are otherwise identical to the original single-kinase, single-phosphatase system, $ P^{A \prime}_{\rm react}$ and $X$ are unchanged (in the low yield limit considered here), despite the fact that now $f^A_{\alpha}= f^A_{\beta}=0$.
      
Thus pseudo-processivity itself is not required for the drop in specificity. Why, then, does $X$ drop as $P^{A \prime}_{\rm react} \rightarrow 1$?  $P^{A \prime}_{\rm react}$ is the likelihood of a successful reaction given proximal $K$ and $A_p$. For $P^{A \prime}_{\rm react} \rightarrow 1$, we require $P_{\rm on} \rightarrow 1$; Equation  (\ref{effective-rates}) shows that  $P^{A \prime}_{\rm react} < P_{\rm on}$ and $P^{A \prime}_{\rm react} \rightarrow 1$ as $P_{\rm on} \rightarrow 1$. We note that  $P^\prime_{\rm cat} \rightarrow 1$ is neither sufficient nor necessary; even with $P^\prime_{\rm cat}=1$, catalysis is largely distributive if rebinding is rare, and even inefficient catalysis can be pseudo-processive at high $P_{\rm on}$. When $P_{\rm on} \rightarrow 1$, there can be many rounds of dissociation and rebinding before modification occurs, favoring inferior substrates that are less likely to be catalyzed the first time. Mathematically (Equation (\ref{effective-rates})), we see that when $P^{\prime}_{\rm react} \rightarrow 1$, the dependence on the factor which distinguishes $A$ and $B$, $P^\prime_{\rm cat}$, is lost. In this low kinase activity limit, therefore, frequent rebinding (due to high $P_{\rm on}$) reduces specificity and can also coincidentally cause pseudo-processive behaviour if an enzyme can modify both sites.

      Secondly, the efficacy of proofreading is not $X$, but the increase in $X$ due to the  second site. Equation (\ref{X-low-psiY}) shows that the contribution from the first site is just as vulnerable to $P_{\rm on}$-driven increases in $P^{A}_{\rm react}$ as the contribution from the second site is to increases in $P^{A \prime}_{\rm react}$. A single-site substrate with the same properties as the first site of the two-site system has specificity
      \begin{equation}
      X_{\rm ss} = \lg(S) + \lg \left( \frac{1+ Y \psi^A \theta^A/S }{1 + Y \psi^A \theta^A} \right),
      \label{X-single-site}
      \end{equation} 
      in which  $\psi^A \theta^A = k^A_{\rm eff}/h^{A \prime}_{\rm eff}$ is the ratio of effective rate constants for phosphorylation and dephosphorylation. In the limit of low kinase activity, $Y \psi^A \rightarrow 0$, $X_{\rm ss} \rightarrow \lg(S) = \lg \left(S_0(1-P_{\rm react}^A) + P_{\rm react}^A\right)$, and the additional specificity due to the second site is $X-X_{\rm ss} \rightarrow \lg \left(S^\prime_0(1-P_{\rm react}^{A \prime}) + P_{\rm react}^{A \prime}\right)$. 
Clearly the contribution of the first site is compromised by $P_{\rm react}^A \rightarrow 1$ in the same way as the contribution of the second site is by   $P_{\rm react}^{A \prime} = f_\alpha^A \rightarrow 1$. It too suffers a loss of selectivity due to rebinding; Equations (\ref{k_eff eqn}) and (\ref{effective-rates}) show that $P_{\rm react}^A$ and $P_{\rm react}^{A \prime}$ have equivalent dependencies on $P_{\rm on}$. Thus the contribution of the second site  does not systematically fall off faster than the first as rebinding becomes more common  (the site with larger $P^A_{\rm cat}$ is more sensitive).      

We note that $X - X_{\rm ss} $ can remain substantial even when  pseudo-processivity is high ($f^A_\alpha \geq \frac{1}{2}$). For example, if  $S^\prime_0 =10$ (the intrinsic selectivity without rebinding is a factor of 10), $X - X_{\rm ss} $ drops from 1 in the limit $f^A_\alpha\rightarrow 0$ to 0.70 at $f^A_\alpha = \frac{1}{2}$, and only drops to 0.50 when $f^A_\alpha = 0.760$ (at which point the specificity is `halved' in the logarithmic sense; $[A_{\rm pp}]$ and    $[B_{\rm pp}]$ are distinguished by a factor of $\sqrt10$ rather than 10). For lower values of   $S^\prime_0$, this halving occurs at lower $f^A_\alpha$, but for higher values  it occurs even later.  Robustness of specificity is therefore clearly dependent on the intrinsic specificity  at low $P_{\rm on}$, but importantly pseudo-processive reactions do not {\it necessarily} preclude proofreading.

We now compare our results to the original work  of Swain and Siggia \cite{Swain2002}. The main results (Equations (4) and (6) of their paper) look quite different,  because they considered a distinct limit. They also considered a system with weak kinase activity, but treated the two stages of phosphorylation asymmetrically. They assumed that the success rate of phosphorylation once the kinase and substrate are bound is low for the first stage (the reaction is ``close to equilibrium"), but potentially not for the second stage. This assumption was made because the authors reasoned that it would be optimal in allowing the full selectivity from the first stage to be manifested, whilst permitting possible processive behaviour. Thus when Swain and Siggia allowed processive phosphorylation, they observed that the selectivity arising from the second stage  was compromised whereas that arising from the first was not. In our case, however, processivity arises from rebinding events which increase the probability of successful phosphorylation for {\it both} stages, compromising both $S$ and $S^\prime$ and incidentally leading to pseudo-processivity. This symmetry  does not arise naturally unless rebinding is explicitly modelled as the cause of pseudo-processivity.  
 
 Swain and Siggia state that proofreading is optimized at low processivity ($f_\alpha^A$ small). Whilst we do not contradict this result, we find that proofreading is more robust than this statement suggests. Specificity can be relatively high even when the majority of phosphorylations are pseudo-processive (in the low kinase activity limit, processivity of dephosphorylation reactions is irrelevant). Proofreading discriminates between two substrates, $A$ and $B$; even when phosphorylation of $A$ is moderately pseudo-processive, $B$ can still be less efficiently phosphorylated. 
      Furthermore, the second stage of phosphorylation is not more strongly affected than the first. 
        	This conclusion is the central result of this work. In what follows, we explore the consequences of finite kinase activity for this result, and then compare to other proposed uses of the dual phosphorylation motif.

        \subsection*{Finite kinase activity for distributive systems}

For finite kinase activity ( $Y \psi^A >0$), the second term in Equation (\ref{X-full}) is non-zero. We first consider the distributive limit of $f^A_\alpha = f^A_\beta =0$, which is obtained if $P_{\rm on} \rightarrow 0$ or by considering a system with two distinct kinases and two distinct phosphatases. In this case,
          \begin{equation}
      \begin{array}{c}
       X_{\rm dis} = \lg \left(S S^\prime \right) + \vspace{1mm}\\
       \lg \left( \frac{\phi^A + \left(Y \psi^A  \right) \theta^A \phi^A/S  + \left(Y \psi^A \right)^2 \theta^A/ S S^\prime}
      {\phi^A + \left( Y \psi^A \right) \theta^A \phi^A   + \left(Y \psi^A \right)^2 \theta^A } \right).
      \end{array}
      \label{X-no-processivity}
      \end{equation} 
       The second term in Equation (\ref{X-no-processivity}) is always negative. It captures the fact that finite concentrations of $A_p$ and $A_{pp}$ tend to suppress specificity, as the phosphorylation transitions $A \rightarrow A_p$ and $A_p \rightarrow A_{pp}$  become saturated for $A$ but not for $B$. If  $[A_{pp}] \approx [A_0]$, then a substantial change in $Y =[K]/[P]$ can hardly change $[A_{pp}]$, whereas the smaller $[B_{pp}] $ will still grow towards $[B_0]$, reducing the difference in yields. Similarly, if $[A_p] > [A]$, then increasing $Y$ can do little to convert more $A$ into $A_p$ whereas it will  have a larger effect on the $B \rightarrow B_p$ transition: large $[A_p]$ thus reduces the difference between substrates due to the first phosphorylation stage. 
       
       Specificity can therefore be compromised by high yields of phosphorylated products. But the efficacy of {\it proofreading} is perhaps best represented by  $X_{\rm dis} - X_{\rm ss} $. As is clear from Equation (\ref{X-single-site}), high kinase activity in a single-site system also suppresses specificity. To make a fair comparison, we therefore consider the two-site and single-site systems at the same yield of product $g$ ($g=[A_p]/[A_0]$ for the single-site system,  $g= [A_{pp}]/[A_0]$ for the two-site system) rather than at the same $Y =[K]/[P]$, as the yield of activated product is after all the output of the system.  Below, we express specificity in terms of the overall yield $g$ and parameters that depend only on the microscopic rate constants (eliminating $[K]$ and $[P]$): $\phi^A = k^A_{\rm eff}/k^{A\prime}_{\rm eff}$, $\theta^A = h^A_{\rm eff}/h^{A\prime}_{\rm eff}$, $S$ and $S^\prime$.
       \begin{equation}
       \begin{array}{c}
       X_{\rm ss} = \lg \left(S \right) + \lg \left((1-g)  +\frac{g}{S} \right), \vspace{2mm} \\
       X_{\rm dis} = \lg \left(S S^\prime \right) + \vspace{2mm} \\
       \lg \left((1-g) \left( \frac{1+Y_d\psi^A \theta^A/S}{1+Y_d \psi^A \theta^A} \right) +\frac{g}{S S^\prime} \right), \vspace{2mm} \\
        Y_d \psi^A \theta^A= \frac{ g \theta^A \phi^A + \sqrt{(g \theta^A \phi^A)^2 + 4g(1-g) \theta^A \phi^A }}{2(1-g)}.
       \label{Yd equation}
        \end{array}
        \label{X(g)}
       \end{equation}
       As $g \rightarrow 1$, the value of $S$ required to achieve a given specificity $X_{\rm ss}$ rises. When $g$ is large, $S= 1/(1-g)$ is required to give $X_{\rm ss} \approx \lg 2 $. This implies $S \geq 1/(1-g)$ is needed to discriminate between substrates by a factor of 2 at high yield, quantifying the magnitude of $S$ required to distinguish substrates at a given $g$. 
       
        \begin{figure}
        \centering
                \includegraphics[width=14pc, angle=-90]{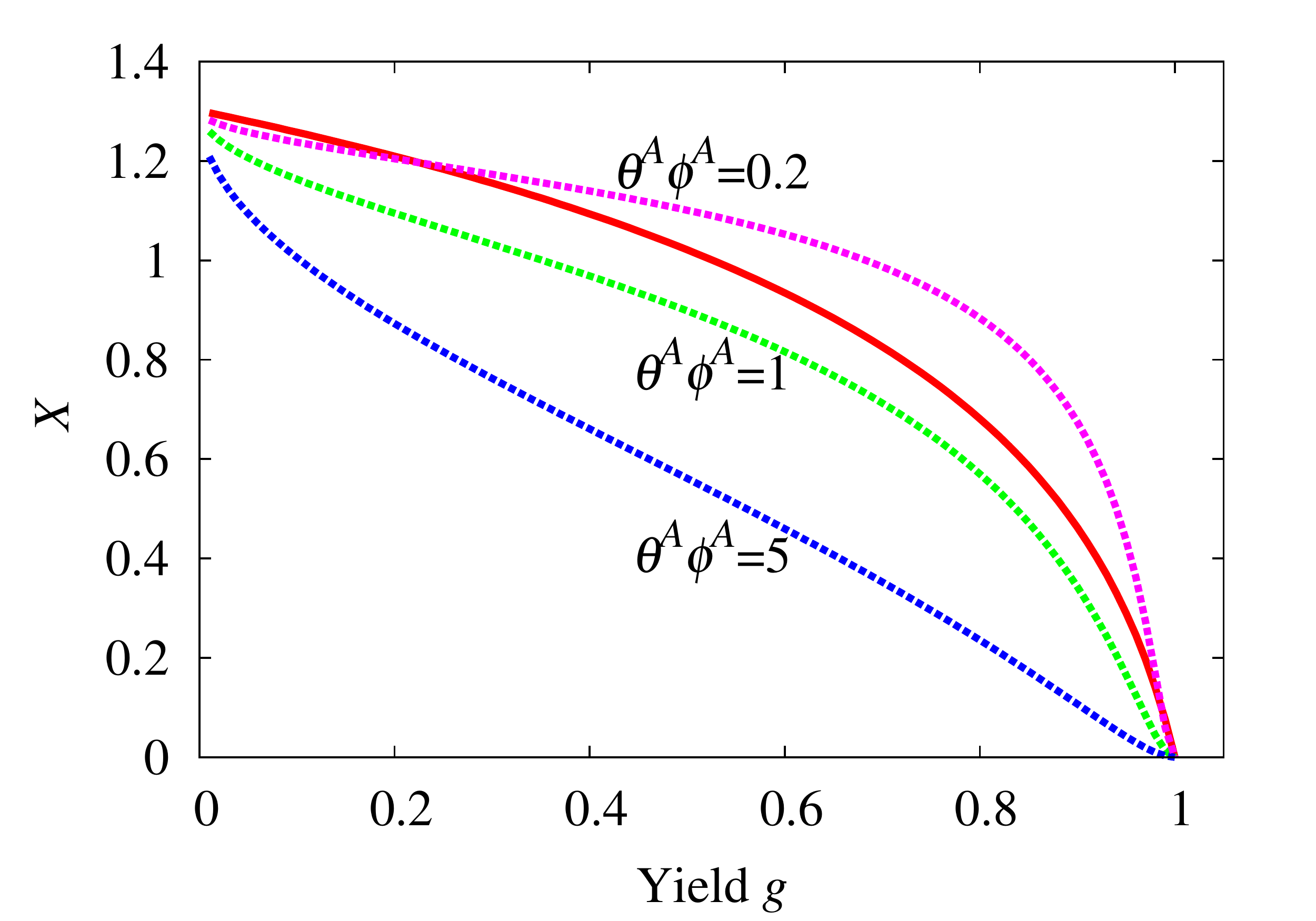}
        \caption{Drop in specificity of a single-site system ($X_ {\rm ss}$, solid line) and the additional specificity of the second site in a distributive system ($X_{\rm dis} - X_{\rm ss}$, dashed lines) with yield $g$. We use intrinsic selectivities  $S,S^{\prime} =20$. For the distributive system, we plot several values of $\theta^A \phi^A = k^A_{\rm eff} h^A_{\rm eff} / k^{A \prime}_{\rm eff} h^{A \prime}_{\rm eff} $ (defined in Figure \ref{model}). $X_{\rm dis} - X_{\rm ss}$ is more robust to high yields when $\theta^A \phi^A$ is small.
                \label{g-fig}}
\end{figure}

 Whether the second site's specificity $X_{\rm dis} -X_{\rm ss}$ is more strongly affected by $g$ than $X_{\rm ss}$ depends on  $Y_d \psi^A \theta^A = [K] k^A_{\rm eff}/ [P] h_{\rm eff}^{A\prime}$, or $[A_p]/[A]$ in the two-site system (Equation (\ref{solution})). If it is negligible, then $X_{\rm dis} - X_{\rm ss} > X_{\rm ss}$ for equal intrinsic selectivities $S=S^\prime$ (see Section S4 of the Supplementary material). Indeed, if $S \geq 1/(1-g)$, $X_{\rm dis} - X_{\rm ss} > \lg S^\prime -\lg 2$ (see Section S4), so the specificity of the second site is  weakly affected by $g$ when $[A_p]/[A]$ is small.
 However, if $Y_d\psi^A \theta^A = [A_p]/[A] \gg 1$, and g is not close to unity, $X_{\rm dis} - X_{\rm ss} \approx \lg(S^\prime/S)$. This is disastrous -- adding the second site eliminates the specificity  from the first. When $[A_p]/[A] \gg 1$, the fully unphosphorylated states are almost unoccupied, so we essentially have a single-site system based on the second stage $[A_p] \rightarrow [A_{pp}]$.
Equation (\ref{X(g)}) shows that, at fixed yield $g$,  $\theta^A \phi^A = k^A_{\rm eff} h^A_{\rm eff} / k^{A \prime}_{\rm eff} h^{A \prime}_{\rm eff}$ determines $Y_d \psi^A \theta^A$. Lower $\theta^A \phi^A$ is advantageous, as $A_p$ is rapidly converted into either $A_{pp}$ or $A$, keeping its concentration low. For $\theta^A \phi^A =1$,  $X_{\rm dis} -X_{\rm ss}$  is compromised  marginally more by $g$ than the $X_{\rm ss}$ (see Section S4). In Figure \ref{g-fig}, we show how $X_{\rm dis} - X_{\rm ss}$ falls off with $g$ for some representative values of $\theta^A \phi^A$, in comparison to $X_{\rm ss}$, illustrating this dependency of  $X_{\rm dis} - X_{\rm ss}$ on $\theta^A \phi^A$.

Overall, finite kinase activity in distributive systems reduces specificity, and the second site's contribution can be more vulnerable to high product yields than that of the first site. In the next section, we will consider pseudo-processivity. First, we study the effects of $P_{\rm on}$ without pseudo-processivity by considering a system with two distinct kinases and two distinct phosphatases. From the previous section, increasing $P_{\rm on}$ tends to reduce
$S$ and $S^\prime$ and hence specificity; here we instead examine the effect of finite  $P_{\rm on}$ on the sensitivity of single- and two-site systems to finite $g$. Equation (\ref{X(g)}) shows that whether the two-site system suffers more from finite yield as $P_{\rm on}$ increases depends on whether $\theta^A \phi^A=k^A_{\rm eff} h^A_{\rm eff} / k^{A \prime}_{\rm eff} h^{A \prime}_{\rm eff}$ grows or shrinks with $P_{\rm on}$.
     
         High values of $P_{\rm on}$ tend to make all reactions equally fast by allowing multiple attempts for intrinsically inefficient reactions  (Equations (\ref{k_eff eqn}) and (\ref{effective-rates})). Consequently, $\theta^A \phi^A \rightarrow 1$ as $P_{\rm on} \rightarrow 1$. Rebinding thus makes systems that are intrinsically robust to finite $g$ (with low  $\theta^A \phi^A$ as $P_{\rm on} \rightarrow 0$) become less so, but makes systems that are intrinsically vulnerable to finite $g$ (with high  $\theta^A \phi^A$ as $P_{\rm on} \rightarrow 0$) become more robust. Rebinding makes it more challenging to evolve a system with low $\theta^A \phi^A$ (and therefore a low concentration of $A_{p}$), and the consequences of rebinding for specificity can be substantial if the intrinsic ($P_{\rm on} \rightarrow 0$) value of $\theta^A \phi^A$  is very different from unity. Nonetheless, proofreading can remain effective for systems with $\theta^A \phi^A \approx 1$ at high yields, as discussed in  Section S4, provided the selectivity is not as small as $S^\prime \sim 1/(1-g)$.

    \subsection*{Finite kinase activity for pseudo-processive systems}
    We now consider finite kinase activity for systems with the potential for pseudo-processivity. $X_{\rm proc}$, the specificity with $f^A_\alpha,f^A_\beta \neq 0$, can be written in terms of $g$ and parameters that depend only on the rate constants: $\phi^A = k^A_{\rm eff} /k^{A \prime}_{\rm eff} $, $\theta^A= h^A_{\rm eff} /h^{A \prime}_{\rm eff} $, $S$, $S^\prime$, $f^A_{\alpha}$ and $f^A_{\beta}$.        
            \begin{equation}
            \begin{array}{c}
       X_{\rm proc} = \lg \left(S S^\prime \right) \vspace{1mm}+\\
         \lg \left((1-g) \left( \frac{1+\frac{Y_p\psi^A \theta^A}{S} \left( 1 + \frac{f^A_\beta S}{\phi^A S^\prime} - \frac{f^A_\alpha f^A_\beta}{S^\prime}\right)}{1+{Y_p\psi^A \theta^A} \left( 1 + \frac{f^A_\beta }{\phi^A} -{f^A_\alpha f^A_\beta}\right)} \right) +\frac{g}{S S^\prime} \right). \vspace{4mm}\\
         {Y_p\psi^A \theta^A} =   \frac{ g\theta^A f^A_\beta +  g \theta^A \phi^A(1-f^A_\alpha f^A_\beta) -(1-g)\phi^A f^A_\alpha}{2(1-g)} + \vspace{1mm} \\
          \frac{\sqrt{(g\theta^A f^A_\beta +  g \theta^A \phi^A(1-f^A_\alpha f^A_\beta) -(1-g)\phi^A f^A_\alpha)^2 + 4g(1-g) \theta^A \phi^A}}{2(1-g)} .
        \end{array}
        \label{X(g)p}
       \end{equation}
       The expression reduces to $X_{\rm dis}$ if $f^A_\alpha = f^A_\beta =0$, so we need only study the consequences of  $f^A_\alpha, f^A_\beta >0$. Firstly, $\partial X_{\rm proc} / \partial f^A_\alpha \geq 0$ (with $g,S,S^\prime,\theta^A, \phi^A,f_\beta^A$ fixed, see Supplementary Section S5). Thus finite $f_\alpha^A$   reduces the effect of finite yield $g$; it is {\it always} better to have a single (potentially pseudo-processive) kinase than two distinct kinases (implying $f^A_\alpha=0$) with otherwise identical parameters. This is because converting $A$ directly to $A_{pp}$ helps to avoid the buildup of $A_p$ which  reduced $X_{\rm dis}$ in the previous section. Note that  $\partial X_{\rm proc} / \partial f^A_\alpha \geq 0$  {\it does not} imply that higher $P_{\rm on}$, which will cause increased $f^A_{\alpha}$, is  always beneficial provided $f^A_{\beta}=0$; increased rebinding will also tend to reduce $S$ and $S^\prime$, and will  influence $\theta^A$ and  $\phi^A$.

 $f_\beta^A$, the degree of pseudo-processivity in dephosphorylation, is more ambiguous. When it appears in  $-f^A_\alpha f^A_\beta$ terms, it too reduces  the buildup of $A_p$. When it appears separately from $f_\alpha^A$, however, it reduces $X_{\rm proc}$. The physical explanation, discussed in detail in Supplementary Section S6, is subtle. Here, we simply note that pseudo-processivity in the {\it dephosphorylation} pathway, rather than in the phosphorylation pathway or rebinding (high $P_{\rm on}$) itself, can compromise specificity and proofreading when $Y_p\psi^A \theta^A f^A_\beta/\phi^A$ and $f^A_\beta/\phi^A$ are large.  
 We outline the parameter space for which this sensitivity to processive dephosphorylation is strong in Section S6, where we show that having phosphatases that are intrinsically less efficient than kinases is sufficient (but not necessary) to inhibit this sensitivity.  Although interesting, we focus on the majority of parameter space where this unwanted behavior is weak.
 
Overall, we find that finite $f^A_\alpha$ never reduces specificity relative to otherwise equivalent systems with distributive phosphorylation. Processive dephosphorylation can compromise specificity, but outside of a regime of strong sensitivity to $f^A_\beta$, potentially pseudo-processive systems are not worse than distributive systems with equivalent microscopic parameters. The specificity provided by the second site in either case is generally more sensitive to higher yield $g$ than that of the first site, due to the possibility of saturating the $A \rightarrow A_p$ transition prior to the $A_p \rightarrow A_{pp}$ transition. It is also harder to avoid this saturation through low values of 
$\theta^A \phi^A =k^A_{\rm eff} h^A_{\rm eff} / k^{A \prime}_{\rm eff} h^{A \prime}_{\rm eff} $ when $P_{\rm on}$ is high. However, in general the earlier results still hold: the loss of specificity with increased $P_{\rm on}$ is primarily associated with rebinding itself (and hence high reaction probabilities), rather than pseudo-processivity; the selectivity of both the first and second sites are compromised by rebinding;  and the additional contribution from the second site can remain significant even when the system is substantially pseudo-processive ($f_{\alpha}^A, f_{\beta}^A \geq \frac{1}{2}$), particularly if intrinsic ($P_{\rm on} \rightarrow 0$) specificities are high. 
               \begin{figure}
        \centering
                \includegraphics[width=14pc, angle=-90]{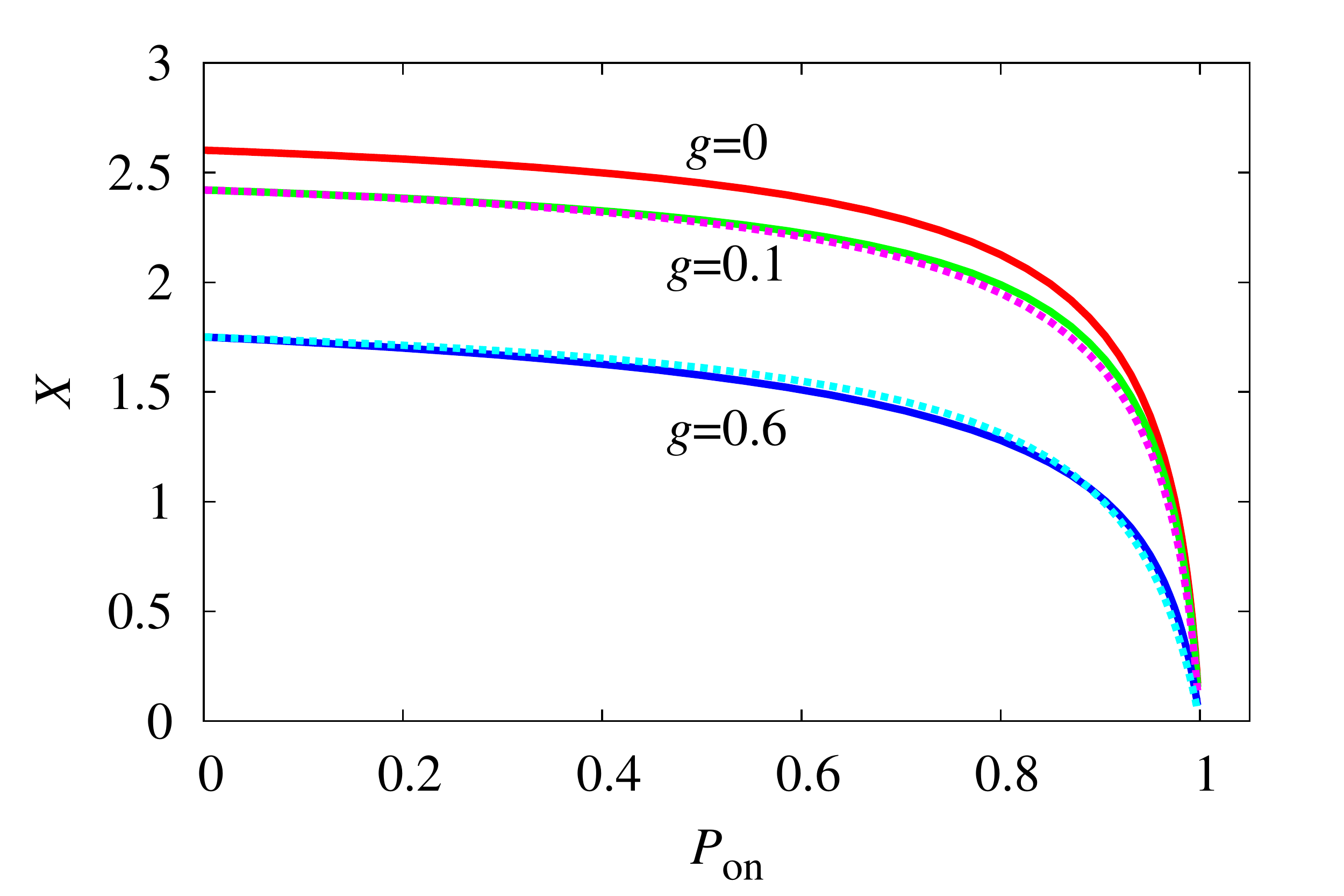}
        \caption{Drop of specificity with $P_{\rm on}$ and yield $g$ for the system outlined in the text, showing the similarity between pseudo-processive systems (solid lines) and distributive systems with two kinases and phosphatases (dashed lines). We consider activated substrate yields of $g= 0$, 0.1 and 0.6. The two systems give identical results for $g= 0$, and so only a single solid line is plotted. 
        }
        \label{calculation-1}
\end{figure}

\begin{figure}
        \centering
                \includegraphics[width=14pc, angle=-90]{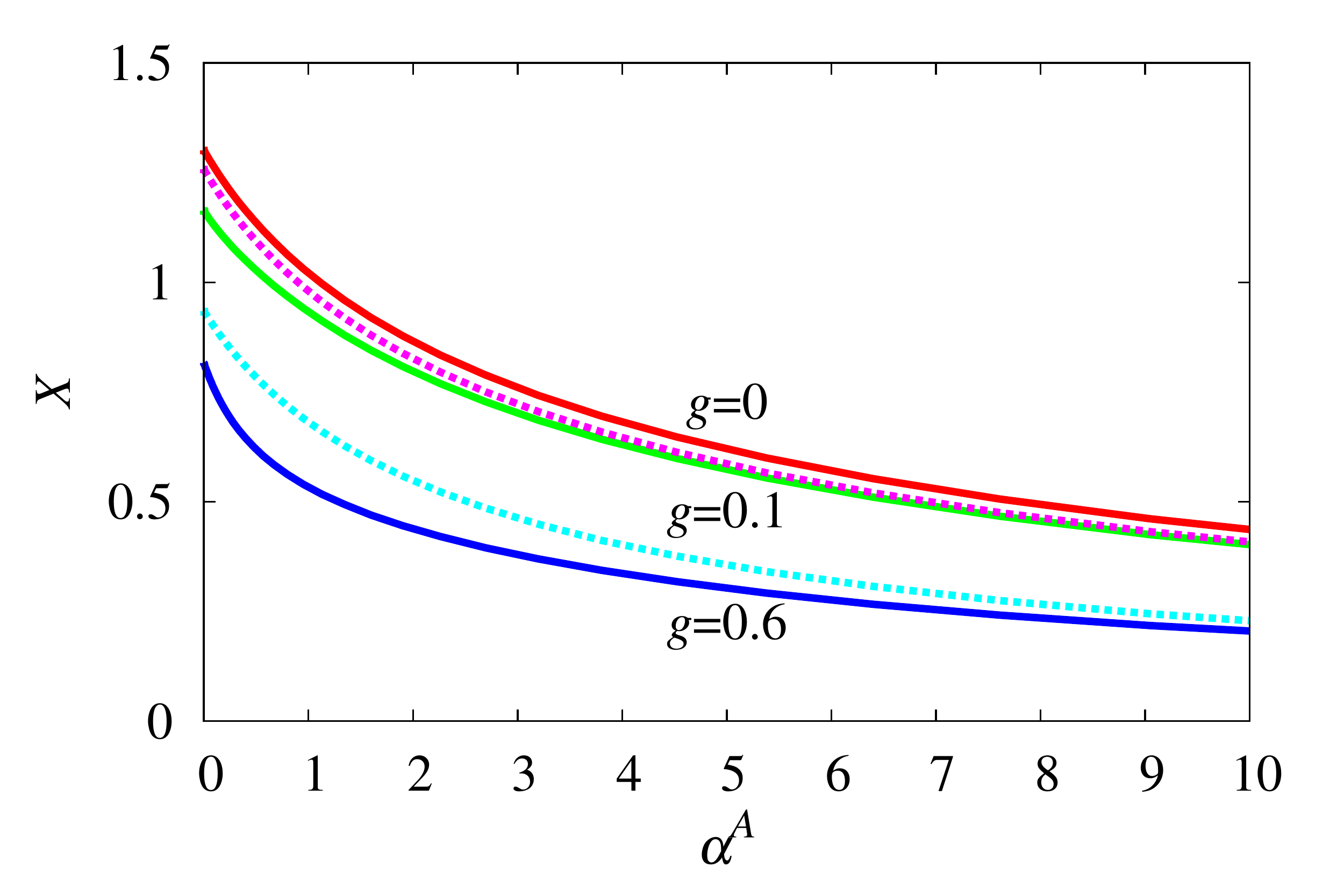}
        \caption{Contributions to $X$ of the first and second sites against $\alpha^A = f_\alpha^A/(1-f_\alpha^A)$, the ratio of processive to non-processive phosphorylations for the system outlined in the text. We plot $X_{\rm ss}$ (dashed lines) and $X_{\rm proc}-  X_{\rm ss}$ (solid lines) for activated substrate yields of $g= 0$, 0.1 and 0.6. $X_{\rm ss}$ is the specificity in a system with only the first phosphorylation site. The two curves are identical for $g= 0$,  so we plot only a single solid line. The contribution of the second site is somewhat more sensitive to $g$, but sensitivity to $P_{\rm on}$ is similar. Specificity, and the contribution of the second site, can remain significant at $\alpha^A > 1$. 
       }
        \label{calculation-3}
\end{figure}

We plot characteristic behavior in Figures \ref{calculation-1}
 and \ref{calculation-3}. Conceptually, we consider  a system with fixed microscopic enzymatic rate constants, but in which the rate of diffusion with respect to binding can be modulated (by adding crowders, for example). In this picture, all catalysis probabilities ($P^A_{\rm cat}$ {\it etc.}) are constant, and $P_{\rm on}$ is variable. We then ask how $X$ varies with $P_{\rm on}$ at fixed $g$ (to provide a fair comparison). 
We take $P^A_{\rm cat}, P^{A \prime}_{\rm cat}, Q_{\rm cat}, Q^\prime_{\rm cat}=0.2$,  $P^B_{\rm cat},P^{B \prime}_{\rm cat} =0.01$ to provide representative plots. Other parameter choices are shown in Supplementary Section S9. We consider yields $g =0$, 0.1 and 0.6. In Figure \ref{calculation-1}, we plot $X_{\rm proc}$ and $X_{\rm dis}$ as a function of $P_{\rm on}$, showing that although specificity drops with $P_{\rm on}$ and increased $g$, it also does so when each stage requires its own kinase and phosphatase and pseudo-processivity is impossible. 

In   Figure \ref{calculation-3}, we plot $X_{\rm proc}-X_{\rm ss}$ and $X_{\rm ss}$ parametrically against the ratio of processive to non-processive reactions, $\alpha^A = f_\alpha^A/(1-f_\alpha^A)$. Both contributions to specificity drop with increased  $\alpha^A$ (which itself rises with $P_{\rm on}$), and increased $g$, but the additional specificity of the second site is somewhat more sensitive to finite $g$.  Figure \ref{calculation-3} further demonstrates that contribution of the second site to specificity can remain appreciable at $\alpha^A \geq 1$ ($f_\alpha^A \geq \frac{1}{2}$), even at fairly high yields of $A_{pp}$. For illustrative purposes we have chosen $S= S^\prime =20$ as $P_{\rm on} \rightarrow 0$. Higher values would make specificity at both stages more robust to increased $\alpha$ and yield $g$.

Phosphorylation kinetics, as well as the steady-state, could be important. Following a sudden activation of upstream kinases, $[A_{pp}]/[A_0]$ and $[B_{pp}]/[B_0]$ always initially rise in a ratio $S S^\prime$ (see Section S7). Thus finite kinase activity does not compromise the difference in phosphorylation rates -- only whether this difference is manifested in the steady-state yield.

\subsection*{Comparison with other proposed advantages of dual phosphorylation}
As outlined in Section S8 of the Supporting Material, the robustness to pseudo-processivity of ultrasensitivity and the use of dual phosphorylation to favor scaffold-mediated pathways 
can be treated with the same simple model. 
Firstly, we can show (as others have \cite{Salazar2009,Takahashi2010}) that ultrasensitivity arising from dual phosphorylation is always small when $f_\alpha \,{\rm and}\, f_\beta \geq \frac{1}{2}$. Ultrasensitivity can be fairly robust when either $f_\alpha \,{\rm or}\, f_\beta \geq \frac{1}{2}$ individually,
provided that the second stage of the processive reaction is intrinsically faster than the first. 
With regard to the use of dual phosphorylation to favor scaffold-mediated pathways, 
the ratio of scaffold-derived $A_{pp}$ to that produced without a scaffold is limited to $1/f_\alpha$, unless factors independent of dual phosphorylation are relevant. 
If the mechanism in the cytosol is purely distributive, the scaffold-derived yield 
can be arbitrarily larger, but  $f_\alpha \approx \frac{1}{2}$ almost completely eliminates the advantage of scaffolds in this context.

Unlike proofreading, these alternative uses are generally compromised by pseudo-processivity itself, rather than rebinding (high $P_{\rm on}$). As with proofreading, we can imagine a system with identical parameters, but containing two distinct species of phosphatases and upstream kinases  that each can only catalyze one step. For the alternatives uses, the effects of increasing $P_{\rm on}$ 
are substantially alleviated if pseudo-processivity is prohibited  in this way (see Section S8). Ultrasensitity (and scaffold-mediated enhancement) require kinases and phosphatases to compete against each other at two separate stages when activating/deactivating substrates in the cytosol, whereas proofreading requires two stages at which substrate $A$ can be discriminated from  $B$. The first requirement can be met  even when $P_{\rm on}$ is high by having distinct kinases and phosphatases for each stage, whereas this does not help to discriminate $A$ from $B$.

\section*{DISCUSSION}
We have studied the effect of rebinding-induced pseudo-processivity on proofreading via dual phosphorylation in the linear regime. Whilst specificity drops as pseudo-processivity increases, this is generally due to a loss of enzymatic selectivity through rebinding, rather than pseudo-processivity itself. High binding probabilities leading to multiple rebinding events reduce the difference in phosphorylation rates between good and poor substrates, and can incidentally increase pseudo-processivity.
We contrast this with other proposed advantages of dual phosphorylation, namely ultrasensitivity and the ability to enhance scaffold-mediated signalling pathways. These alternative uses for dual phosphorylation are specifically compromised by pseudo-processivity. This distinction is not academic --  it might be easier for a cell to prevent pseudo-processive behavior ({\it e.g.} via a finite refractory period of a kinase after phosphorylation \cite{Dushek2011,Takahashi2010}) than to prevent rebinding after a failed reaction.  We find that eliminating pseudo-processive {phosphorylation} in this way would always be beneficial for ultrasensitivity and the ability to enhance scaffold-mediated signalling, but never for proofreading (Sections S5 and S8). However, pseudo-processivity itself in the {\it dephosphorylation} pathway can compromise specificity under certain conditions.
 Although rebinding might be difficult to control through evolution, the reduction in both ultrasensitivity and specificity with $P_{\rm on}$ could be  tested in vitro by varying the concentration of an inert crowding agent \cite{Aoki2011}. The distinction between $P_{\rm on}$ and pseudo-processivity would also be testable with kinases modified to reduce nucleotide release rates. 

The efficacy of proofreading is primarily related to the additional benefit in specificity obtained by adding a second site. We find that even when reactions become  significantly pseudo-processive due to rebinding, the addition of the second site can still provide a substantial relative boost to specificity, meaning that proofreading is still useful. In fact, as pseudo-processivity can only occur in parallel with a reduced intrinsic selectivity for a single site, proofreading via multi-site phosphorylation is even more important in maintaing specificity. 

Proofreading is based on the {\it difference} between two pathways, and a poor
substrate can still be less efficiently phosphorylated even if some  discrimination is lost through rebinding (and finite product yields). The degree to which this is true depends, of course, on the intrinsic discrimination without rebinding. The other uses of dual phosphorylation considered here, however, depend on the properties of a single pathway and are  fundamentally limited by moderate pseudo-processivity in that pathway. We would therefore argue that if a signalling cascade is observed to be significantly pseudo-processive in experiment, its dual phosphorylation motif is most likely used for proofreading.

We have considered an extremely simple model without spatial resolution. To test this simplification, we simulate a lattice model in Section S3, reproducing basic results. We also demonstrate that our approach is consistent with limits of a reaction-diffusion description (Section S2). We have also neglected long-lived enzyme/substrate complexes due to the increased number of relevant parameters and non-linearities in equations. Analytic results in the unsaturated linear regime are valuable for three reasons. Firstly, the biophysical principles underlying our conclusions are still relevant in the non-linear regime;  our basic findings are thus likely to be widely applicable. Indeed, we have considered finite complex concentrations for a few systems in Section S12 of the Supplementary Material; moderate concentrations of enzyme/substrate complexes have only a weak effect, and  we do not see evidence that reactant saturation invalidates our previous findings on proofreading.
 Secondly, detailed analysis of the linear regime is an important first step in comprehending the full non-linear system, and the analytic results presented here will help to frame the the findings of future work into rebinding and pseudo-processivity in the general case. Finally, although some authors have argued for substantial saturation in kinase cascades \cite{Huang1996, Ferrell1997},  recent work has suggested that MAPK cascades can function in regimes in which the reactants are not strongly saturated \cite{Aoki2011}. Our analysis in the linear regime is therefore not only instructive, but of direct biological relevance.
Nonetheless, the effect of pseudo-processivity in non-linear systems remains an important open question. To explore the accuracy of our modeling of a crowded environment, explicit simulations (analogous to recent work on transcription factors \cite{Brackley2013}) would also be beneficial.

\section*{Acknowledgements}
We thank Chris Govern for a careful reading of the manuscript. This work is part of the research program of the ``{Stichting voor Fundamenteel Onderzoek der Materie} (FOM)", which is financially supported by the ``{Nederlandse} Organisatie voor Wetenschappelijk Onderzoek (NWO)". TEO was funded by University College, Oxford.

\section*{Supporting citations}
References \citep{Elf2004, Montroll1965, Levchenko2000} appear in the Supporting Material.

\makeatletter 
\renewcommand{\thefigure}{S\@arabic\c@figure} 
\renewcommand{\thetable}{S\@arabic\c@table} 
\renewcommand{\theequation}{S\@arabic\c@equation} 

\begin{appendix}
\section{Requirements for unsaturated kinetics}
For all reactants to be unsaturated, we require that the total time spent with reactants in close proximity/bound is short compared to the time between encounters. Consider the $A \rightarrow A_p$ reaction. The average time between encounters with kinases for any given $A$ molecule is ${1}/{k_{\rm D} [K]}$, and the average time between encounters with an $A$ molecule for a given kinase is $1/k_{\rm D} [A] > 1/k_{\rm D} [A_0]$. These two times need to be large compared to the average time that it takes for the system to resolve -- either the reaction $A \rightarrow A_p$ must occur, or the reactants must diffuse apart. This time can be calculated by solving differential equations for a system initiated in $K \circ A$, with absorbing boundary conditions when the proteins diffuse apart or catalysis occurs. 
\begin{equation}
\begin{array}{c}
\frac{\rm d}{{\rm d} t} [K \circ A] = - (k_{\rm a} + k_{\rm esc}) [K \circ A]  + k_{\rm d} [KA] \vspace{3mm}\\
 \frac{\rm d}{{\rm d} t} [K A] = - (k_{\rm cat} + k_{\rm d}) [K A]  + k_{\rm a} [K \circ A] 
\end{array}
\end{equation}
These coupled differential equations can be solved by standard methods, yielding
\begin{equation}
\begin{array}{c}
[K \circ A](t) = \frac{1}{\lambda_+ - \lambda_-} \vspace{1mm}\\
\left( (\lambda_+ + k_{\rm d} + k_{\rm cat}) \exp(\lambda_+ t) - (\lambda_- + k_{\rm d} + k_{\rm cat})\exp(\lambda_- t) \right), \vspace{3mm}\\
\left[K A \right](t) = \frac{k_{\rm a}}{\lambda_+ - \lambda_-} \left( \exp(\lambda_+ t) - \exp(\lambda_- t) \right), 
\end{array}
\end{equation}
where 
\begin{equation}\begin{array}{c}
\lambda_{\pm} = -\frac{k_{\rm d} + k_{\rm cat} +k_{\rm a} + k_{\rm esc}}{2} \pm  \vspace{1mm}\\ \frac{\sqrt{(k_{\rm d} + k_{\rm cat} +k_{\rm a} + k_{\rm esc})^2 - 4(k_{\rm esc}k_{\rm d}+k_{\rm esc}k_{\rm cat}+k_{\rm a}k_{\rm cat})}}{2}. 
\end{array}
\end{equation}
The average time prior to either escape or catalysis can then be calculated through
\begin{equation}
\langle t \rangle = - \int_0^{\infty} {\rm d}t \,  t \frac{\rm d}{{\rm d} t} \left( [K \circ A] +  [K A] \right).
\end{equation}
Performing the integral yields
\begin{equation}
\langle t \rangle =  \frac{k_{\rm d} + k_{\rm cat} + k_{\rm a}} {k_{\rm esc}k_{\rm d} + k_{\rm esc}k_{\rm cat} +k_{\rm a}k_{\rm cat} },
\end{equation}
as quoted in the main text. For unsaturated kinetics to hold, we require that the equivalent quantities for all reactions are small compared to all encounter times. It is worth noting that, if $k_{\rm esc} > k_{\rm cat}$,
\begin{equation}
\langle t \rangle <  \frac{k_{\rm d} + k_{\rm cat} + k_{\rm a}} {k_{\rm cat}k_{\rm d} + k_{\rm cat}k_{\rm cat} +k_{\rm a}k_{\rm cat}} = \frac{1}{k_{\rm cat}}.
\end{equation}
Similarly, if  $k_{\rm esc} < k_{\rm cat}$
\begin{equation}
\langle t \rangle <  \frac{k_{\rm d} + k_{\rm cat} + k_{\rm a}} {k_{\rm esc}k_{\rm d} + k_{\rm esc}k_{\rm cat} +k_{\rm a}k_{\rm esc}} = \frac{1}{k_{\rm esc}}.
\end{equation}
Consequently, $\langle t \rangle \leq {\rm max}(1/k_{\rm cat}, 1/k_{\rm esc})$.

\section{The model's description of rebinding}
\subsection{Mapping to a continuum model of diffusion}
The model's description of rebinding is very simple, which allows it to be analysed quantitatively and understood qualitatively. Despite this simplicity, it can be directly mapped to the results of a previous analysis of rebinding in which diffusion is explicitly treated with a conventional continuous diffusion equation.\cite{Gopich2014} In that work, the authors showed that the rate of change of concentration of a substrate $A$ with $N$ phosphorylation states due to the action of a kinase $K$ can be approximated by
\begin{equation}
\frac{{\rm d} [{\bf A}]}{{\rm d} t} = - [K] {\bf M} [{\bf A}] 
\end{equation}
in the dilute limit, where $[{\bf A}]$ is a vector containing the concentrations of each phosphorylation state, and ${\bf M}$ is a constant matrix. ${\bf M}$ is  related to ${\bf M}^{\rm rl}$, a matrix whose only non-zero elements are ${\bf M}^{\rm rl}_{i,i} = -{\bf M}^{\rm rl}_{i+1,i} = \kappa_i$, the rate constants for phosphorylation reactions if diffusion were infinitely fast, by 
\begin{equation}
{\bf M} = k_{\rm diff}{\bf M}^{\rm rl} ({\bf M}^{\rm rl}  + k_{\rm diff} {\bf I})^{-1}.
\end{equation}
Here $k_{\rm diff}$ is Smoluchowski's diffusion-limited rate constant. For our system, in which there are three phosphorylation states and two intrinsic phosphorylation rate constants $\kappa$ and $\kappa^\prime$,
\begin{equation}
{\bf M}^{\rm rl} = 
\left(
\begin{array}{c c c}
\kappa & 0 & 0\\
-\kappa & \kappa^\prime & 0 \\
0 &-\kappa^\prime & 0
\end{array}
\right),
\end{equation}
which implies
\begin{equation}
{\bf M} = 
\left(
\begin{array}{c c c}
\frac{\kappa k_{\rm diff}}{\kappa + k_{\rm diff}} & 0 & 0\\
-\frac{\kappa k_{\rm diff}}{\kappa + k_{\rm diff}} \left(1 - \frac{\kappa^\prime}{\kappa^\prime + k_{\rm diff}} \right)&\frac{\kappa^\prime k_{\rm diff}}{\kappa^\prime + k_{\rm diff}}  & 0 \\
-\frac{\kappa k_{\rm diff}}{\kappa + k_{\rm diff}} \left(\frac{\kappa^\prime}{\kappa^\prime + k_{\rm diff}} \right) &-\frac{\kappa^\prime k_{\rm diff}}{\kappa^\prime + k_{\rm diff}}  & 0
\end{array}
\right).
\label{Gopich}
\end{equation}
The appearance of the term in the lower left hand corner corresponds to the possibility of rebinding-induced pseudo-processivity, as $A$ can be converted directly to $A_{pp}$.

In our simple model, no matter how efficient the reactions once the enzymes are in close proximity, overall rate constants are limited by diffusion to $k_{\rm D}$, and thus it plays the same role as $k_{\rm diff}$ in the description of Gopich and Szabo.\cite{Gopich2014} In the limit of infinitely fast diffusion, rate constants in our model would be given by a rate of reaction given close proximity ($k_a P_{\rm cat}$, $k_a P^\prime_{\rm cat}$), multiplied by $k_{\rm D}/k_{\rm esc}$ (which essentially gives the probability that the reactants are in close proximity). Thus a natural mapping between the quantities appearing in Equation (\ref{Gopich}) and those in our model is
\begin{equation}
\begin{array}{c}
k_{\rm diff} \rightarrow k_{\rm D}.\\
\kappa \rightarrow k_a P_{\rm cat} k_{\rm D}/k_{\rm esc}.\\
\kappa^\prime \rightarrow k_a P^\prime_{\rm cat} k_{\rm D}/k_{\rm esc}.
\label{mapping}
\end{array}
\end{equation}
We note that in reaction-diffusion descriptions such as Ref. \onlinecite{Gopich2014}, quantities such as the reaction-limited rate $\kappa$
(with units of M$^{-1}$s$^{-1}$) arise naturally in the definition of the model, whereas for descriptions in which there is an explicit treatment of a contact state or a state of close proximity, the reaction rate given contact/close proximity $k_a$ (with units of s$^{-1}$) is natural. This is because in the reaction-diffusion picture, reactions occur at a two-dimensional surface, rather than from some close-proximity state of finite volume as is implicitly assumed in models such as the one used here. Equation (\ref{mapping}) shows how the two descriptions are related. 

Substituting into Equation (\ref{Gopich}), rearranging, and using the quantities defined in the main text, we find
\begin{equation}
{\bf M} = 
\left(
\begin{array}{c c c}
k_{\rm eff}  & 0 & 0\\
-k_{\rm eff} (1- f_\alpha)& k_{\rm eff}^\prime& 0 \\
-k_{\rm eff} f_\alpha &-k_{\rm eff}^\prime  & 0
\end{array}
\right).
\label{Gopich-us}
\end{equation}
The resultant rate equations for changes in states of $A$ due to kinase action are identical to those in Figure 3 of the main text. The same result holds for other combinations of substrate and enzyme. Thus we conclude that our simple model can be directly mapped to the results of Gopich and Szabo for systems in which the reactants are dilute.\cite{Gopich2014} 

\subsection{Increasing rebinding within the model}
Within the framework of Gopich and Szabo\cite{Gopich2014}, rebinding is made more likely by reducing $k_{\rm diff}$ relative to $\kappa$ and $\kappa^\prime$, thereby slowing diffusion relative to intrinsic reaction rates. One could directly implement this in our model by reducing $k_{\rm D}$, and keeping $k_{\rm a}$, $k_{\rm esc}/k_{\rm D}$, $P_{\rm cat}$ and $P_{\rm cat}^\prime$ constant, as can be seen from Equation (\ref{mapping}). 

Our model, however, is also intended to be relevant to systems involving crowded environments (in which the crowders slow diffusion). It is well-known that crowding affects not only the reaction kinetics but also the dissociation constant of binding, at least in part due to depletion forces. There is therefore no reason, in our system, to require that $k_{\rm D}/k_{\rm esc}$ is held constant. Furthermore, although the discussion in the main text assumes that $k_{\rm a}$ does not change as rebinding becomes more likely, one might postulate that depletion interactions could also modulate this parameter. However, the exact details of how the parameters $k_{\rm D}$, $k_{\rm esc}$ and $k_{\rm a}$ change is unimportant: as we show in the main text, the results are actually independent of $k_{\rm D}$, and only depend on $k_{\rm esc}$ and $k_{\rm a}$ through $P_{\rm on}$. As all results in the main text are discussed in terms of the consequences of varying $P_{\rm on}$, our conclusions do not depend on how $k_{\rm D}$, $k_{\rm esc}$ and $k_{\rm a}$ change individually to achieve said variation in $P_{\rm on}$.

These considerations, however, serve to highlight that the model is only the simplest possible description of a complex process.
As outlined in the discussion of the main text, explicit simulation of multisite phosphorylation in crowded environments will establish the accuracy of the picture presented here,
and establish whether the phenomenological parameters can be directly related to the underlying physics. We do show in Section \ref{simulations}, however, that our results are robust to simulation with a lattice model.

\section{Simulations of a lattice-based model}
\label{simulations}
To prove that the results of the simple model presented in the text are robust to the details of diffusion, we simulate a lattice-based model of phosphorylation and dephosphorylation for two substrates $A$ and $B$. We treat the proteins as existing on a finite cubic lattice, with diffusion modelled as `hops' between adjacent lattice sites -- one could take these hops to model jumps between adjacent regions enclosed by crowders. When proteins are on the same lattice site, reactions can occur.  All proteins are assumed to hop to a given adjacent site at the same rate $r_{\rm hop}$, and bind at the same rate $r_{\rm bind}$ if they are on the same lattice site. As in the simple model in the main text, we ignore the finite lifetime of protein complexes, thereby prohibiting saturation -- binding events  instantly result in catalysis or resolution back into the reactants. As a result, enzymes and substrates in the same cell undergo catalysis with a rate given by $r_{\rm bind}$ multiplied by the appropriate catalysis probability. As an example, we consider  $P^{A}_{\rm cat}=0.1$, $P^{A \prime}_{\rm cat}=0.2$, $P^{B}_{\rm cat}=0.01$, $P^{B \prime}_{\rm cat}=0.02$, $Q_{\rm cat}=0.1$ and $Q^{\prime}_{\rm cat}=0.3$. 

We consider a range of reactant concentrations  as outlined in Table \ref{table-simulations}, and ratios $r_{\rm bind}/r_{\rm hop} = 1 \rightarrow 10^{-3}$. Reactant concentrations are deliberately chosen to preserve the ratio of kinases to phosphatases $Y$. For each set of parameters, we perform 6 simulations of $1.6\times 10^{11}$ steps (following $4\times 10^9$ steps of initialization) using the event-driven algorithm of Elf and Ehrenberg\cite{Elf2004} to measure the specificity $\log([A_{pp}]/[B_{pp}])$. We also simulate an identical  set of systems, but with two distinct species of kinase and phosphatase, each able to catalyze only one stage of phosphorylation/dephosphorylation.

\begin{table}
\begin{center}
\begin{tabular}{c c  c  c  c c}
Identity label & Lattice size & \multicolumn{4}{c}{Total number of}\\
& & $A$ & $B$ & $X$ & $P$ \\
\hline
(a) &  50x50x50 & 100 & 100 & 28 & 64\\
(b) & 20x20x20 & 100 & 100 & 28 & 64\\
(c) & 70x70x70 & 25 & 25 & 7 & 16\\
(d) & 50x50x50 & 50 & 50 & 112 & 256\\
\end{tabular}
\end{center}
\caption{\footnotesize{Lattice dimensions and total numbers of proteins used in simulations. In all cases, $Y=0.4375$.}}
\label{table-simulations}
\end{table}

\begin{figure}[!]
\includegraphics[width=15pc, angle=-90]{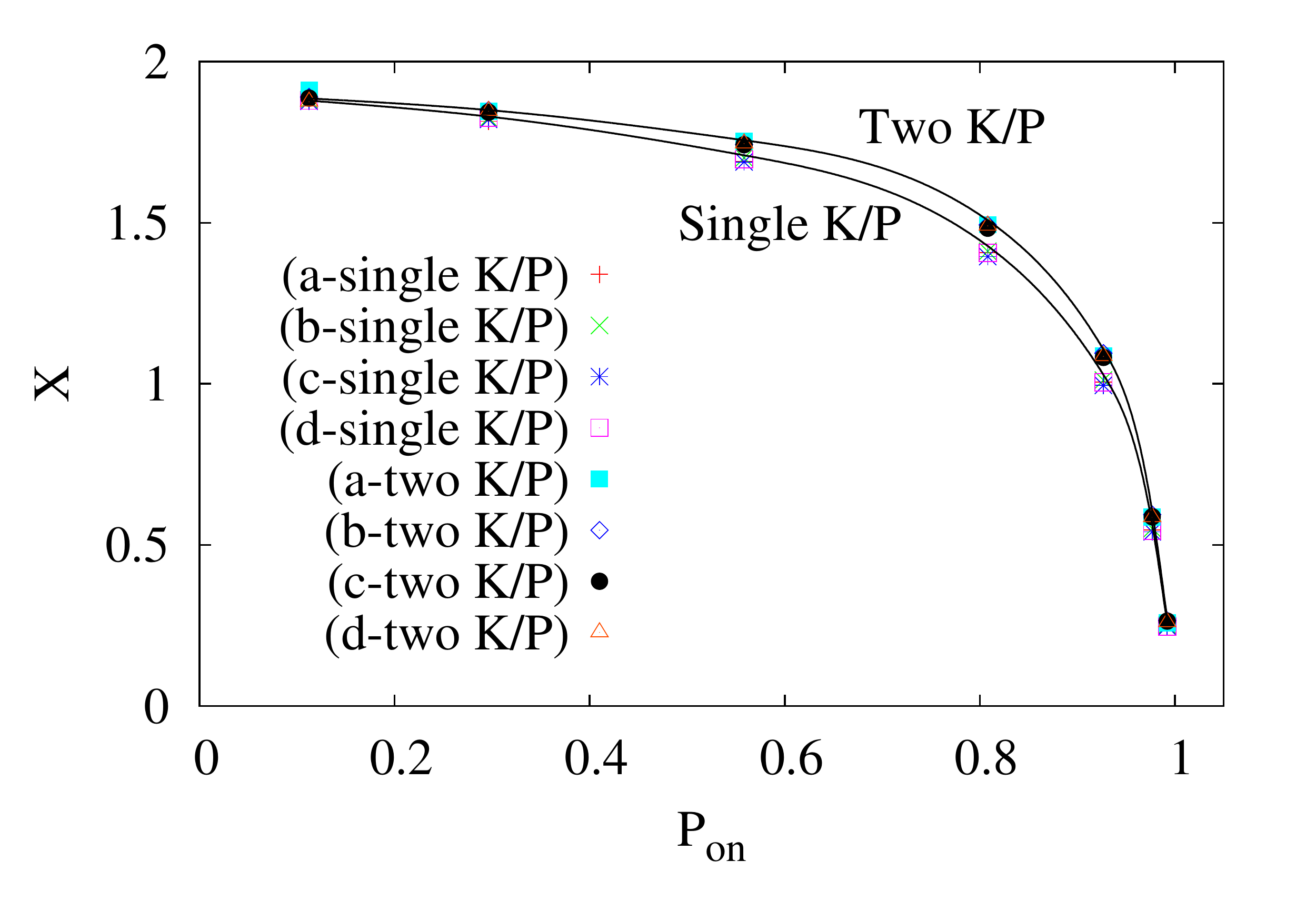}
\caption{\footnotesize{Specificity $X$ as a function of $P_{\rm on}$ (calculated as $P_{\rm on}^{\rm lattice}$ as defined in Equation (\ref{Ponlattice})) obtained from lattice simulations for a range of reactant concentrations. Labels correspond to those in Table \ref{table-simulations}. Data for systems with only one species each of kinase and phosphatase (single K/P), and with two distinct species (two K/P), are shown. Solid lines show the predictions for the simple model in the main text  (Equation (9) of the main text) with the same probabilities of catalysis, demonstrating close agreement (plotted lines correspond to cubic spline interpolations between analytic calculations at the same $P_{\rm on}$ values used in simulation). Estimated measurement errors are smaller than the symbols used to represent the data.}
\label{lattice-results}}
\end{figure}

To compare the results with the simple model in the main text, it is important to estimate the effective $P_{\rm on}$ for these simulations. Given an enzyme and a substrate in an interior cell within the lattice, the probability of a binding event occurring before one of the two reactants hops away is
\begin{equation}
P_{\rm on}^{0} = \frac{r_{\rm bind}}{r_{\rm bind} +12 r_{\rm hop}},
\end{equation}
as there are 6 possible hops for each protein. If a hop occurs, however, it is then possible that another hop immediately brings the reactants back onto the same lattice site. It is clear that the locations of the two proteins remained highly correlated, and it is not reasonable to compare such an event to the escape from close proximity which enters the simple model. We should therefore renormalize $P_{\rm on}^{0} $ to account for these events, giving
\begin{equation}
P_{\rm on}^{\rm lattice} \approx \frac{P_{\rm on}^{0}}{1 - 0.34056(1- P_{\rm on}^{0})}.
\label{Ponlattice}
\end{equation}
Here 0.34056 is the probability that a random walk on an infinite 3D cubic lattice returns to its starting point,\cite{Montroll1965} which is a good approximation for the probability that two proteins will hop apart and then back together quickly.  $P_{\rm on}^{\rm lattice}$ follows from summing the probabilities that binding occurs after $n \geq 0$ occasions on which the proteins hop apart and back together again.

Specificity is plotted against  $P_{\rm on}^{\rm lattice}$ for all systems simulated in Figure \ref{lattice-results}, and compared to the predictions of the simple model for the same parameters and effective $P_{\rm on}$. It is evident that systematic differences are small, meaning that the underlying biophysics inferred from the simple model is still meaningful in this more complicated description of diffusion.


\section{Specificity in a distributive two-site system at high yield $g$}
\subsection{$X_{\rm dis} -X_{\rm ss}$ with $\theta^A \phi^A\rightarrow 0$} 
We start from Equation (13) of the main text. As $\theta^A \phi^A\rightarrow 0$, we find
 \begin{equation}
 X_{\rm dis} = \lg(S S^\prime (1-g) +g).
 \end{equation}
  Thus , taking $X_{\rm ss} = \lg(S(1-g) +g)$ from Equation (13) of the main text,
    \begin{equation}
 X_{\rm dis} -X_{\rm ss} = \lg S^\prime + \lg \left( \frac{S S^\prime (1-g) +g}{S S^\prime (1-g) +gS^\prime} \right).
 \end{equation}
 This in turn implies
    \begin{equation}
 X_{\rm dis} -X_{\rm ss} > \lg S^\prime + \lg \left( \frac{S  (1-g) }{S  (1-g) +g} \right).
 \end{equation} 
For large $S(1-g)$, the specificity of the second site is barely affected by $g$. For $S(1-g) = 1$ (the point at which the first site provides discrimination by only a factor of two),  $X_{\rm dis} -X_{\rm ss} > \lg S^\prime - \lg 2$, and the additional specificity of the second site is only slightly compromised.

To compare $ X_{\rm dis} -X_{\rm ss} $ and $X_{\rm ss}$ directly for $S=S^\prime$, consider
\begin{equation}
\begin{array}{c}
 10^{X_{\rm dis} -X_{\rm ss}} -10^{X_{\rm ss}} \\
\\
=S\left(\frac{(1-g) +\frac{g}{S^2}}{(1-g) + \frac{g}{S}}\right) -S\left((1-g)  +\frac{g}{S}\right) \\
\\
=S \frac{g\left(1-\frac{2}{S} +\frac{1}{S^2}\right) - g^2 \left(1-\frac{2}{S} +\frac{1}{S^2}\right)}{(1-g) + \frac{g}{S}}.
\end{array}
\end{equation}
As the fractional yield cannot exceed unity, $g^2 \leq g$ and so the above expression is positive.

\subsection{$X_{\rm dis} -X_{\rm ss}$  with $\theta^A \phi^A=1$ }
 We start from Equation (13) of the main text. When $g \approx 1$, and $\theta^A \phi^A=1$, we obtain
 \begin{equation}
 X_{\rm dis} \approx \lg(S S^\prime) +\lg \left( (1-g) (1- g +1/S) +1/SS^\prime \right).
 \end{equation}
 Thus, taking $X_{\rm ss} \approx \lg(S(1-g) +1)$ from Equation (13) of the main text,
    \begin{equation}
 X_{\rm dis} -X_{\rm ss} =\lg \left( S^\prime (1-g) +\frac{1}{S(1-g) +1} \right).
 \end{equation}
 It is clear the additional specificity of the second site is slightly more affected than the first by large $g$. This difference is largest when $S, S^\prime \sim 1/(1-g)$; {\it i.e.}, at the very limit of distinguishing one substrate over another. For $S=S^\prime = 1/(1-g)$, the single site system has a specificity of $\lg 2$ and the two-site system has a specificity of $\lg 3 < 2 \lg 2$. For larger $S$ and $S^\prime$ relative to $1/(1-g)$, the difference between the effectiveness of the two stages is smaller. 
 
 \section{Proving that $\partial X/ \partial f^A_\alpha >0$}
 \label{derivative proof}
 We start from Equation (14) of the main text. First of all, we differentiate $Y \psi^A \theta^A$ with respect to $f^A_\alpha$, holding all other parameters constant.
 \begin{equation}
 \begin{array}{c}
\frac{ \partial Y \psi^A \theta^A}{ \partial f^A_\alpha} = - \frac{g\theta^A \phi^A f^A_\beta +(1-g) \phi^A}{2(1-g)} \times\\
\\
\left( 1+ \frac{g\theta^A f^A_\beta +  g \theta^A \phi^A(1-f^A_\alpha f^A_\beta) -(1-g)\phi^A f^A_\alpha}{\sqrt{(g\theta^A f^A_\beta +  g \theta^A \phi^A(1-f^A_\alpha f^A_\beta) -(1-g)\phi^A f^A_\alpha)^2 + 4g(1-g) \theta^A \phi^A}} \right).
 \end{array}
 \label{partialYpsitheta}
 \end{equation}
 Given that $0<f^A_\alpha,f^A_\beta,g<1$, we thus conclude that $\frac{ \partial Y \psi^A \theta^A}{ \partial f^A_\alpha} < 0$. The sign of  $\partial X/ \partial f^A_\alpha $ is the same as the sign of the derivative of the quotient involving $Y_d \psi^A \theta^A$ in Equation (14) of the main text. The sign of this derivative is in turn the same as the sign of
  \begin{equation}
 \begin{array}{c}
 \left(1+{Y_p\psi^A \theta^A} \left( 1 + \frac{f^A_\beta }{\phi^A } -{f^A_\alpha f^A_\beta}\right)\right) \times \\
\\
\left( \left(\frac{1}{S} \frac{ \partial Y \psi^A \theta^A}{ \partial f^A_\alpha} \right) \left(1 + \frac{f^A_\beta S}{\phi^A S^\prime} - \frac{f^A_\alpha f^A_\beta}{S^\prime} \right) -\frac{f_\beta^A}{S^\prime} \frac{Y \psi^A \theta^A}{S} \right)\\
\\
-\left(1+\frac{Y_p\psi^A \theta^A}{S} \left( 1 + \frac{f^A_\beta S}{\phi^A S^\prime} - \frac{f^A_\alpha f^A_\beta}{S^\prime}\right)\right) \times \\
\\
\left( \left(\frac{ \partial Y \psi^A \theta^A}{ \partial f^A_\alpha} \right) \left(1 + \frac{f^A_\beta }{\phi^A} - {f^A_\alpha f^A_\beta} \right) -f_\beta^A Y \psi^A \theta^A \right).
 \end{array}
 \end{equation}
 Gathering terms, this expression is equal to
   \begin{equation}
 \begin{array}{c}
-\frac{ \partial Y \psi^A \theta^A}{ \partial f^A_\alpha} \left((1-\frac{1}{S}) +\frac{f^A_\beta}{\phi} (1-\frac{1}{S^\prime}) -f^A_\alpha f^A_\beta(1-\frac{1}{SS^\prime})   \right)\\
\\
+ f_\beta^A Y \psi^A \theta^A  \left( 1 - \frac{1}{SS^\prime}\right)\\
\\
+ \frac{f_\beta^A}{S} (Y \psi^A \theta^A )^2 \left( 1- 1/S^\prime +  \frac{f^A_\beta S}{\phi^A S^\prime}(1-1/S)  \right).
 \end{array}
 \label{gathered terms}
 \end{equation}
 Given that all parameters are positive, and $S,S^\prime > 1$, the second and third terms of the above expression are trivially positive. The first is slightly more subtle. As derived in Equation (\ref{partialYpsitheta}), $-\frac{ \partial Y \psi^A \theta^A}{ \partial f^A_\alpha}$ is positive, and so the whole expression in Equation (\ref{gathered terms}) is definitely positive if  
 \begin{equation}
 \left(1-\frac{1}{S} \right) +\frac{f^A_\beta}{\phi} \left(1-\frac{1}{S^\prime} \right) -f^A_\alpha f^A_\beta \left(1-\frac{1}{SS^\prime} \right) >0.
 \end{equation}
 To proceed further, we note that $f_\alpha \phi \leq 1$ (this is discussed in more detail in Section \ref{sec-scaffold}). Thus the crucial term is necessarily greater than
 \begin{equation}
 \left(1-\frac{1}{S} \right) - \frac{f^A_\alpha f^A_\beta}{S^\prime} \left(1-\frac{1}{S} \right),
 \end{equation}  
 which is clearly positive as ${f^A_\alpha f^A_\beta}/{S^\prime} < 1$. Therefore $\partial X/ \partial f^A_\alpha >0$ and a system with pseudo-processive phosphorylation is necessarily more sensitive than a system without pseudo-processive phosphorylation and otherwise identical parameters.  
 
         \begin{figure}[!]
\centerline{\includegraphics[width=20pc]{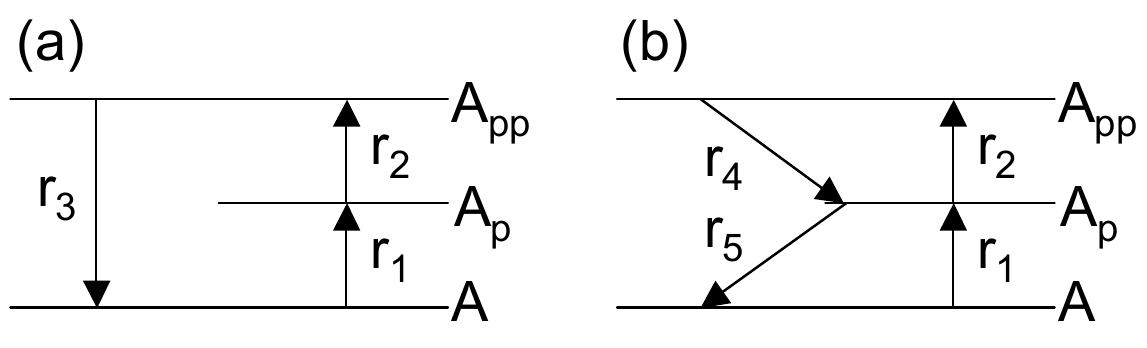}}
\caption{\footnotesize{(a) Illustration of a system in which dephosphorylation is always pseudo-processive. (b) Illustration of a distributive system. These figures define rates $r_i$ that are used for discussion in the text. We use $r_i$ for simplicity -- each represents a rate constant multiplied by the concentration of the relevant enzyme.}
\label{cyclic illustration}}
\end{figure}

   \begin{figure*}
        \subfloat[]
              {\includegraphics[width=15pc, angle=-90]{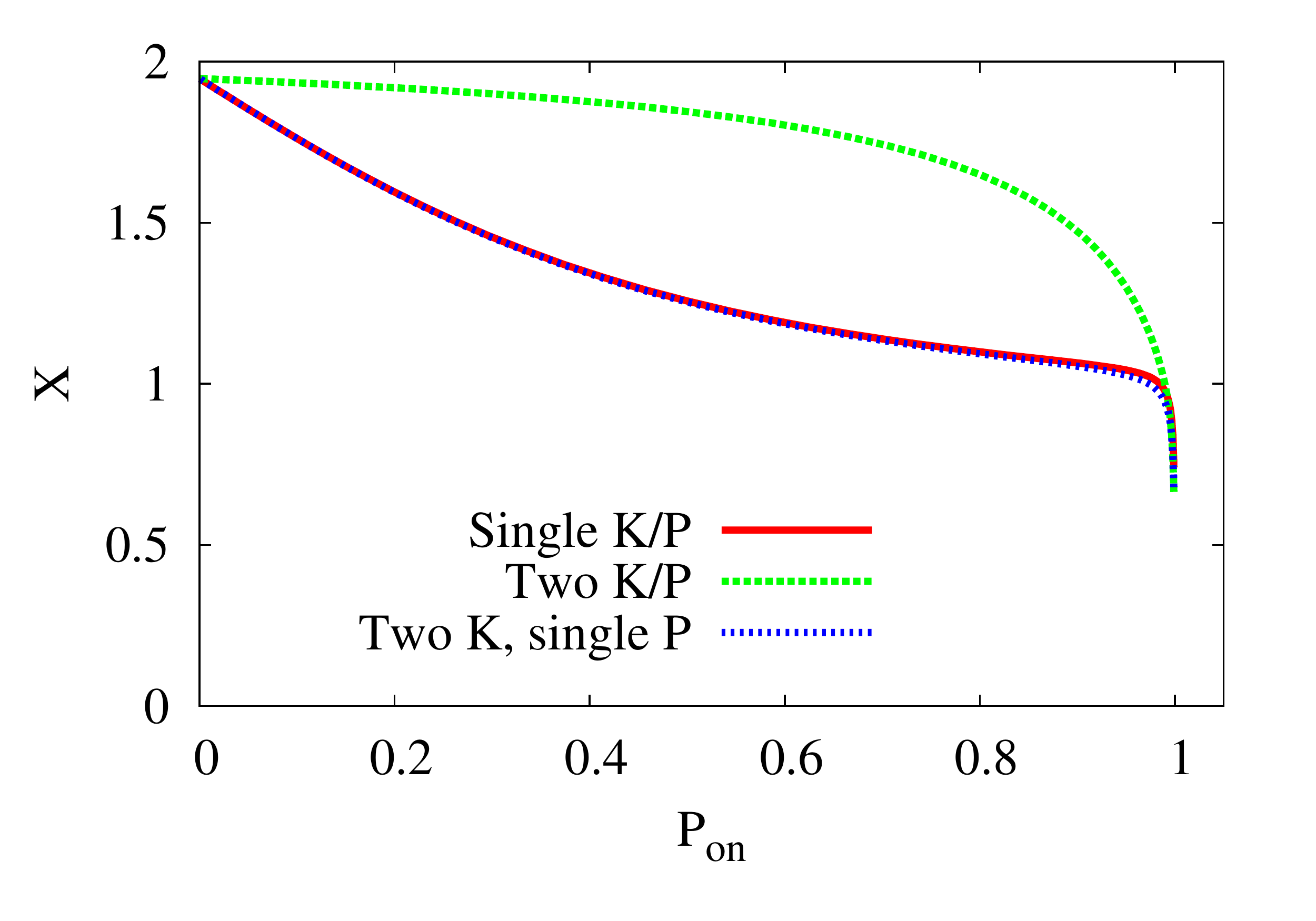}}        
        ~ 
        \subfloat[]{
              \includegraphics[width=15pc, angle=-90]{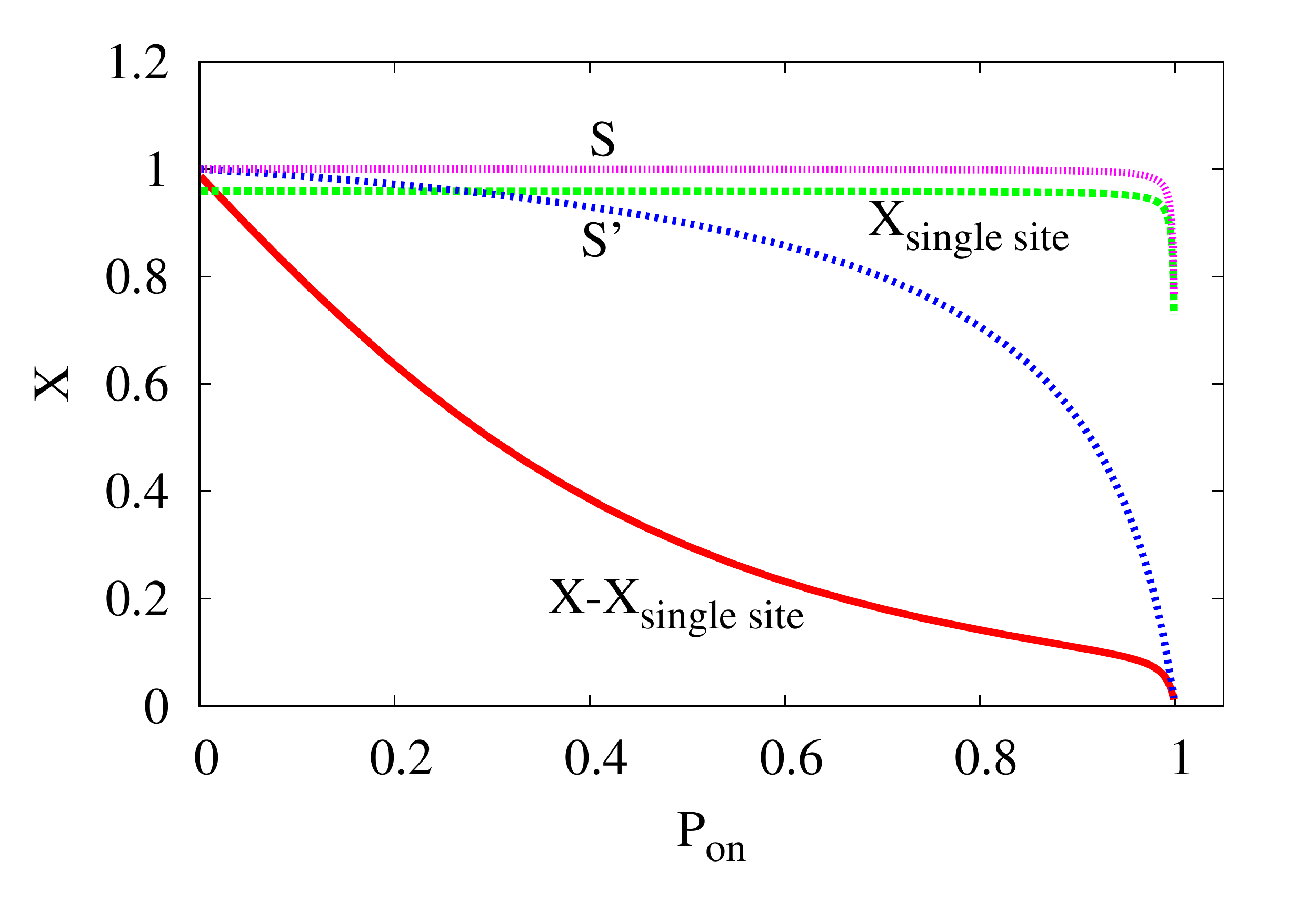}}        
        ~ 
        \caption{\footnotesize{A extreme example of proofreading being compromised by processivity in the dephosphorylation pathway.  We take $P^A_{\rm cat} = 0.001$, $P^{A\prime}_{\rm cat} = 0.3$, $P^B_{\rm cat} = 0.0001$, $P^{B\prime}_{\rm cat} = 0.03$ and $Q_{\rm cat} =Q^{\prime}_{\rm cat}= 0.8$, with a fixed yield of $[A_{pp}]/[A_0]=0.1$. (a) Specificity $X$ as a function of $P_{\rm on}$ for the system, for an equivalent system in which two separate kinases are needed for the two phosphorylation events, and an equivalent system in which two separate kinases {\it and} two separate phosphatases are required. (b) Comparing contributions to specificity from the two phosphorylation sites. $X_{\rm ss}$, the specificity for a system with only the first phosphorylation site, and  $X-X_{\rm ss}$, the additional contribution from he second site, are plotted. For comparison, intrinsic specificity factors $S$ and $S^\prime$ are shown.}
        \label{proofreading_non_robust}}
\end{figure*}

\section{Proofreading systems with strong sensitivity to pseudo-processivity in the dephosphorylation pathway}   
\label{dephos_sensitivity}
For a  pseudo-processive system, Equation (14) of the main text shows that large values of $Y_p \psi^A \theta^A f_\beta^A/\phi^A =Y_p f^\beta_A k^{A \prime}_{\rm eff} / h^{A \prime}_{\rm eff} $ can compromise specificity. 
What is the physical content of this ratio? $Y_p  k^{A \prime}_{\rm eff} / h^{A \prime}_{\rm eff} $ is the relative probability that $A_{p}$ is converted into $A_{pp}$ rather than $A$, and $f^\beta_A$ is the probability that a given dephosphorylation event of $A_{pp}$ modifies both sites.   When both terms are large, one can find a steady state in which substrates $A$ typically pass through the following cycle: $A \rightarrow A_{p} \rightarrow A_{pp} \rightarrow A$, whilst almost never undergoing $A_p \rightarrow A$ or $A_{pp} \rightarrow A_p$. An idealized cycle is illustrated in Figure \ref{cyclic illustration}, where it is contrasted with a system without pseudo-processivity. It might seem strange that dephosphorylation can be almost entirely pseudo-processive, implying an extremely efficient second stage of dephosphorylation, but that $A_p \rightarrow A$ should be negligible compared to $A_p \rightarrow A_{pp}$. This can be achieved if dephosphorylation reactions are very efficient ($f_\beta^A \sim 1$), but phosphatases are at a very low concentration relative to kinases ($Y_p \psi^A \theta^A /\phi^A =Y_p k^{A \prime}_{\rm eff} / h^{A \prime}_{\rm eff} >1$). 
Thus $A_p \rightarrow A$ can be very unlikely compared to $A_p \rightarrow A_{pp}$, because phosphatases rarely come into close proximity with substrates, even though the reaction is intrinsically more efficient.

In this section, we are not asking whether the rebinding-influenced rates of phosphorylation are different for $A$ and $B$ (that information is contained in the factors $S$ and $S^\prime$, and is independent of pseudo-processivity), but considering whether the difference in phosphorylation rates is actually manifest in the overall yields. To do that, it is helpful to analyse the how the steady-state yield of $[A_{pp}]$ depends on the transition rates labelled in Figure \ref{cyclic illustration}. In this part of the analysis it is simpler to use abstract rates $r_i$ rather than the full expressions derived for these rates in the main text.  For pseudo-processive dephosphorylation (Figure \ref{cyclic illustration}\,(a)),
\begin{equation}
\frac{[A_{pp}]}{[A_0]}= \frac{1}{1+r_3/r_2 + r_3/r_1}.
\end{equation}
For the purely distributive system (Figure \ref{cyclic illustration}\,(b)),
\begin{equation}
\frac{[A_{pp}]}{[A_0]}= \frac{1}{1+r_4/r_2 + r_4 r_5/r_1 r_2}.
\end{equation}
In the distributive case, we see that $\frac{[A_{pp}]}{[A_0]}$ is insensitive to $r_2$, the rate of the second step of phosphorylation, when $r_2 \gg r_4$ and $r_1 r_2 \gg r_4 r_5$, or equivalently when ${[A_{pp}]}\sim{[A_0]} \sim 1$. It is insensitive to $r_1$, the rate of the first stage of phosphorylation, when $r_1 r_2/r_4 r_5  \gg 1$, or equivalently when $[A_{pp}] \gg [A]$. When sensitivity to $r_1$ and $r_2$ is lost, specificity for $A$ over $B$ is lost as the two substrates are only differentiated through these rates. To be more concrete, if $[A_{pp}
]$ hardly changes when $r_2$ is reduced, then the discrimination between $A$ and $B$ through their differing values of $r_2$ is not fully manifest in the overall yields.
 Thus for the distributive system, we have reiterated the findings of the main text for finite yields of phosphorylated substrates; namely, that finite yields of $A_p$ and $A_{pp}$ can reduce specificity.

How does the system with pseudo-processive dephosphorylation differ? In this case, there are four regimes in which $[A_{pp}]/[A_0]$ can become insensitive to phosphorylation rates.
\begin{enumerate}
\item  $\frac{[A_{pp}]}{[A_0]}$ is insensitive to $r_1$ if $r_1 \gg r_3$, regardless of $r_2$. This corresponds to $[A_{pp}] \gg [A]$. 
\item $\frac{[A_{pp}]}{[A_0]}$ is insensitive to $r_2$ if $r_2 \gg r_3$, regardless of $r_1$. This corresponds to $[A_{pp}] \gg [A_p]$. 
\item $\frac{[A_{pp}]}{[A_0]}$ is insensitive to $r_1$ if $r_1 \gg r_2$, regardless of $r_3$. This corresponds to $[A_{p}] \gg [A]$. 
\item $\frac{[A_{pp}]}{[A_0]}$ is insensitive to $r_2$ if $r_2 \gg r_1$, regardless of $r_3$. This corresponds to $[A] \gg [A_p]$. 
\end{enumerate}
The first three of the cases listed above are apparently analogous to the yield saturation effects observed for the distributive system. In the pseudo-processive case, however, sensitivity to $r_2$ is lost when $[A_{pp}] \gg [A_p]$, whereas this is not necessarily true in the distributive case; a low value of  $r_1/r_5 = [A_{p}]/[A]$ can compensate for $r_2 \gg r_4$ (in this case, $[A_{pp}] \gg [A_p]$ but $[A_{pp}] \lesssim [A]$). In other words, in the distributive case, even if $[A_{pp}] \gg [A_p]$, sensitivity to $r_2$ is not lost until $[A_{pp}] \gg [A]$, but in the pseudo-processive case, sensitivity to $r_2$ is always lost when $[A_{pp}] \gg [A_p]$, even if $[A_{pp}] < [A]$. In the final limit given above, $r_2 \gg r_1$, $[A_{pp}]$ loses sensitivity to  $r_2$,  but the only condition on yields is $[A] \gg [A_p]$. Again, we see that  $[A_{pp}]$ can be insensitive to $r_2$ even when $[A_{pp}] < [A_0]$ in the system with pseudo-processive dephosphorylation, whereas this is impossible in the distributive system. As insensitivity of $[A_{pp}]$ to $r_2$ implies a reduced ability to discriminate between $A$ and $B$, the contribution of $r_2$ to the steady-state specificity for $A$ over $B$ can be lost at much lower yields of $A_{pp}$ in a system which cycles substrates through $A \rightarrow A_{p} \rightarrow A_{pp} \rightarrow A$ than in a distributive one.

The loss of sensitivity to $r_2$ in the system with fully pseudo-processive dephosphorylation, which leads to a reduction in specificity, can be understood intuitively. For a system that cycles through three states $A \rightarrow A_{p} \rightarrow A_{pp} \rightarrow A$, 
\begin{equation}
A = \frac{1/r_1}{\sum_i 1/r_i}, A_p = \frac{1/r_2}{\sum_i 1/r_i} , A_{pp} = \frac{1/r_3}{\sum_i 1/r_i}.  
\end{equation}
Physically, the relative yields are given directly by the lifetimes of the states, because this is the average amount of time each substrate spends in each state during a cycle. $r_2$ determines the lifetime of $A_p$: if $1/r_2$ is small compared to any of the other lifetimes, then the system spends only a short amount of time in $A_p$ and reducing this time further has almost no effect. For the distributive system, substrates do not move in the cycle $A \rightarrow A_{p} \rightarrow A_{pp} \rightarrow A$, and so this argument breaks down. In this case, even when $[A_p]$ is low, increasing $r_2$ can still be helpful in increasing $[A_{pp}]$. There is always a competition between $r_2$ and $r_5$ to determine whether the majority of molecules that reach the $[A_p]$ state are converted into $[A]$ or $[A_{pp}]$; this competition is no less important just because the lifetime of $[A_p]$ is short.

 We emphasize that in this section we are not attempting to map directly between specific realizations of the pseudo-processive case in Figure \ref{cyclic illustration}\,(a) and the distributive case in Figure \ref{cyclic illustration}\,(b). This would be inappropriate unless $r_i$ are carefully adjusted, as $r_i$ depend on kinase and phosphatase concentrations, which must be set to obtain the same yield of $A_{pp}$ in both cases. Instead, we are simply demonstrating that in a generic cyclic system with processive dephosphorylation such as Figure \ref{cyclic illustration}\,(a), $[A_{pp}]/[A_0]$  can lose sensitivity to $r_2$ at lower ratios of $[A_{pp}]/[A]$ than in a generic distributive system such as  Figure \ref{cyclic illustration}\,(b).

Pseudo-processive dephosphorylation can therefore make it easier  to lose sensitivity of $[A_{pp}]/[A_0]$ to the rates of phosphorylation (particularly the second stage), and thereby lose the ability to provide specificity for $A$ over $B$. Of course, our partially pseudo-processive systems are not as idealized as Figure \ref{cyclic illustration}\,(a), but the principle remains valid, and the contribution to specificity of the second site can be lost if substrates typically follow a loop  $A \rightarrow A_{p} \rightarrow A_{pp} \rightarrow A$ in which the $A_{p} \rightarrow A_{pp}$ transition is fast. This explains why specificity can be reduced if we convert a system with a distinct phosphatase for each stage of dephosphorylation into one with only a single phosphatase without changing any other parameters of the system, including the rebinding-influenced selectivity factors $S$ and $S^\prime$.

One might ask whether pseudo-processive phosphorylation can have a similar effect, but it cannot. By symmetry, pseudo-processive phosphorylation can reduce specificity in the dephosphorylation pathway in an analogous fashion,  but that is not our concern here. We show in Section \ref{derivative proof} that if we convert a  system with a distinct kinase for each stage of phosphorylation into one with only a single kinase without changing anything else, the specificity can {\it only} improve. In the language of this section, if a system undergoes the cycle $A \rightarrow A_{pp} \rightarrow A_{p} \rightarrow A$, we find that $A_{pp}$ only loses sensitivity to phosphorylation rates when $[A_{pp}] \gg [A]$ or  $[A_{p}] \gg [A]$.

Having established the underlying physical mechanism, we return to Equation (14) of the main text to outline when the effect is substantial. The relevant term is large when $f^A_\beta /\phi^A \gtrsim 1$ and $Y_p \psi^A \theta^A \gtrsim 1$. Additionally, if $\phi^A \sim 1$,   $f^A_\alpha$ must be small (or the $-f^A_\alpha f_\beta^A$ term in Equation (14) of the main text can provide a counteracting effect). These requirements are easiest to fulfill at high yield $g$ (when $Y_p \psi^A \theta^A$ is generally large), and when $P^{A \prime}_{\rm cat}, Q^{A \prime}_{\rm cat} > P^A_{\rm cat}$, as 
\begin{equation}
\frac{f^A_\beta }{\phi^A} = Q^{A \prime}_{\rm react} \frac{P^{A \prime}_{\rm react}}{P^{A}_{\rm react}}.
\end{equation}  
$Q^{A \prime}_{\rm cat} < P^A_{\rm cat}$ (or more generally, intrinsically inefficient phosphatases compared to kinases) prevents this loss of specificity due to pseudo-processive dephosphorylation. If $Q^{A \prime}_{\rm cat} < P^A_{\rm cat}$, then we can consider two regimes.
\begin{enumerate}
\item At low $P_{\rm on}$,  the difference in intrinsic efficiencies is manifest in reaction rates and $Q^{A \prime}_{\rm react} < P^A_{\rm react}$. Thus $\frac{f^A_\beta }{\phi^A}  = Q^{A \prime}_{\rm react} \frac{P^{A \prime}_{\rm react}}{P^{A}_{\rm react}} < 1$ and the relevant effect is not large. 
\item As $P_{\rm on} \rightarrow 1$,   $\frac{f^A_\beta }{\phi^A}  \rightarrow 1$, which would suggest that the effect in question becomes more substantial. However, in this case $-f^A_\alpha f_\beta^A = -Q^{A \prime}_{\rm react} {P^{A \prime}_{\rm react}} \rightarrow -1$, tending to counteract the growth in the relevant term in Equation (14) of the main text. 
\end{enumerate}

When $f^A_\beta /\phi^A \gtrsim 1$ and $Y_p \psi^A \theta^A \gtrsim 1$, proofreading can be extremely sensitive to pseudo-processivity in the dephosphorylation pathway. An extreme example is given in Figure \ref{proofreading_non_robust}\,(a) -- at low but non-zero values of $P_{\rm on}$, $X$ drops rapidly. As can be seen from Figure \ref{proofreading_non_robust}\,(b), this drop is due to the additional benefit of the second phosphorylation site being rapidly lost. Furthermore, the drop is explicitly due to processivity -- the results for an identical system but with separate enzymes for each stage of phosphorylation/dephosphorylation are also plotted in Figure \ref{proofreading_non_robust}\,(a), showing a drop in specificity only at high $P_{\rm on}$. The results for a system in which only the phosphorylation process requires separate enzymes (also shown in Figure \ref{proofreading_non_robust}\,(a)) match the original case, clearly indicating that it is processivity itself in the dephosphorylation pathway that is to blame.

\section{Phosphorylation kinetics}
It might be argued that the kinetics of phosphorylation should be considered as well as the steady-state yield. External signals are time-varying, and may not be stable for long enough for the steady state to be reached. Alternatively, it may be that a cell need only respond decisively to the initial transients produced by a signal, rather than waiting for the steady state. In this section we show that $A_{pp}$ and ${B_{pp}}$ are initially produced following a sudden activation of upstream kinases in the ratio 
\begin{equation}
\frac{[A_{pp}]/[A_0]}{[B_{pp}]/[B_0]}  = S S^\prime,
\end{equation}
regardless of the degree of pseudo-processivity and the kinase/phosphatase activity ratio. The initial differential equations summarized in Figure 3 of the main text can be solved directly.
\begin{widetext}
\begin{equation}
\begin{array}{c}
\frac{[A_{pp}(t)]}{[A_0]}  = \frac{[A^\infty_{pp}]}{[A_0]} \left(1 + \frac{\lambda^A_- \exp(\lambda^A_+ t)}{\lambda^A_+ - \lambda^A_-}-  \frac{\lambda^A_+ \exp(\lambda^A_- t)}{\lambda^A_+ - \lambda^A_-}\right)
+ k^A_{\rm eff} [K] f^A_\alpha   \frac{1}{\lambda^A_+ - \lambda^A_-} \left( \exp(\lambda^A_+ t) - \exp(\lambda^A_- t)\right),\\
\\
\lambda^A_{\pm} = -\frac{[P](h_{\rm eff}^A + h_{\rm eff}^{A \prime}) + [K](k_{\rm eff}^A + k_{\rm eff}^{A \prime})}{2}\pm \\
\\
\frac{\sqrt{([P](h_{\rm eff}^A + h_{\rm eff}^{A \prime}) + [K](k_{\rm eff}^A + k_{\rm eff}^{A \prime}))^2 - 4[P]^2h_{\rm eff}^Ah_{\rm eff}^{A \prime}-4[K]^2 k_{\rm eff}^A k_{\rm eff}^{A \prime} -4[P][K](k_{\rm eff}^A h_{\rm eff}^A(1-f^A_\alpha f^A_\beta) +k_{\rm eff}^{ A \prime} h_{\rm eff}^A f^A_\beta +k_{\rm eff}^A h_{\rm eff}^{A\prime}f^A_\alpha)}}{2},
\end{array}
\label{t-solution}
\end{equation}
\end{widetext}
for initial conditions in which $[A_{pp}] = [A_p]=0$ (corresponding to sudden activation of upstream kinases), in which $[A_{pp}^\infty]$ is the steady-state yield given in Equation (5) of the main text. The solution for $[B_{pp}]$ follows directly. For $f^A_{\beta}=f^A_{\alpha}=0$, 
\begin{equation}
\frac{{\rm d}{[A_{pp}(t)]/[A_0]}}{{\rm d}t} \approx  \frac{[A^\infty_{pp}]}{[A_0]}  \lambda^A_+ \lambda^A_- t
\end{equation}
in the limit $t \rightarrow 0$. Thus at short times,
\begin{equation}
\lg \left(\frac{[A_{pp}(t)]/[A_0]}{[B_{pp}(t)]/[B_0]} \right)= \lg \left(\frac{[A^\infty_{pp}]/[A_0]}{[B^\infty_{pp}]/[B_0]} \right) +\lg\left( \frac{\lambda^A_+ \lambda^A_-}{\lambda^B_+ \lambda^B_-} \right).
\label{tdepf=0}
\end{equation}
The first term on the right-hand side of Equation (\ref{tdepf=0}) is simply $X$ as calculated in Equation (12) of the main text. In the $f^A_{\beta}=f^A_{\alpha}=0$ limit it can be shown from the definitions of $\lambda_{\pm}$ in Equation (\ref{t-solution}) that
\begin{equation}
\frac{\lambda^A_+ \lambda^A_-}{\lambda^B_+ \lambda^B_-}  = \frac{\phi^A +  Y \psi^A   \theta^A \phi^A + \left(Y \psi^A \right)^2 \theta^A }
{      \phi^A 
      +  \frac{Y \psi^A}{S^\prime} 
      {\theta^A \phi^A}
      + \frac{\left(Y \psi^A \right)^2}{ S S^\prime} {\theta^A}
      }     .
\end{equation}
Thus the second term on the right-hand side of Equation (\ref{tdepf=0})  cancels with the $Y\psi$-dependent term arising from the ratio of steady-state yields, leaving
\begin{equation}
\lg \left(\frac{[A_{pp}(t)]/[A_0]}{[B_{pp}(t)]/[B_0]} \right)= \lg (S S^\prime) 
\end{equation}
at small $t$. 

When $f_\alpha, f_\beta \neq 0$, an equivalent cancellation occurs in the first  term of Equation (\ref{t-solution}). However, for substantial $f_\alpha$, the growth of $[A_{pp}(t)]$ at low $t$ is governed by a linear contribution from the second term in Equation (\ref{t-solution}). The contribution of this term at short times is 
\begin{equation}
\frac{{\rm d}{[A_{pp}(t)]/[A_0]}}{{\rm d}t} \approx  k^A_{\rm eff} [K] f^A_\alpha.  
\end{equation} 
Thus as $t \rightarrow 0$,
\begin{equation}
\lg \left(\frac{[A_{pp}(t)]/[A_0]}{[B_{pp}(t)]/[B_0]} \right)= \lg \left( \frac{k^A_{\rm eff} f^A_\alpha}{k^B_{\rm eff} f^B_\alpha}\right) = \lg(S S^\prime).
\end{equation}

In general, Equation (\ref{t-solution}) therefore gives initial growth in the fractional concentrations of $A_{pp}$ in $B_{pp}$ with a ratio of $SS^\prime$. For otherwise identical parameters, non-zero $f_\alpha$ typically allows the concentration $[A_{pp}]$ to grow closer to $[A^\infty_{pp}]$ before deviations from this ratio are large, as the growth rate is finite for $t\rightarrow 0$ in this case, but not if the system is distributive.  

\section{Other advantages of dual phosphorylation}
  As with proofreading, it is easiest to understand the results at the level of $\theta$, $\phi$, $\psi$, $Y$, $f_\alpha$ and $f_\beta$, rather than the rate constants for
  individual reaction steps. For clarity, we remind readers of the definitions of these quantities (in this section, we do not need to distinguish substrates $A$ and $B$). 
  \begin{itemize}
  \item $\theta = h_{\rm eff}/ h_{\rm eff}^\prime$, the ratio of effective reaction rate constants for the first and second stages of dephosphorylation.
  \item $\phi = k_{\rm eff}/ k_{\rm eff}^\prime$, the ratio of effective reaction rate constants for the first and second stages of phosphorylation.
  \item $\psi = k_{\rm eff}/ h_{\rm eff}$, the ratio of effective reaction rate constants for the first  stage of phosphorylation and the first stage of dephosphorylation.
  \item $Y = [K]/[P]$, the ratio of kinase and phosphatase concentrations.
  \item $f_\alpha$ ($f_\beta$) is the fraction of phosphorylation (dephosphorylation) reactions that lead directly to $A_{pp}$ from $A$ ($A$ from $A_{pp}$). 
  \item In this section it is often helpful to consider $\alpha = f_\alpha/(1- f_\alpha)$ and $\beta = f_\beta/(1- f_\beta)$, ratios of processive to non-processive reactions for phosphorylation and dephosphorylation.
  \end{itemize}

\subsection{Enabling scaffold-mediated enhancement of signalling}
\label{sec-scaffold}
Kocieniewski {\it et al.} studied dual phosphorylation in the context of scaffolds.\cite{Kocieniewski2012} 
  In their paper, they argued that scaffolds can hold upstream and downstream kinases 
  in close proximity, allowing phosphorylation of both residues to happen in quick succession (another pseudo-processive mechanism).  
  Scaffold-induced pseudo-processivity increased yields of activated
  kinases relative to the scaffold-free case when phosphorylation in solution was modelled as distributive. In fact, the authors went as far as to say
  that scaffolds can {\em only} enhance signalling by making otherwise distributive phosphorylation mechanisms effectively processive. 
  
  Here we explore the robustness of this amplification mechanism to a partially pseudo-processive phosphorylation pathway in the cytosol. To do that, 
  we first need to decide how the scaffold should be modelled. We make the following assumptions.  
  \begin{itemize}
   \item We assume {\it all} upstream kinases are bound to a scaffold, and that there are no spare scaffolds to interfere with the process.
   \item We assume that once a downstream kinase binds to the scaffold, it quickly reaches a doubly phosphorylated state and then detaches immediately.
   \item We assume that binding of the downstream kinase (regardless of phosphorylation state) to the scaffold has the same rate constant as the 
   relevant phosphorylation step ($k_{\rm eff}$, $k^\prime_{\rm eff}$) in the cytosol.
     \item We assume that dephosphorylation is cytosolic and unaffected by the scaffold.
  \end{itemize}
  The first assumption negates issues such as the ``prozone effect'',\cite{Levchenko2000} which tend to make scaffolds less effective. The second assumption implies that the 
  scaffolds are maximally efficient in causing double phosphorylation. The second and third assumptions, however, do preclude two mechanisms by which scaffolds can enhance the activation level
  of a substrate.
  \begin{itemize}
   \item If a scaffold binds to the kinases faster than the intrinsic rate of reaction between the kinases in the cytosol  ($k_{\rm eff}$ and $k^\prime_{\rm eff}$),  
   it can potentially accelerate phosphorylation in an enzyme-like manner, allowing kinases to come together and react faster than they
   would in the absence of the scaffold.
   \item If the release of the phosphorylated downstream kinase from the scaffold is slow, the total level of activated downstream kinases can be enhanced because the scaffold
   can either shield the downstream kinase from phosphatases, or provide a platform for `recharging' it if it is dephosphorylated on the scaffold. Either mechanism can lead
   to an enhanced degree of activation for downstream kinases, although the additional activated kinases are attached to the scaffold.
  \end{itemize}
  These alternative enhancement mechanisms, however, are independent of the idea that scaffolds can  enhance signalling by making
  otherwise distributive phosphorylation mechanisms effectively processive. In fact, they do not even require the dual phosphorylation motif. We therefore
  limit the system in this way to restrict our investigation to the question at hand. Given that these effects are precluded, this simplified scaffold model is in some sense maximally 
  efficient; a downstream kinase only has to bind to a scaffold to be instantly doubly phosphorylated, and it can do this as fast as the substrate naturally undergoes single phosphorylation in
  the cytosol.
  
    We shall consider the steady-state signal when only phosphorylation via the scaffold pathway can occur, and  compare that to the cytosolic-only case
  (the same procedure that was followed by Kocieniewski {\it et al.} \cite{Kocieniewski2012}). In the absence of scaffolds, we consider the same system as illustrated in Figure 3 of the main text, and so the steady-state is identical to Equation (5) of the main text (although we use $f_\alpha^{\rm cyt}$ instead of $f_\alpha$ to emphasise that this is pseudo-processive phosphorylation in the cytosol). For scaffold-only phosphorylation, we again have an identical system to Figure 3 of the main text, except that $f_\alpha = f_\alpha^{\rm scaff} = 1$ (all phosphorylations are processive).     All other parameters are identical in the scaffold-mediated and scaffold-free systems under our assumptions.    Thus Equation (5) of the main text can be used again, with $f_{\alpha}$ replaced by unity instead of $f_\alpha^{\rm cyt}$. We can then trivially calculate the ratio of 
 the yields that can be obtained exclusively from the scaffold-mediated pathway relative to exclusively from the cytoplasmic pathway,
  \vspace{0.1mm}
  \begin{widetext}
  \begin{equation}
   \frac{[A_{pp}^{\rm scaf}]}{[A_{pp}^{\rm cyt}]} = \left(\frac{(Y \psi) \phi+ (Y \psi)^2 \theta}{(Y \psi) \phi f^{\rm cyt}_\alpha + (Y \psi)^2 \theta} \right)
   \left(\frac{\phi + (Y \psi) (\theta f_\beta + f^{\rm cyt}_\alpha \phi+ \theta \phi (1-f^{\rm cyt}_{\alpha} f_{\beta})) + (Y \psi)^2 \theta}
   {\phi + (Y \psi) (\theta f_\beta + \phi+ \theta \phi (1-f_{\beta})) + (Y \psi)^2 \theta} \right).
   \label{scaf/cyt}
     \end{equation}
  \end{widetext}
The first term on the left hand side is greater than unity, and monotonically drops as $Y\psi$ is increased. The second term is less clear. However, if $1-f_\beta \theta > 0$, the fraction is unity in the limits $Y \psi \rightarrow 0$ and $Y\psi \rightarrow \infty$ and less than unity for finite $Y \psi$. In this case, the overall expression necessarily has its largest value for $Y \psi \rightarrow 0$ (in the $Y \psi \rightarrow \infty$ limit, $A$ is always doubly phosphorylated, regardless of whether the scaffold-based or cytosolic mechanism is considered).
  
  We will now argue that $1-f_\beta \theta > 0$ holds for all systems in our model. Using $\theta = h_{\rm eff}/h_{\rm eff}^\prime$ and the definitions in Equations (3-4) of the main text, we see that $f_\beta \theta = Q_{\rm cat} Q_{\rm on}/(1-Q_{\rm on}(1-Q_{\rm cat})) = Q_{\rm react} <1$. At a more intuitive level, $f_{\beta}$ is the fraction of dephosphorylation reactions that are processive -- this must  be small if $\theta = h_{\rm eff}/h_{\rm eff}^\prime$ is large, as large $\theta$ requires the second phosphorylation reaction to be inefficient relative to the first.   By identical reasoning, $1-f_\alpha \phi > 0$. 
  
   Considering the optimal low yield (low $Y \psi$)) limit,
  \begin{equation}
   \frac{[A_{pp}^{\rm scaf}]}{[A_{pp}^{\rm cyt}]} = \frac{1}{f^{\rm cyt}_\alpha + Y \psi \theta/\phi}, 
   \label{scaf/cyt2}
     \end{equation}
where in the denominator we have retained the two lowest order terms in $Y$, as we wish to consider the possibility that the lowest order term is zero. We remind the reader that $\theta$, $\phi$ and $\psi$ are ratios of effective rate constants for the stages of phosphorylation and dephosphorylation.  $1/f_\alpha^{\rm cyt}$ fundamentally limits this expression, which otherwise could be arbitrarily large for low enough yield (low $Y$). When phosphorylation in the cytosol is 10\% processive, the scaffold-mediated enhancement 
  is limited to a factor of 10, and when it is 50\% processive the yield is only doubled at most. This use of dual phosphorylation, therefore, is 
  not very robust to moderate levels of kinase processivity in solution. This is perhaps not very surprising: if scaffolds confer an advantage by making the reaction
  processive, this advantage is limited if the reaction is partially processive anyway.
  
  We note that unlike proofreading, this reduction in efficiency is due to processivity in the cytosol itself, rather than high rebinding probabilities. We could imagine, as in the main text,
  that we have two distinct upstream kinases (both bound to the scaffold when appropriate) and two distinct phosphatases. In this system, $f^{\rm cyt}_\alpha= f_\beta=0$ but the reaction on the scaffold is still effectively processive. The result is that the limit placed on ${[A_{pp}^{\rm scaf}]}/{[A_{pp}^{\rm cyt}]}$ by $1/f_\alpha^{\rm cyt}$ at low $Y$ vanishes, regardless of whether rebinding events occur frequently or not. The ratio $\psi \theta/\phi = k_{\rm eff}^\prime/h_{\rm eff}^\prime$ can either increase or decrease with rebinding probability, depending on whether cytosolic phosphorylation or dephosphorylation is intrinsically more efficient. These changes reflect increases or decreases in the overall cytosolic yield with increased rebinding. These variations are not a fundamentally limiting factor in the same sense as $f^{\rm cyt}_\alpha$, and ${[A_{pp}^{\rm scaf}]}/{[A_{pp}^{\rm cyt}]}$ can be pushed arbitrarily high by reducing $Y$ for any value of $\psi \theta/\phi$.

\subsection{Providing an ultrasensitive response}
\label{sec-ultrasensitivity}
Huang and Ferrel \cite{Huang1996} originally pointed out that double phosphorylation, when the mechanism is distributive, can lead to a sharper transition from low to high yield of activated substrate (as 
  $Y=[K]/[P]$ is varied)
  than kinases  that need only a single activation. Such an effect has been dubbed `ultrasensitivity'. 
  For completeness, we now consider the robustness of the ultrasensitive response to partial pseudo-processivity in the language of the model presented in this paper. 
  We note that other authors have reached similar conclusions elsewhere.\cite{Salazar2009,Takahashi2010} In assuming that all enzymes are unsaturated, we have no 
  contribution to ultrasensitivity from various mechanisms involving sequestration of products or enzymes, simplifying the analysis greatly, and allowing focus on the usefulness of the dual phosphorylation motif itself.

    A common, though not necessarily ideal, measure for the sharpness of the transition is to fit the curve to a Hill function, and label transitions as ultrasensitive
  if the coefficient found is greater than unity. The fits can be quite poor, however, meaning that the result may not accurately describe the transition. Here, we will directly consider
  the relative change in $Y$ (which would be the change in $[K]$ at constant $[P]$) required to raise the yield of $A_{pp}$ from 10\% to 90\%. The use of 10\% and 90\% is of course
  somewhat arbitrary, but none of the results presented here are particularly sensitive to this choice.
  
  To be precise, let us consider $M_{0.9,0.1}=\lg \left(\frac{Y_{0.9}}{Y_{0.1}}\right)$, where $Y_g$ is the value of $Y$ required to give a yield of $[A_{pp}]/[A_0]=g$.
  Note that a lower value of this metric implies more ultrasensitivty.
  In our simple model, this metric has the advantage of being easy  to calculate from Equation (5) of the main text
 \vspace{0.1mm}
  \begin{widetext}
    \begin{equation}
  \begin{array}{c}
   M_{g_1,g_2}= \lg \left(\frac{Y_{g_1}}{Y_{g_2}}\right) = 
   \lg \left( \frac{g_1(1-g_2)}{g_2(1-g_1)}\frac{\theta \phi (1-f_{\alpha} f_{\beta})+\phi f_{\alpha}  +\theta f_{\beta} - \phi f_{\alpha}/g_1 +
   \sqrt{(\theta \phi (1-f_{\alpha} f_{\beta})+  \phi f_{\alpha} +  \theta f_{\beta}-  \phi f_{\alpha}/g_1)^2 + 4 \phi \theta (1-g_1)/g1} }
   {\theta \phi (1-f_{\alpha} f_{\beta})+\phi f_{\alpha}  +\theta f_{\beta} - \phi f_{\alpha}/g_2 +
   \sqrt{(\theta \phi (1-f_{\alpha} f_{\beta})+  \phi f_{\alpha} +  \theta f_{\beta}-  \phi f_{\alpha}/g_2)^2 + 4 \phi \theta (1-g_2)/g_2}} \right).
  \end{array}
  \label{ultrasensitivity}
   \end{equation}
\end{widetext}

It is not immediately obvious how Equation (\ref{ultrasensitivity}) behaves as a function of its parameters. It is easy to see, however, that $\psi = k_{\rm eff}/h_{\rm eff}$ 
is irrelevant: the ratio of rate constants for phosphorylation and dephosphorylation sets the value of $Y$ at the transition midpoint, but does not affect its width.
   It is helpful to note that, in the fully processive limit,
   $f_\alpha = f_\beta =1$, the measure reduces to
      \begin{equation}
   M_{g_1,g_2}^{\rm proc} = 
   \lg \left(\frac{g_1(1-g_2)}{g_2(1-g_1)} \right). 
   \label{ultrasensitivity-p}
   \end{equation}
   In the case of $g_1=0.9$ and $g_2=0.1$, $M_{g_1,g_2}^{\rm proc} =\lg 81 \approx 1.91$, the well-known value for a hyperbolic response. Such a value would also be obtained for 
   an unsaturated system involving a single phosphorylation site. In the distributive case, we obtain
   \begin{equation}
   \begin{array}{c}
   M_{g_1,g_2} = 
    M_{g_1,g_2}^{\rm proc} + \\
    \\
   \lg \left( \frac{\theta \phi +
   \sqrt{(\theta \phi )^2 + 4 \phi \theta (1-g_1)/g_1} }
   { \theta \phi +
   \sqrt{( \theta \phi )^2 + 4 \phi \theta (1-g_2)/g_2} }\right).
  \label{ultrasensitivity-np}
  \end{array}
 \end{equation}
  The second term is always less than or equal to zero for $g_1>g_2$, meaning that the metric $M_{g_1,g_2}$ can be reduced below the processive (or single phosphorylation site) value, 
  corresponding to a sharper transition. Furthermore, the degree of reduction depends on the ratio of $\theta \phi$ to  $(\theta \phi)^2$ -- the metric drops monotonically (the transition 
  gets sharper) as $\theta \phi \rightarrow 0$. In the limit of $\theta \phi =0$, we obtain 
     \begin{equation}
   M^{\rm min}_{g_1,g_2} = 
  \frac{1}{2} M_{g_1,g_2}^{\rm proc}. 
  \label{ultrasensitivity-nplimit}
       \end{equation}
   For $g_1=0.9$ and $g_2=0.1$, $M_{g_1,g_2}^{\rm min} =\lg 9$. In the limit of $\theta \phi \rightarrow \infty$, we obtain $M_{d_1,d_2}^{\rm max} =M_{g_1,g_2}^{\rm proc}$. 
   The reason for this dependence on $\theta \phi = k_{\rm eff} h_{\rm eff} / k^\prime_{\rm eff} h^\prime_{\rm eff}$ is that this ratio determines whether the transition is effectively `cooperative'. At low $\theta \phi$, the system moves directly
   from an $A$-dominated situation to a $A_{pp}$-dominated case as $Y$ is varied because low $\theta$ and $\phi$ mean that the first stage of phosphorylation and dephosphorylation are slower than the 
   subsequent stage, and therefore the concentration of the singly phosphorylated $A_p$ intermediate is kept low. As the competition between $A$ and $A_{pp}$ involves two $Y$-dependent antagonistic pairs of reactions,  it is easy to see why this system is ultrasensitive to $Y$ in this limit. In the other limit, when $\theta \phi \rightarrow \infty$, the transition has negative cooperativity and $A_{pp}$
   forms at the expense of $A_{p}$, which itself supercedes $A$ at even lower $Y$. In this limit, 
   we are effectively reduced to a single site phosphosylation/dephosphorylation system involving $A_p$ and $A_{pp}$,  and we recover the low sensitivity to $Y$. 
   
       \begin{figure*}
        \subfloat[$P_{\rm cat} = 0.2$, $P^\prime_{\rm cat} = 0.1$, $Q_{\rm cat} = 0.3$, $Q^\prime_{\rm cat} = 0.1$]{
                \includegraphics[width=15pc,angle=-90]{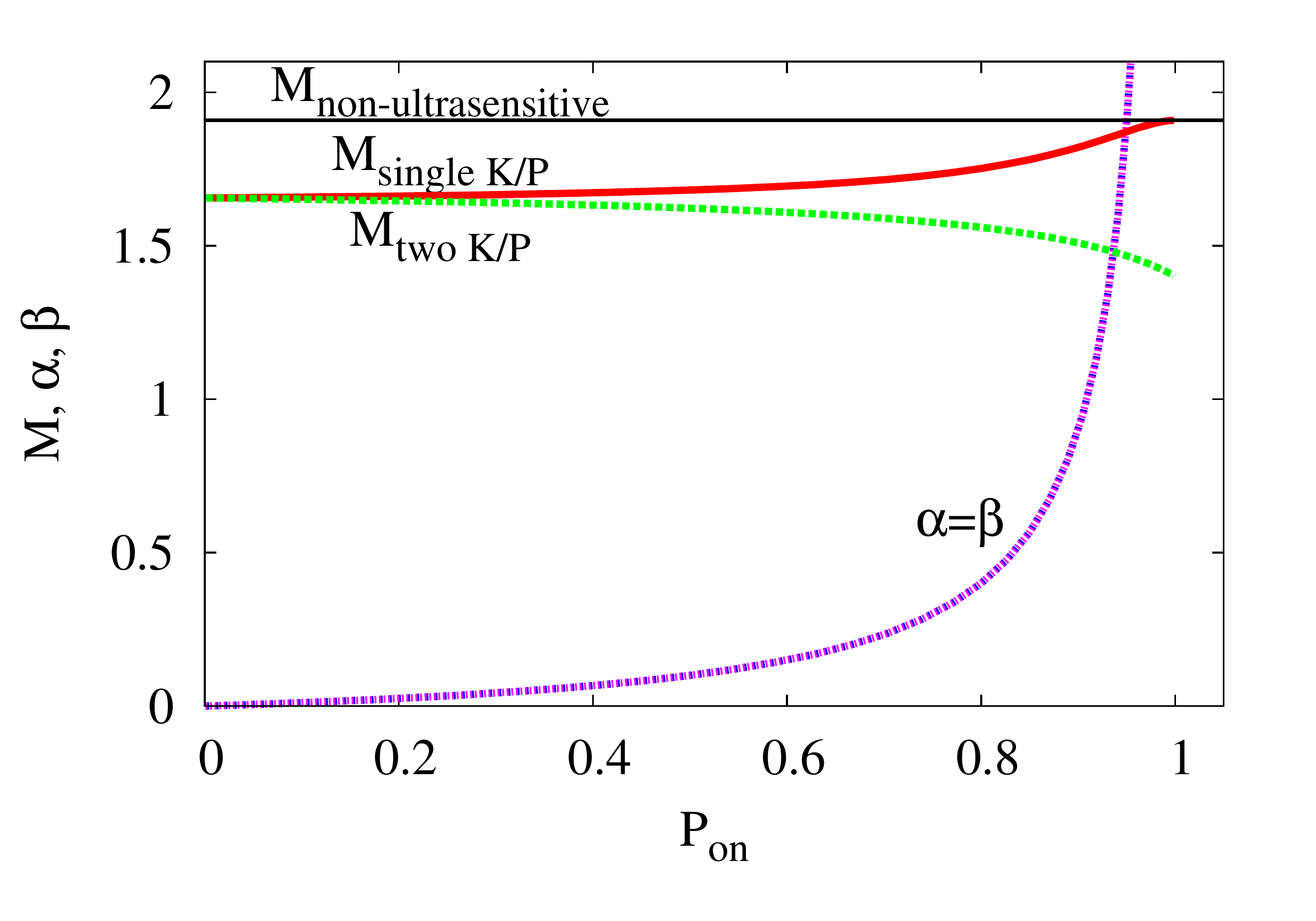}         
        }
        ~ 
        \subfloat[$P_{\rm cat} = 0.1$, $P^\prime_{\rm cat} = 0.2$, $Q_{\rm cat} = 0.1$, $Q^\prime_{\rm cat} = 0.3$]{
                \includegraphics[width=15pc,angle=-90]{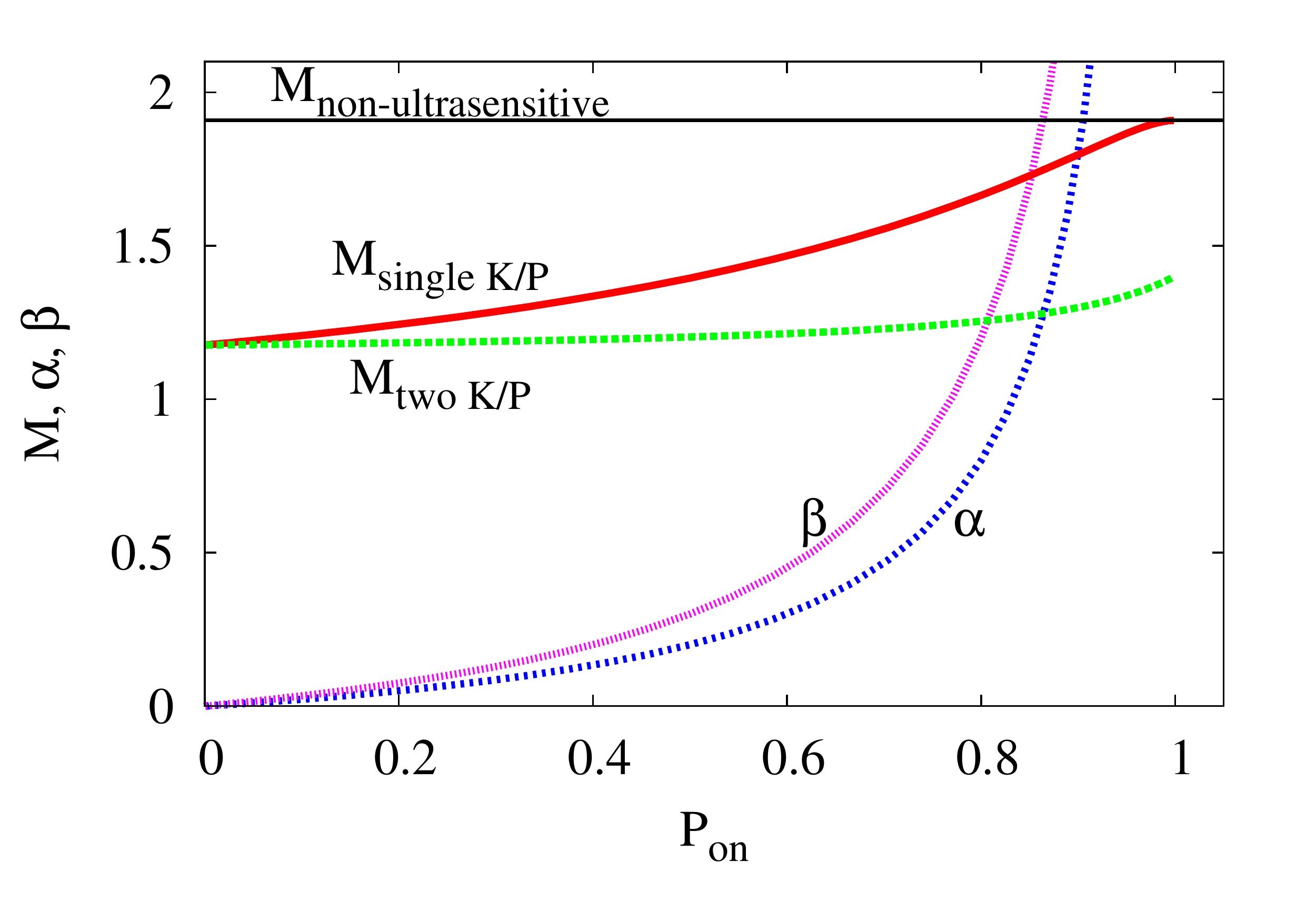}         
        }
        ~ 
          
        \subfloat[$P_{\rm cat} = 0.1$, $P^\prime_{\rm cat} = 1$, $Q_{\rm cat} = 0.1$, $Q^\prime_{\rm cat} = 0.1$]{
                \includegraphics[width=15pc,angle=-90]{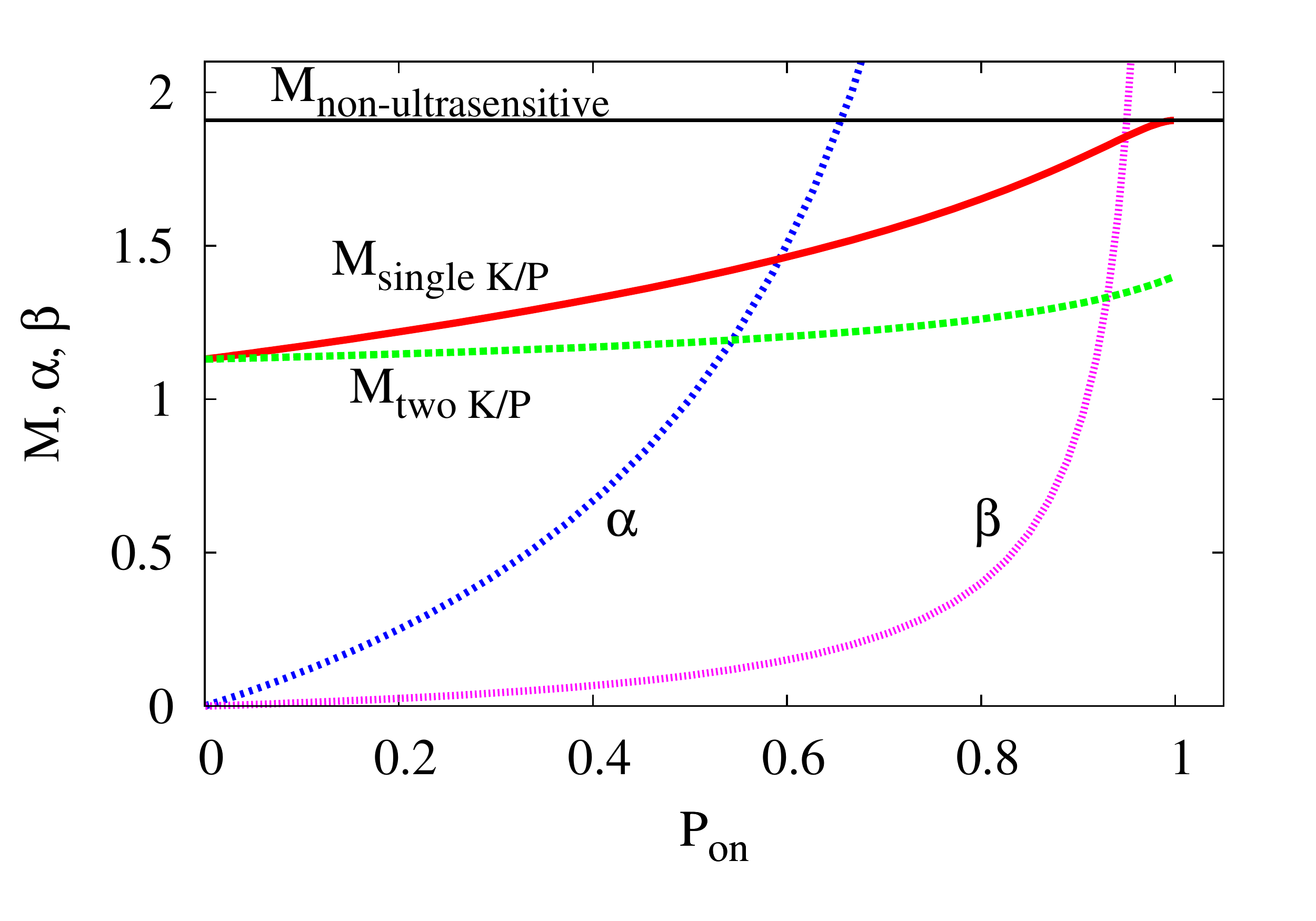}         
        }
        ~ 
        \subfloat[$P_{\rm cat} = 0.1$, $P^\prime_{\rm cat} = 1$, $Q_{\rm cat} = 0.1$, $Q^\prime_{\rm cat} = 1$]{
                \includegraphics[width=15pc,angle=-90]{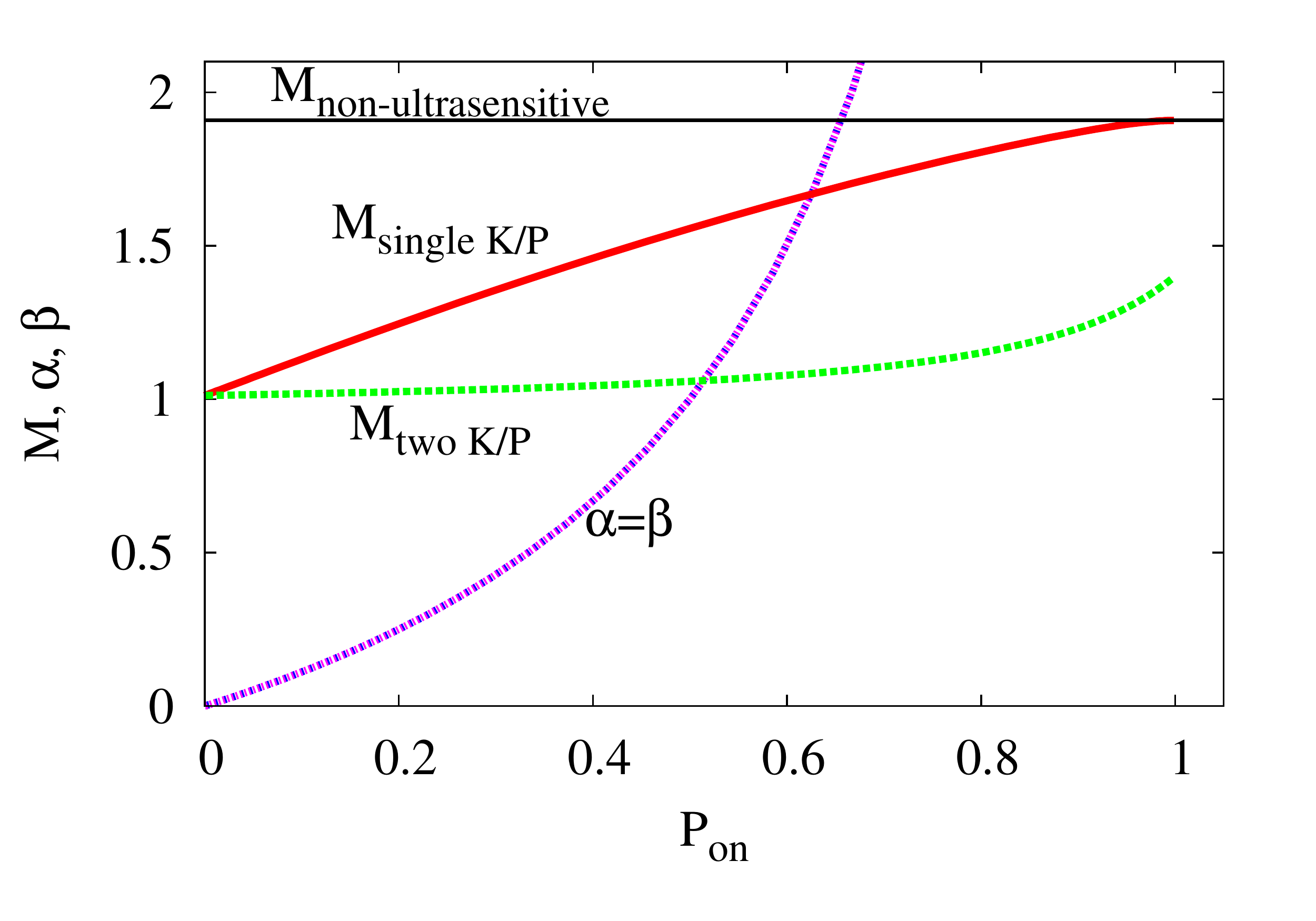}         
        }
        ~ 
        \caption{\footnotesize{Ultrasensitivity metric $M_{0.9,0.1}$ for a number of systems (parameters given in the individual labels). $M_{\rm single\,K/P}$ are for systems with only one kinase and phosphatase, $M_{\rm two\,K/P}$ are for equivalent systems in which processivity is eliminated by requiring distinct enzymes for each stage of phosphorylation and dephosphorylation. Processivity factors $\alpha = f_\alpha/(1-f_\alpha)$ and $\beta = f_\beta/(1- f_\beta)$ are also shown as a function of $P_{\rm on}$ (due to symmetry, these two curves overlap in (a) and (d)).The line at $M \approx 1.91$ is the value obtained for a system with a single phosphorylation site, or equivalently a fully processive system. 
        \label{ultra-curves}}
        }
\end{figure*}

        \begin{figure}
\includegraphics[width=20pc]{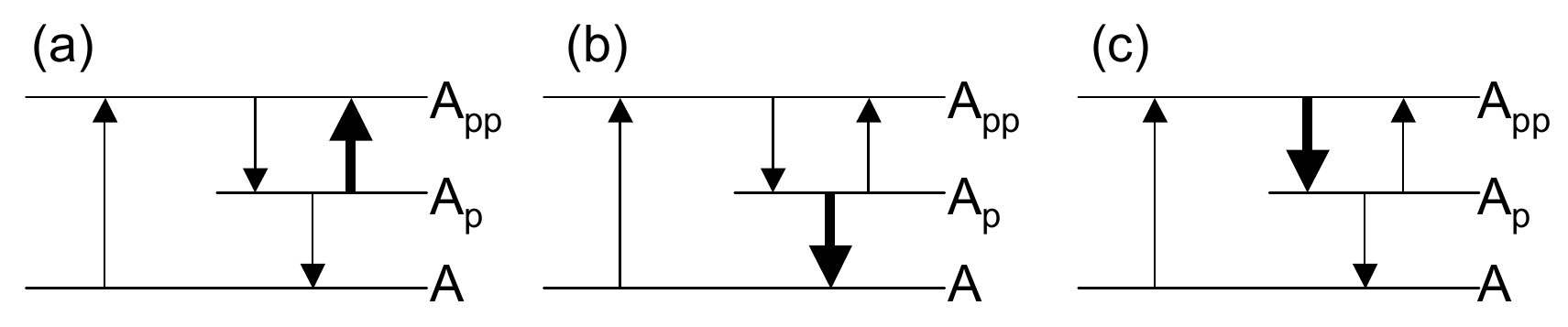}
\caption{\footnotesize{ Illustration of potential robustness of ultrasensitivity to processivity in the phosphorylation pathway, provided that dephosphoryltion is distributive. In these diagrams, no transition from $A$ to $A_p$ is shown, because we take a limit in which all phosphorylations are processive.  (a) A robust scenario, in which $\phi$ is small, meaning that $A_p \rightarrow A_{pp}$ is a fast transition that competes with $A_p \rightarrow A$ and  that $A_p$ is rapidly drained. (b) A non-robust scenario in which $\phi$ not small, but $\theta$ is. Consequently, at values of $Y$ at which $[A] \sim [A_{pp}]$, the intermediate state $A_p$ is overwhelmingly resolved into $A$  and both phosphorylation {\it and} dephosphorylation might as well be processive. (c) A non-robust scenario in which both $\phi$ and $\theta$ are not small. In this case, the intermediate $A_p$ is drained slowly and hence builds up a large concentration, inhibiting ultrasensitivity.}
\label{ultra-robust}}
\end{figure}
   
   We have therefore identified two preconditions for ultrasensitivity. We need an intermediate state ($A_p$), from which the system can progress either to $A_{pp}$ or $A$, so that 
   there can be two stages of phosphorylation at which the relative concentrations of upstream kinases and phosphatases ($Y$) can play a role, yet we need the concentration of this intermediate to 
   remain low.
   
    Of most interest to us, of course, is the partially pseudo-processive case. By differentiating Equation (\ref{ultrasensitivity})
    with respect to $f_\alpha$ and $f_\beta$, and using $\phi f_\alpha, \theta f_\beta <1$ as justified in Section \ref{sec-scaffold}, it is possible to show that $M_{g_1,g_2}$ increases monotonically with $f_\alpha$ and $f_\beta$ at fixed $\theta$ and $\phi$. This implies that a pseudo-processive system is always less ultrasensitive than an alternative system without pseudo-processivity but with the same $\theta$ and $\phi$ (for example, a system with identical parameters but two distinct kinases and two distinct phosphatases). Physically, non-zero $f_\alpha$ and $f_\beta$ allow the system to circumvent the essential intermediate $A_p$ state.
    
 When analyzing proofreading, we considered its efficacy for a given set of catalytic probabilities as the diffusive properties were varied (through $P_{\rm on}$), to reflect the consequences of slower diffusion with fixed underlying chemistry. We shall do the same here for ultrasensitivity; as with proofreading, we shall assume that reactions are differentiated by the unbinding rate of reactants.  This calculation is not the same as varying  $f_\alpha$ and $f_\beta$ at fixed $\theta$ and $\phi$, as all four quantities are in principle functions of $P_{\rm on}$.   
    
Equation (\ref{ultrasensitivity}) can be written explicitly in terms of $P_{\rm on}$ and the catalytic properties, but the result is not simple to interpret -- it is more helpful to think in terms of $\theta$, $\phi$, $f_\alpha$ and $f_\beta$. As $P_{\rm on}$ is increased $f_\alpha$ and $f_\beta$ grow, tending to compromise ultrasensitivity as discussed above. Separately, $\theta, \phi = k_{\rm eff} / k^\prime_{\rm eff} , h_{\rm eff}/h^\prime_{\rm eff}  \rightarrow 1$, as reaction rates converge in the limit of frequent rebinding. Low  $\theta$ and $\phi$ are advantageous for ultrasensitivity, so this effect can be either helpful or unhelpful, depending on the value of  $\theta \phi$ as $P_{\rm on} \rightarrow 0$.

Examples of this behavior are shown in Figure \ref{ultra-curves}, where $M_{0.9,0.1}$ is plotted against $P_{\rm on}$, along with $\alpha = f_\alpha/(1-f_\alpha)$ and $\beta = f_\beta/(1-f_\beta)$ for various values of the catalytic probabilities. Also shown in these graphs is the value of
   $M^{\rm}_{0.9,0.1}$ that would be obtained for identical microscopic parameters, but if a distinct kinase/phosphatase were needed for each step so that processivity could not occur. In this case, the only consequence of changing $P_{\rm on}$ is to change $\theta$ and $\phi$. 
   
   In all four cases, increasing $P_{\rm on}$ reduces ultrasensitivity (increases $M_{0.9,0.1}$). For (b)-(d), $\theta \phi<1$ as $P_{\rm on} \rightarrow 0$, and so $P_{\rm on}$ compromises specificity both through non-zero $f_{\alpha}$ and $f_{\beta}$, and through increasing $\theta \phi$. Through comparison to the curves for the system with distinct kinases and phosphatases when only the $\theta \phi$ dependence is present, it is clear that the effects of non-zero  $f_{\alpha}$ and $f_{\beta}$ are stronger.
 
 In (a), $\theta \phi >1$ as $P_{\rm on} \rightarrow 0$, and so larger values of $P_{\rm on}$ might be expected to have two opposing effects on ultrasensitivity. Indeed, for the system with two separate kinases and phosphatases, one sees that ultrasensitivity rises ($M_{0.9,0.1}$ drops) as  $P_{\rm on} \rightarrow 1$ and $\theta \phi \rightarrow 1$. Despite this, ultrasensitivity drops ($M_{0.9,0.1}$ rises) monotonically with $P_{\rm on}$ for the system with only a single kinase and phosphatase, indicating that the undesirable effect of non-zero  $f_{\alpha}$ and $f_{\beta}$ dominates. It is possible to choose catalytic probabilities such that $M_{0.9,0.1}$  initially falls with $P_{\rm on}$, before eventually rising to the non-ultrasensitive limit. To do this, however, requires such a large value of $\theta \phi$ in the limit  $P_{\rm on}\rightarrow 0$ that the degree of ultrasensitivity is always very small.

 It is interesting to note that lower values of $\phi$ allow ultrasensitivity to be moderately robust to increases in  $\alpha$ (pseudo-processivity in the phosphorylation pathway) provided $\beta$ (pseudo-processivity in the dephosphorylation pathway) is small. 
   This phenomenon can be seen clearly in Figure \ref{ultra-curves}\,(c), in which the system retains the majority of its ultrasensitivity at the point $\alpha=1$.
   To explain this observation, consider Figure \ref{ultra-robust}, which
   represents a system with $\alpha \rightarrow \infty$ and $\beta=0$. Although there is no intermediate state during phosphorylation, there is one present in dephosphorylation, and thus there are 
   still two stages in which phosphorylation can compete with dephosphorylation, permitting ultrasensitivity.
   As we have pointed out previously, establishing ultrasensitivity involves keeping the concentration of the intermediate low. If $\phi$ is low, the intermediate will necessarily be 
   drained quickly relative to the processive phosphorylation of $A$ to $A_{pp}$, and hence the concentration of $A_p$ remains low regardless of $Y$, permitting a sharp transition. If
   $\phi$ is not low, one of two things can happen:
   \begin{itemize}
    \item If $\theta$ is low, then at a concentration ratio $Y$ near the transition from $A$- to $A_{pp}$-dominated systems, the intermediate state during dephosphorylation has almost
    no effect, because it is overwhelmingly converted into $A$ rather than $A_{pp}$.
    Hence it is not effective in establishing ultrasensitivity, just like in fully a processive system in which intermediates are effectively always committed to end in $A_{pp}$ or $A$.
    \item If $\theta$ is high, then the intermediate state $A_p$ drains slowly, becoming very prevalent at the transition from $A$- to $A_{pp}$-dominated systems
    and precluding ultrasensitivity. 
   \end{itemize}
   These two cases, along with the low $\phi$ example, are illustrated schematically in Figure \ref{ultra-robust}.
   Mathematically, one can see that if $f_\beta = 0$ in Equation (\ref{ultrasensitivity}), the first term inside the square root is proportional to $\phi^2$, and thus can be small compared to the second term if $\phi$ is small. Such a result is advantageous as the second term in the square root is the one responsible for reducing $M_{g1,g2}$ below $M_{g1,g2}^{\rm proc}$, as it is larger in the denominator than the numerator. Although this result is intriguing, we note that one would require a very low intrinsic value of $\phi$ (that obtained in the limit $P_{\rm on} \rightarrow 0$) to maintain ultrasensitivity at large $\alpha$, especially as the higher values of $P_{\rm on}$ that cause  $\alpha$ to grow also cause $\phi\rightarrow 1$
  
  Low values of $\theta$ have an equivalent effect for $\beta \rightarrow \infty$ provided $\alpha $ is small. It is not possible, however, to preserve ultrasensitivity through lower values
  of $\theta$ and $\phi$ when both $\beta$ and $\alpha  \gtrsim 1$. Mathematically, when both $\alpha$ and $\beta$ are large,  the first term inside the square root in Equation (\ref{ultrasensitivity}) contains terms proportional to $\theta \phi$, and the argument given for moderate $\alpha$, low $\phi$ is invalid. For example, $\theta$ and $\phi$ are both low in Figure \ref{ultra-curves}\,(d), but  over half the ultrasensitivity present
  at $\alpha=\beta=0$ is lost by the time $\alpha=\beta=1$. Indeed, Equation (\ref{ultrasensitivity}) can be numerically minimized with respect to $\theta$ and $\phi$ at $\alpha=\beta=1$; the result is   $M_{0.9,0.1} = 1.50$ (obtained with $\theta\phi \ll 1$). Such a system is not strongly ultrasensitive; the transition from $g=0.1$ to $g=0.9$ is only 2.6 times more rapid than in the single-site case, and we also note that low values of $\theta \phi$ are more difficult to achieve when rebinding is frequent.  By the time  $\alpha=\beta=4$, the minimal value of $M_{0.9,0.1}$ is 1.72 and the transition from $g=0.1$ to $g=0.9$ is only 1.5 times more rapid than in a single-site system. These mathematical limits are quite distinct from the consequences of finite pseudo-processivity for proofreading; the efficacy of discrimination between substrates and proofreading is only capped by the biophysical limitations of the intrinsic selectivity parameters $S$ and $S^\prime$ , rather than mathematical constraints.
   
   It is important to note that the loss of ultrasensitivity is predominantly associated with 
   pseudo-processivity, rather than the basic rebinding kinetics as it was with proofreading. Although there is some variation in $M_{0.9,0.1}$ with $P_{\rm on}$ for the system with distinct enzymes (because high rebinding probabilities can influence $\theta$ and $\phi$ by making less favourable reactions more likely to be successful), at least for the illustrated systems it is  a small effect compared to the change with finite $f_\alpha$ and $f_\beta$.

\section{More examples of proofreading with pseudo-processivity}


    \begin{figure*}
        \centering
        \subfloat[]{
                \includegraphics[width=15pc, angle=-90]{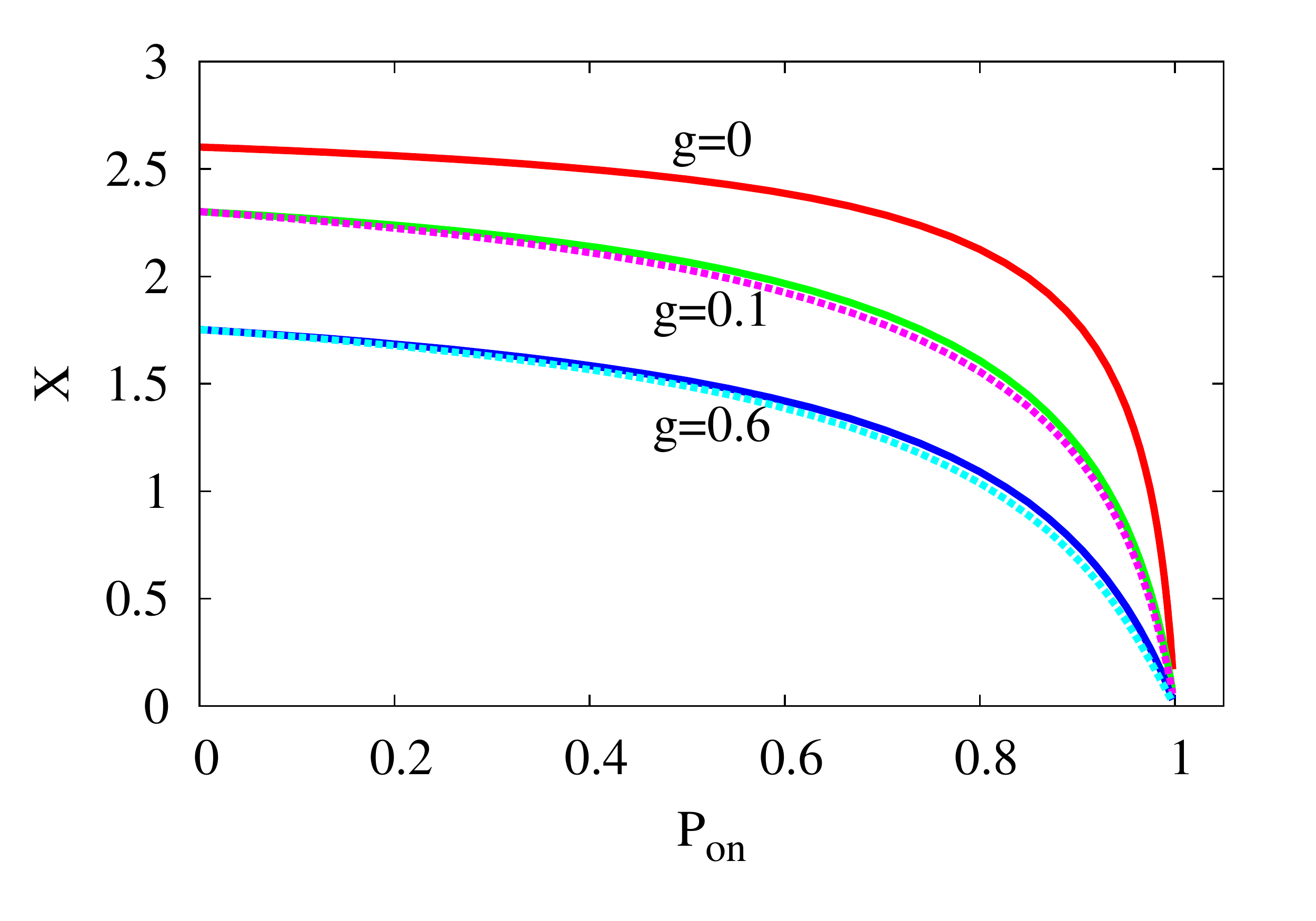}       
        }%
        ~ 
         \subfloat[]{
               \includegraphics[width=15pc, angle=-90]{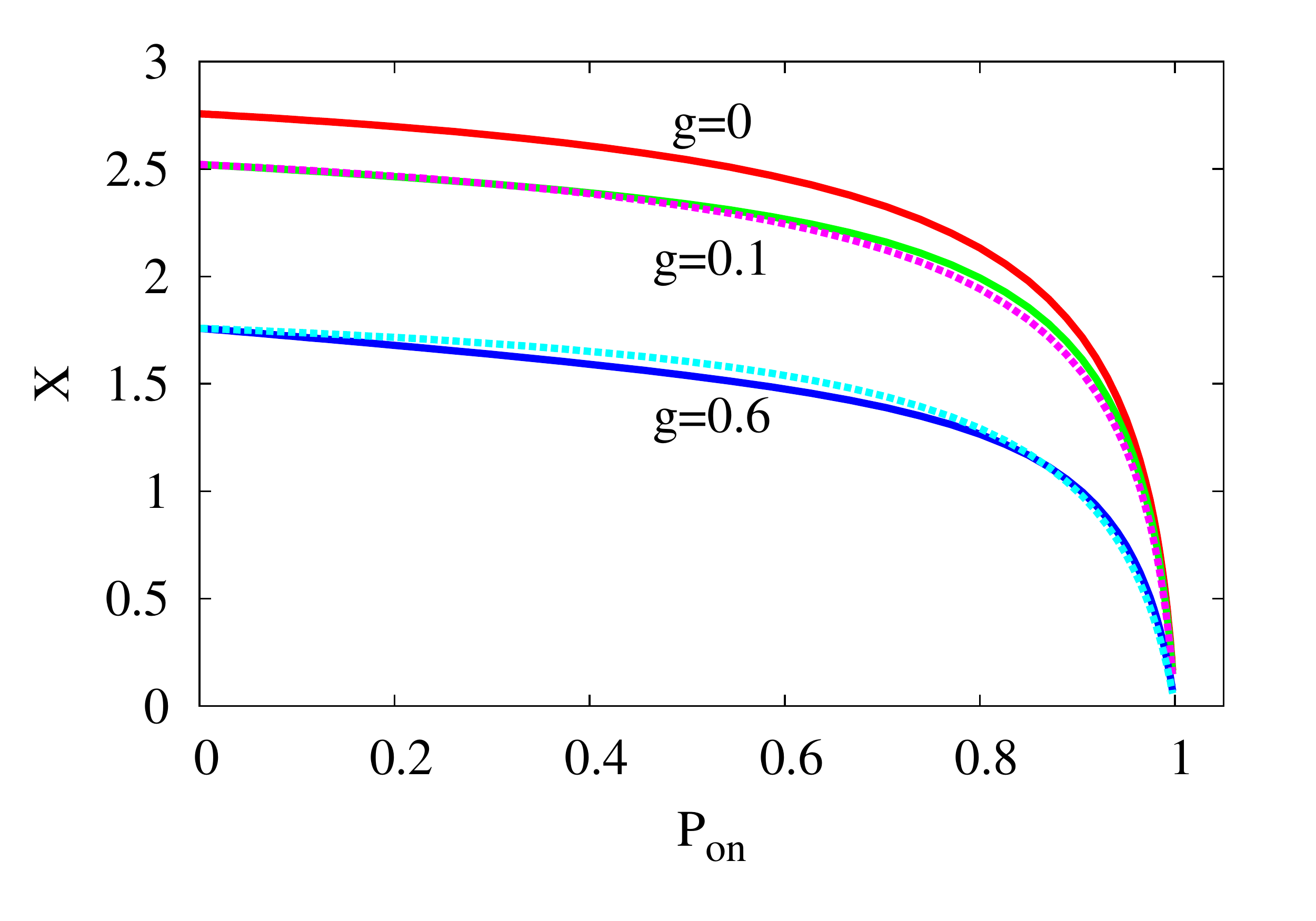}
        }
        ~ 
          
        \subfloat[]{
                \includegraphics[width=15pc, angle=-90]{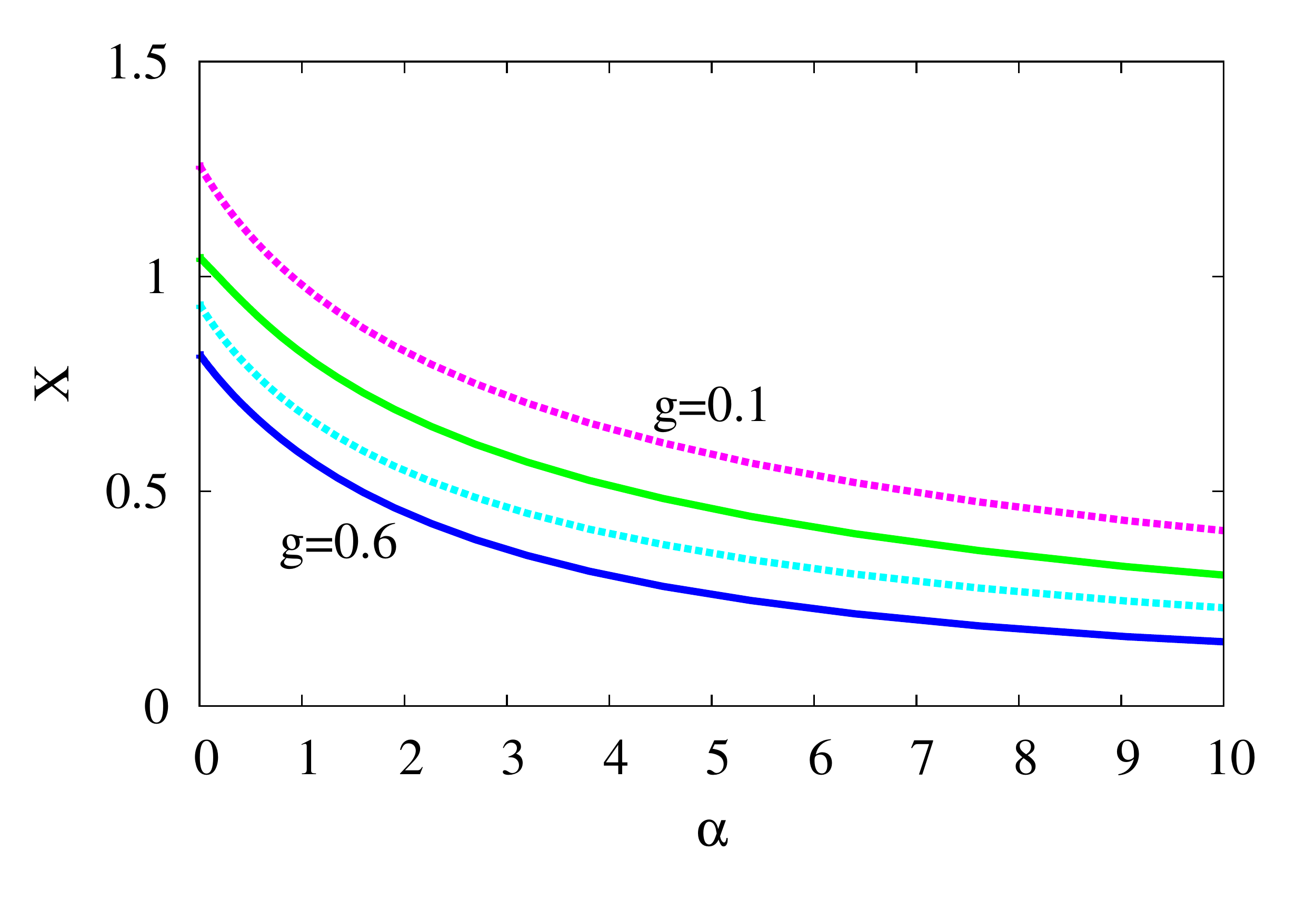}         
        }%
        ~ 
        \subfloat[]{
                \includegraphics[width=15pc, angle=-90]{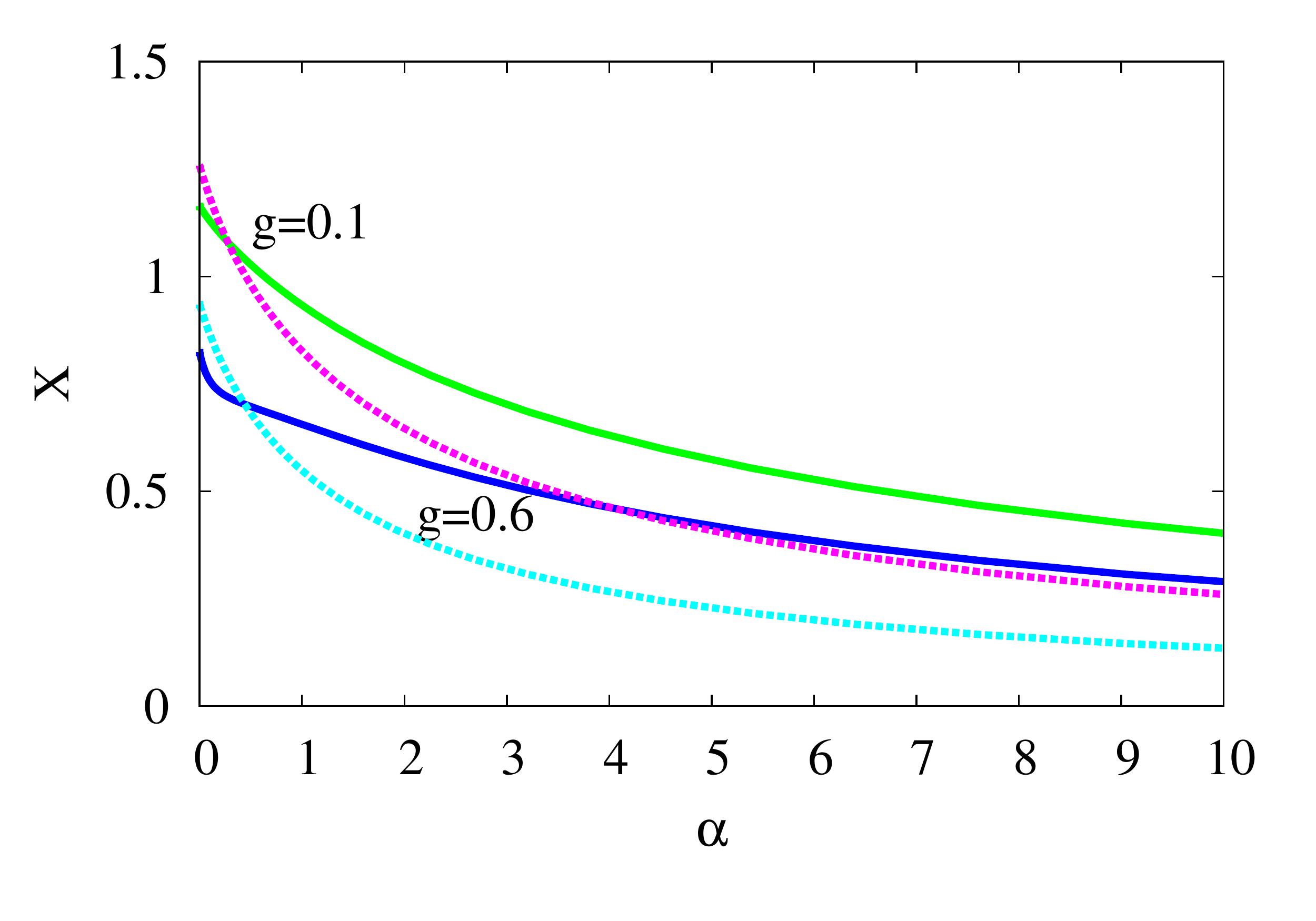}
        }
        ~ 
        \caption{\footnotesize{Other examples of the effect of pseudo-processivity on proofreading, outside the regime explored in Section \ref{dephos_sensitivity} (weak residual effects of the behaviour discussed in Section \ref{dephos_sensitivity}  can be seen in (d)). In (a) and (b), we plot specificity $X$ as a function of $P_{\rm on}$ at fixed $[A_{pp}]/[A_0]$ for potentially pseudo-processive systems (solid lines) and distributive systems with two kinases and phosphatases (dashed lines), at $g=0,0.1$ and 0.6. Only a single solid line is plotted for $g=0$, when the two systems produce identical results. In (c) and (d) we plot the specificity $X_{\rm ss}$ (dashed lines) given a system with only the first phosphorylation site,  and the additional specificity provided by the second site $X - X_{\rm ss}$ (solid lines), against the processivity ratio $\alpha = f_\alpha /(1 - f_\alpha)$ (which itself is a function of $P_{\rm on}$). Only the potentially pseudo-processive system with a single kinase and phosphatase is considered, and only results for $g=0.1$ and $g=0.6$ are presented for clarity.
 For (a) and (c) we take $P^{A}_{\rm cat}=0.4$, $P^{A \prime}_{\rm cat}=0.4$, 
$P^{B}_{\rm cat}=0.02$, $P^{B \prime}_{\rm cat}=0.03$, $Q_{\rm cat}=0.02$, and $Q^{\prime}_{\rm cat}=0.06$, for (b) and (d) we take $P^{A}_{\rm cat}=0.4$, $P^{A \prime}_{\rm cat}=0.2$, 
$P^{B}_{\rm cat}=0.02$, $P^{B \prime}_{\rm cat}=0.007$, $Q_{\rm cat}=1$ and $Q^{\prime}_{\rm cat}=1$. } 
        \label{other_proof}}
\end{figure*}

In this section we show data for additional choices of system parameters, to demonstrate that the results in the main text are not atypical. Results for two additional sets of parameters are given in Figure \ref{other_proof}. We plot $X$ against $P_{\rm on}$, with the results when distinct kinases and phosphatases are required for each step also
shown for comparison. We also compare the specificity gained from the first and second phosphorylation stages, $X_{\rm ss}$ and $X- X_{\rm ss}$, and plotting both against  $\alpha$.
Although the results are numerically distinct, they do not contradict our claims from the main text; 
namely that although specificity decreases with increased processivity, this is usually a result of a reduction in intrinsic specificity due to rebinding, that specificity
can be robust to moderate processivity in the $A$ phosphorylation pathway, and that the contribution of the second site to specificity can also remain substantial. Higher intrinsic specificities would give increased tolerance to finite rebinding probabilities and yields.




\section{Pseudo-processivity with enzymatic interactions differentiated by $k_a$ or $k_{\rm cat}$}
In the main text, we considered enzymatic reactions that were only differentiated by the rate at which reactants unbound from each other. Alternative limits might be that
reactants are distinguished by their intrinsic catalytic constants, or their intrinsic binding rates. In this section we show that our arguments also hold in these cases.

     \begin{figure*}
        \centering
        \subfloat[]{
                \includegraphics[width=15pc, angle=-90]{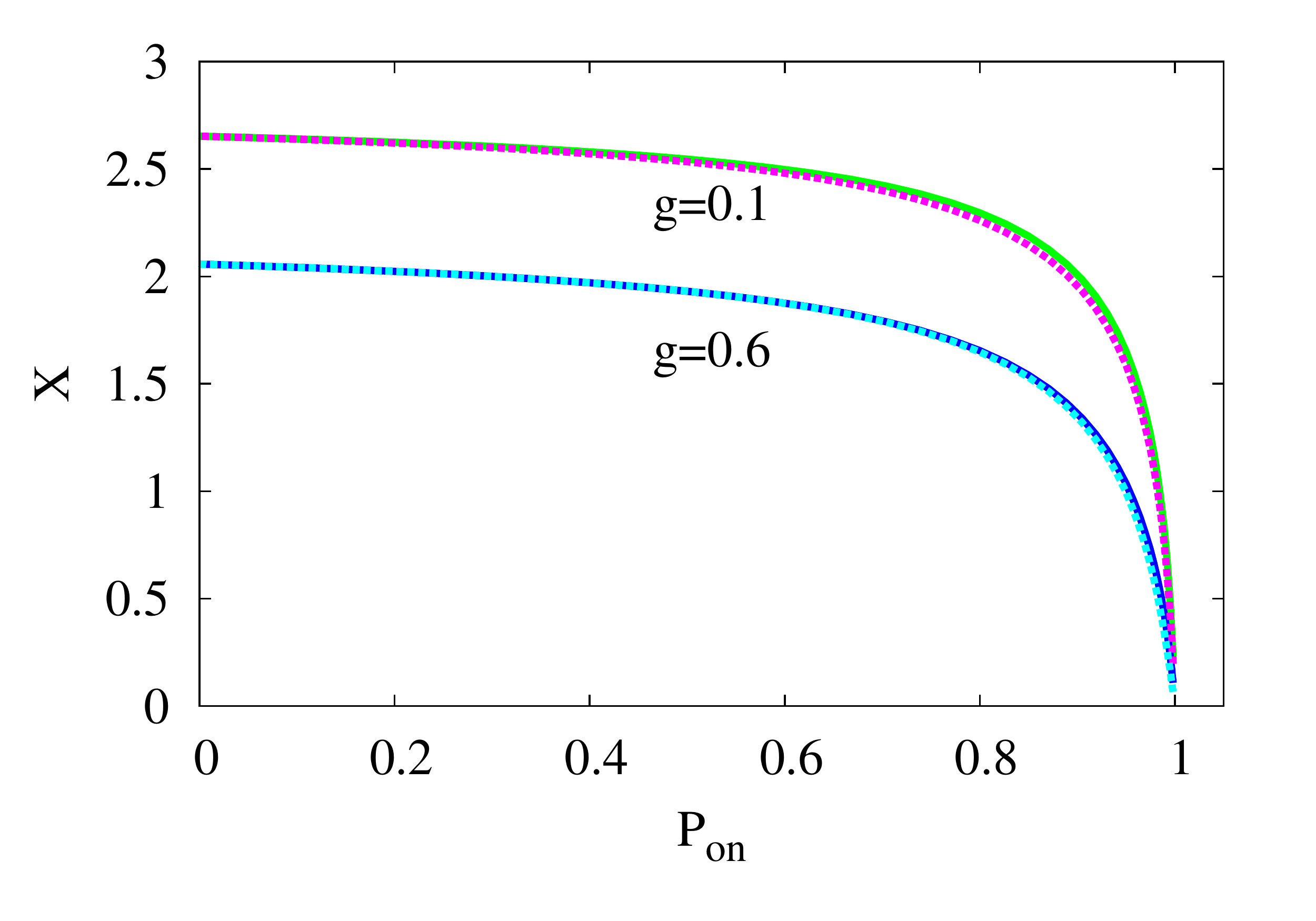}         
        }%
        ~ 
        \subfloat[]{
                \includegraphics[width=15pc, angle=-90]{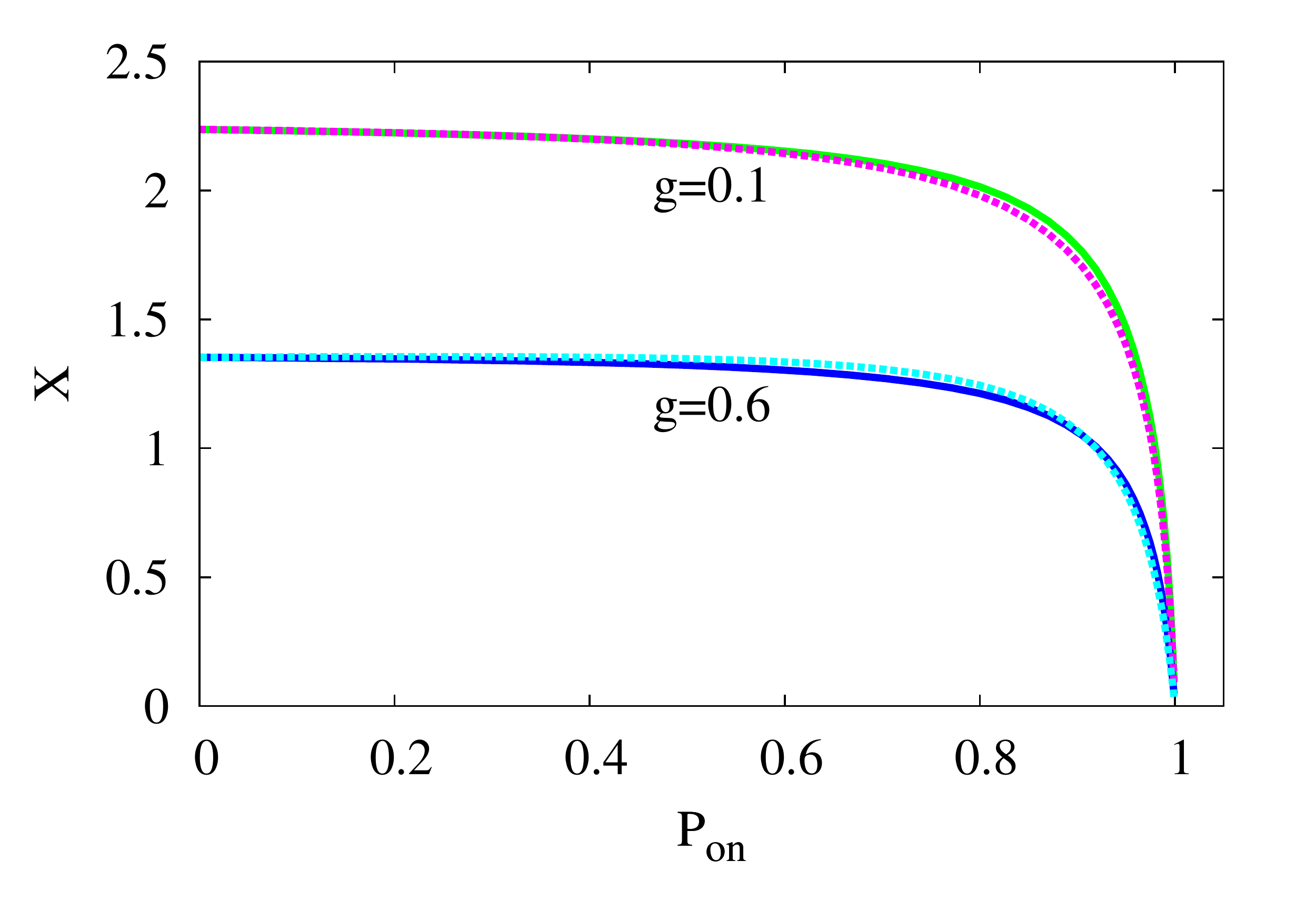}
        }
        ~ 
          
                  \subfloat[]{
                \includegraphics[width=15pc, angle=-90]{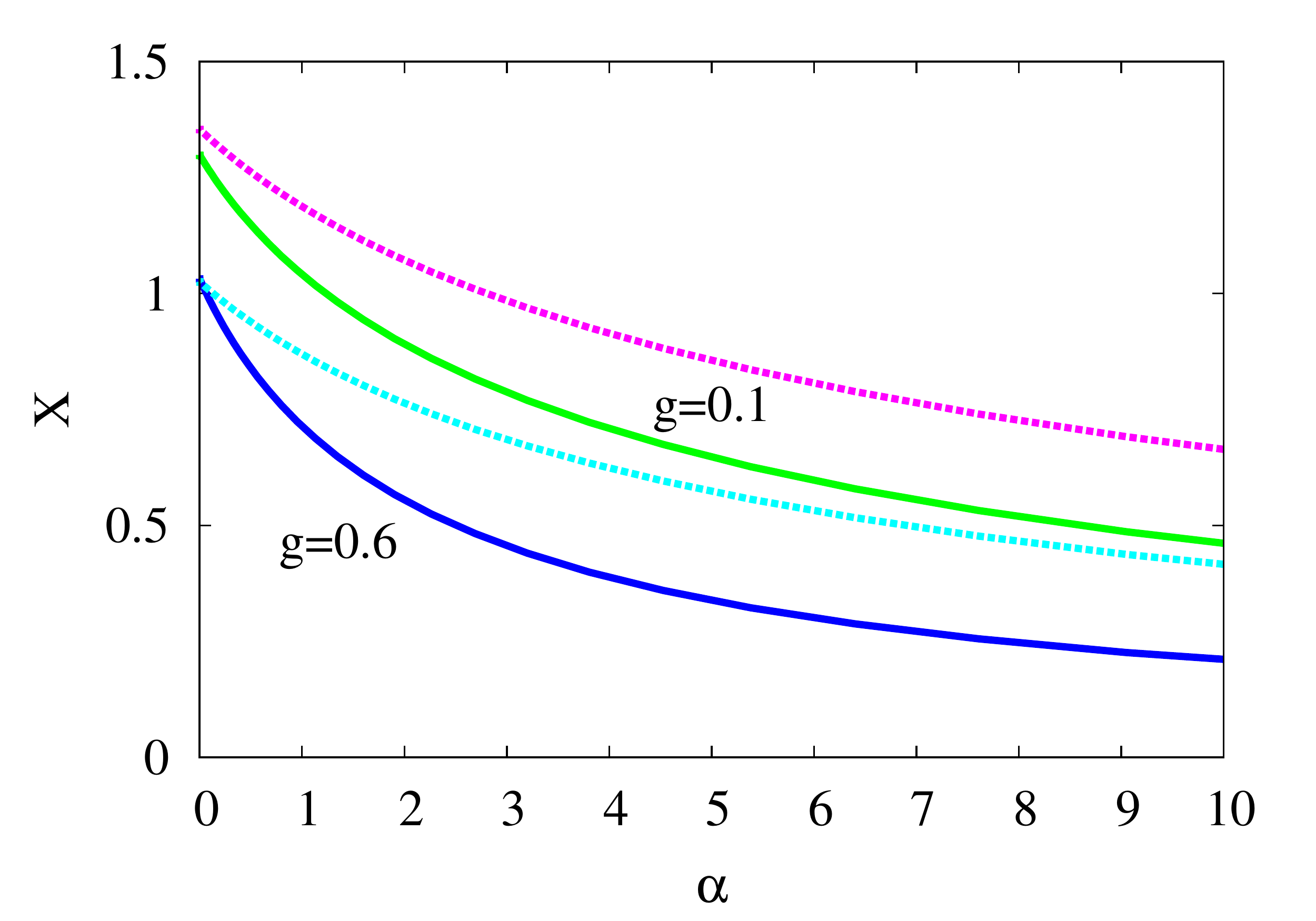}         
        }%
        ~ 
        \subfloat[]{
                \includegraphics[width=15pc, angle=-90]{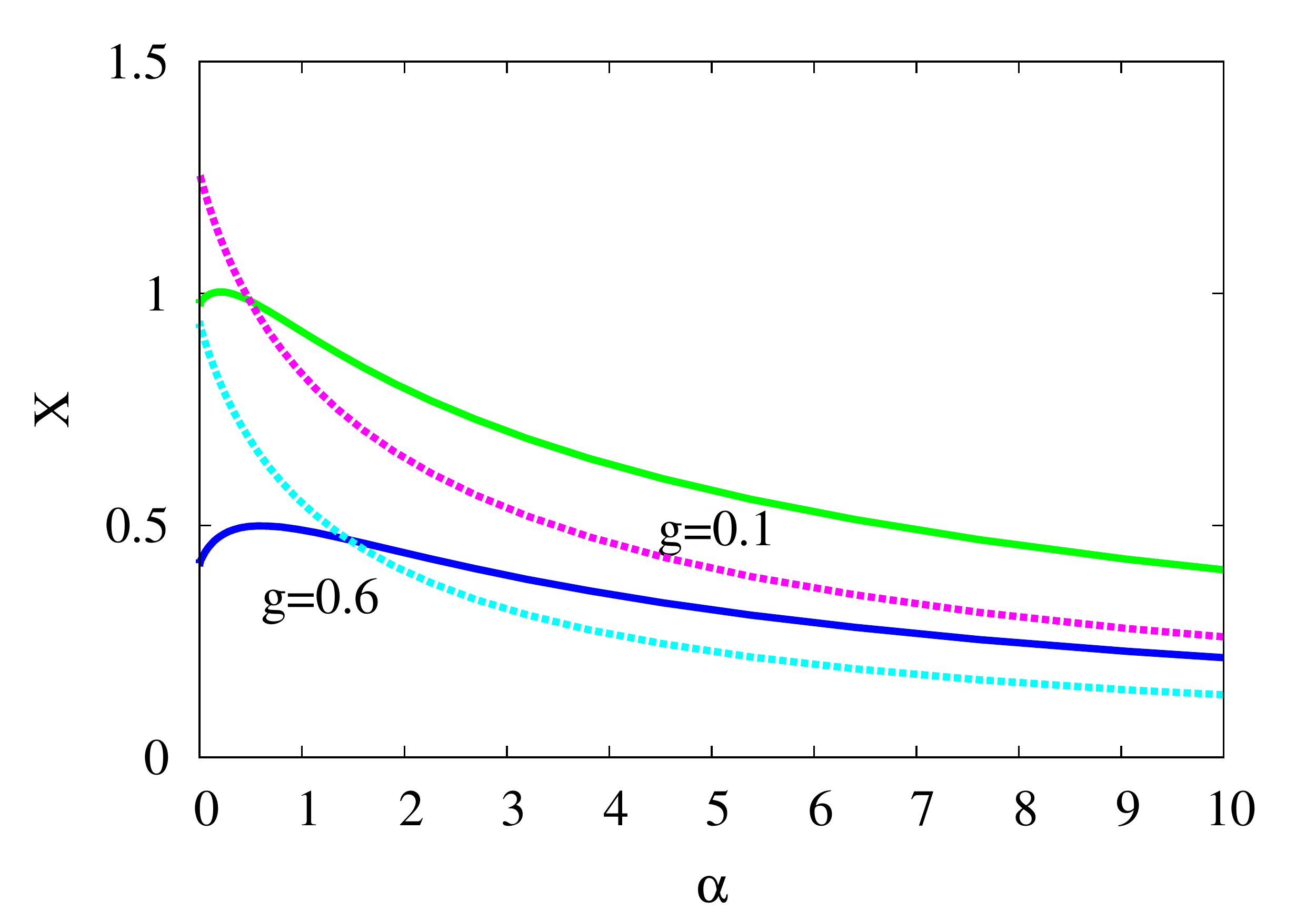}
        }
        ~ 
        \caption{\footnotesize{Example results for systems in which enzymatic efficiencies are differentiated 
        by binding rates.  In (a) and (b), we plot specificity $X$ as a function of $P^A_{\rm on}$ at fixed yield $g=[A_{pp}]/[A_0] =0.1,0.6$  for two distinct sets of parameters. We consider systems in which pseudo-processive behavior is possible (solid lines) and equivalent systems with with two kinase and phosphatase species in which reactions are necessarily distributive (dashed lines). In (c) and (d) we plot the specificity $X_{\rm ss}$ (dashed lines) given a system with only the first phosphorylation site,  and the additional specificity provided by the second site $X - X_{\rm ss}$ (solid lines), against the processivity ratio $\alpha = f_\alpha /(1 - f_\alpha)$ (which itself is a function of $P_{\rm on}$). Only the potentially pseudo-processive system with a single kinase and phosphatase is considered. For (a) and (c) we take relative on rates $k^{A}_{a}=0.5$, $k^{A \prime}_a=1$, 
$k^{B}_{a}=0.02$, $k^{B \prime}_{\rm cat}=0.04$, $h_{a}=0.1$, $h^{\prime}_{a}=0.1$ and $P_{\rm cat}=0.1$. For (b) and (d) we take $k^{A}_{a}=1$, $k^{A \prime}_a=0.5$, 
$k^{B}_{a}=0.1$, $k^{B \prime}_{\rm cat}=0.05$, $h_{a}=3$, $h^{\prime}_{a}=1$ and $P_{\rm cat}=0.2$. In (c), the specificity of the second site drops faster than the first as the second site is phosphorylated more efficiently, and thus feels the effect of $P_{\rm on}$  sooner (as is also observed for  systems in which enzymatic efficiencies are differentiated 
        by unbinding rates or catalytic rates). In (d), the slight non-monotonic behavior is associated with decreased $\theta^A \phi^A$ as rebinding becomes significant, making the unwanted effects of finite $[A_p]$ less severe. Such behaviour can also be seen for  systems in which enzymatic efficiencies are differentiated 
        by unbinding rates or catalytic rates, if $\theta^A \phi^A >1$ as $P_{\rm on} \rightarrow 0$.
        \label{onrate}}}
\end{figure*}

Firstly, if reactions are distinguished by $k_{\rm cat}$, the arguments of the main text follow exactly as before. This is because in this unsaturated model 
the effect of modulating $k_{\rm d}$ is only felt via $P_{\rm cat}$, and the same is true for $k_{\rm cat}$. Given a set of probabilities  $P_{\rm cat}$, $P^\prime_{\rm cat}$, 
$Q_{\rm cat}$ and $Q^\prime_{\rm cat}$, our results are equally valid whether differences in these quantities are due to intrinsic catalytic rates, intrinsic unbinding rates,
or a combination of the two. We note that when the system is not in the second-order regime and enzymes/substrates are saturated, we would expect distinct behaviour in some cases.

If the enzymatic interactions are differentiated by $k_{\rm a}$, the situation is a little different. Now $P_{\rm cat} = P^\prime_{\rm cat} =Q_{\rm cat} = Q^\prime_{\rm cat}$, but 
$P_{\rm on}$, $P^\prime_{\rm on}$, $Q_{\rm on}$ and $Q^\prime_{\rm on}$ are all distinct (and different for $A$ and $B$). The overall specificity $X$ is still given by 
 \begin{equation}
      \begin{array}{c}
       X = \lg \left(S S^\prime \right) + \\
       \\
      \lg \left( \frac{
      \phi^A 
      +  \frac{Y \psi^A}{S^\prime} 
      \left(\frac{f^A_\alpha \phi^A}{S} + f^A_\beta \theta^A 
      + \frac{S^\prime-f^A_\alpha f^A_\beta}{S} {\theta^A \phi^A}\right)
      + \frac{\left(Y \psi^A \right)^2}{ S S^\prime} {\theta^A}
      }
      {\phi^A +  Y \psi^A  \left(f^A_\alpha \phi^A + f^A_\beta \theta^A + \left(1-f^A_\alpha f^A_\beta \right) \theta^A \phi^A \right)  + \left(Y \psi^A \right)^2 \theta^A }\right), 
      \end{array}
      \label{X-full}
      \end{equation}
but now the selectivity factors and rate constant ratios are
\begin{equation}
 \begin{array}{c}
      S = \left(\frac{P^A_{\rm on}}{P^B_{\rm on}} \right) \left( \frac{1- P^B_{\rm on}(1- P_{\rm cat})}{1- P^A_{\rm on}(1- P_{\rm cat})}\right), \\
      \\
      S^\prime = \left( \frac{{P^{A \prime}_{\rm on}}} {{P^{B \prime}_{\rm on}}}\right) \left(\frac{1- {P^{B \prime}_{\rm on}(1- {P}_{\rm cat}})}{1- {P^{A \prime}_{\rm on}}(1- {P_{\rm cat}})}\right).\\
      \\
      \theta^A =\left(\frac{Q^A_{\rm on}}{Q^{A\prime}_{\rm on}} \right) \left( \frac{1- Q^{A\prime}_{\rm on}(1- P_{\rm cat})}{1- Q^A_{\rm on}(1- P_{\rm cat})}\right),\\
      \\
      \phi^A = \left(\frac{P^A_{\rm on}}{P^{A\prime}_{\rm on}} \right) \left( \frac{1- P^{A\prime}_{\rm on}(1- P_{\rm cat})}{1- P^A_{\rm on}(1- P_{\rm cat})}\right),\\
      \\
      \psi^A = \left(\frac{P^A_{\rm on}}{Q^{A}_{\rm on}} \right) \left( \frac{1- Q^{A}_{\rm on}(1- P_{\rm cat})}{1- P^A_{\rm on}(1- P_{\rm cat})}\right).\\
      \\
      \end{array}
      \label{S-equation}
    \end{equation}
$S$ and $S^\prime$ should be compared with Equation (7) of the main text -- note in particular that the first factor now depends on $P_{\rm on}$ values rather than $P_{\rm cat}$ values.

Unlike our previous analyses, there is no single $P_{\rm on}$ for all reactions that can be varied 
    while all other parameters are kept fixed to model increased crowding, and $S$ and $S^\prime$ cannot be expressed as succinctly in terms of $P^A_{\rm react}$ and $P^{A\prime}_{\rm react}$ as in Equation (8) of the main text. Instead, we can fix $P_{\rm cat}$, 
    $k^{A}_a$, $k^{A \prime}_a$, $h^{A}_a$, $h^{A \prime}_a$, $k^{B}_a$, $k^{B \prime}_a$, $h^{B}_a$ and $h^{B \prime}_a$  and vary $k_{\rm esc}$, which leads to simultaneous and coordinated variation of all binding probabilities in a way that reflects  modulated diffusion. If this is done, it can be seen that $S$ and $S^\prime$ respond to lower $k_{\rm esc}$ (and hence increased rebinding probabilities) in a similar way to the original system, falling from $S_0, S_0^\prime = k^A_{a}/k^B_{a}, k^{A \prime}_{a}/k^{B \prime}_{a}$ in the limit of large $k_{\rm esc}$ to unity as $k_{\rm esc} \rightarrow 1$.
    \begin{equation}
    \begin{array}{c}
    S = S_0 \frac{k^{B}_a + k_{\rm esc}}{k^A_a + k_{\rm esc}} ( 1 - P^A_{\rm react}(1 - 1/P_{\rm cat}))   \vspace{2mm}\\
    +   P^A_{\rm react}(1 - 1/P_{\rm cat})),\\
    \\
    S^\prime = S^\prime_0 \frac{k^{B \prime}_a + k_{\rm esc}}{k^{A\prime}_a + k_{\rm esc}} ( 1 - P^{A\prime}_{\rm react}(1 - 1/P_{\rm cat}))   \vspace{2mm}\\
    +   P^{A\prime}_{\rm react}(1 - 1/P_{\rm cat})).\\
    \end{array}
    \end{equation}
    Here, $ P^A_{\rm react}$ and $ P^{A \prime}_{\rm react}$ are defined as probabilities that given reactions occur once that the reactants are in close proximity, exactly as in Equations (3) and (4) of the main text. These quantities both tend to unity as $k_{\rm esc} \rightarrow 0$ with all other parameters fixed ({\it i.e.}, as rebinding becomes increasingly certain).
    \vspace{6mm}
    
   $\theta^A$, $\phi^A$ and $\psi^A$ all tend to unity as $k_{\rm esc} \rightarrow 0$; this was also true in the alternative case considered in the main text, in which substrates were distinguished by unbinding rates. Therefore, given that all relevant parameters of Equation (\ref{X-full}) respond to increased rebinding similarly to the system in the main text, and that the actual expression for $X$ in terms of these parameters in Equation (\ref{X-full}) is identical to Equation (9) of the main text, it seems reasonable that our conclusions will not be qualitatively affected. In any case, the precise behavior will always depend on the details of the system; we are not concerned with these details, only the overall trend. As an example that this trend is preserved, we plot  the behavior of two systems in which enzymatic reactions are distinguished by $k_{\rm a}$ in Figure \ref{onrate}, showing results that do not contradict our previous arguments.

\section{Pseudo-processivity with two independent phosphorylation sites}
Throughout this work we have considered a strictly ordered phosphorylation process, so that there is only one intermediate phosphorylation state $A_p$. This has been done for 
the sake of simplicity, but in general systems will have two intermediate states, corresponding to either one of the two residues being phosphorylated. In this section we 
discuss the behavior of the system if the sites are phosphorylated independently, an alternative limit.

To do this, it is necessary to introduce a second singly-phosphorylated intermediate, $A^p$. In this section, we will describe the fully phosphorylated state as $A^p_p$.
Given our unsaturated and noiseless assumptions, the differential equations governing the four species are
 \vspace{0.1mm}
  \begin{widetext}
\begin{equation}
 \begin{array}{c}
  \dot{[A]} = -k_{D}(P_1 + P_2) [K][A] + k_D[P](Q_1^\prime [A_p] +Q_2^\prime [A^p] +(Q_2 Q_1^\prime + Q_1 Q_2^\prime)[A_p^p] ),\\
\\
    \dot{[A_p]} = -k_D P_2^\prime [K][A_p] - k_D Q_1^\prime [P][A_p] + k_D[K][A] P_1(1-P_2^\prime) + k_D[P][A_p^p] Q_2(1-Q_1^\prime), \\
 \\
    \dot{[A^p]} = -k_D P_1^\prime [K][A^p] - k_{D}[P]Q_2^\prime [A^p] + k_D[K][A] P_2(1-P_1^\prime) + k_D[P][A_p^p] Q_1(1-Q_2^\prime),\\
  \\
\\
  {\rm and} \hspace{3mm}\left[ A_0 \right] = [A]+ [A_p] +[A^p] +[A_{p}^p].
  \end{array}
  \label{system_multisite}
  \end{equation}
  \end{widetext}
  For clarity we have not labelled constants to indicate that they belong to the phosphorylation cycle of $A$ rather than $B$.
  When $K$ and unphosphorylated $A$ come into close proximity, three things can happen initially. Either site 1 can be phosphorylated, or site 2, or the two proteins could diffuse apart.
  $P_1$ is the probability of the the first of those, and $P_2$ the second. $Q_1$ and $Q_2$ are equivalents for dephosphorylation when $A_p^p$ and $P$ come into close
  proximity. Primed quantities represent the probabilities that phosphorylation/dephosphorylation reactions occur given that the system is an intermediate state ($P_{1,2}^\prime > P_{1,2}$
  as from an intermediate state there is no competition from the other site for phosphorylation). These probabilities are given by
  \begin{equation}
    \begin{array}{c}
   P_1 = \frac{k_a P_{{\rm cat}1}}{k_{\rm esc} + k_a P_{\rm cat 1} + k_a P_{\rm cat 2}},\\
   \\
   P^\prime_1 = \frac{k_a P_{{\rm cat}1}}{k_{\rm esc} + k_a P_{\rm cat 1}}\\
   \\
      P_2 = \frac{k_a P_{{\rm cat}2}}{k_{\rm esc} + k_a P_{\rm cat 1} + k_a P_{\rm cat 2}},\\
   \\
   P^\prime_2 = \frac{k_a P_{{\rm cat}2}}{k_{\rm esc} + k_a P_{\rm cat 2}}\\
   \\
      Q_1 = \frac{k_a Q_{{\rm cat}1}}{k_{\rm esc} + k_a Q_{\rm cat 1} + k_a Q_{\rm cat 2}},\\
   \\
   Q^\prime_1 = \frac{k_a Q_{{\rm cat}1}}{k_{\rm esc} + k_a Q_{\rm cat 1}}\\
   \\
      Q_2 = \frac{k_a Q_{{\rm cat}2}}{k_{\rm esc} + k_a Q_{\rm cat 1} + k_a Q_{\rm cat 2}},\\
   \\
   Q^\prime_2 = \frac{k_a Q_{{\rm cat}2}}{k_{\rm esc} + k_a Q_{\rm cat 2}},\\
    \end{array}
    \label{multisite}
  \end{equation}
where we have again assumed for simplicity that all intrinsic binding rates given close proximity are equal to $k_a$. The steady state solution of Equations (\ref{multisite}) can be found directly, although the result is  unwieldy and we have not found a simple way of expressing it such as in Equation (5) of the main text for the sequential system. In the special case that the two sites are equivalent, a simple expression can be found
\begin{equation}
 \begin{array}{c}
  \frac{[A]}{[A_0]} = \frac{\phi + (Y \psi) \theta f_\beta}{\phi + (Y \psi) (\theta f_\beta + \phi f_\alpha + \theta \phi (1-f_{\alpha} f_{\beta})) + (Y \psi)^2 \theta},\vspace{1mm}\\
    \frac{[A_p]}{[A_0]} = \frac{[A^p]}{[A_0]} = \frac{(Y \psi) \theta \phi (1-f_{\alpha}  f_{\beta})/2}{\phi + (Y \psi) (\theta f_\beta + \phi f_\alpha + \theta \phi (1-f_{\alpha} f_{\beta})) + (Y \psi)^2 \theta},\vspace{1mm}\\
    \frac{[A^p_{p}]}{[A_0]} = \frac{(Y \psi) \phi f_\alpha + (Y \psi)^2 \theta}{\phi + (Y \psi) (\theta f_\beta + \phi f_\alpha + \theta \phi (1-f_{\alpha}f_\beta)) + (Y \psi)^2 \theta}.\\
  \end{array}
  \label{solution_identical}
  \end{equation}
In this expression	
\begin{equation}
 \begin{array}{c}
 \phi = \frac{k_{{\rm eff}} }{k^\prime_{{\rm eff}} }= \frac{2P_1}{P_1^\prime} = \frac{2 (k_{\rm esc} + k_a P_{\rm cat,1})}{k_{\rm esc} + 2k_a P_{\rm cat,1}}, \vspace{2mm} \\
  \theta= \frac{h_{{\rm eff}} }{h^\prime_{{\rm eff}} }= \frac{2Q_1}{Q_1^\prime} = \frac{2 (k_{\rm esc} + k_a Q_{\rm cat,1})}{k_{\rm esc} + 2k_a Q_{\rm cat,1}},\vspace{2mm} \\
   \psi = \frac{k_{{\rm eff}} }{h_{{\rm eff}} }= \frac{P_1}{Q_1} =\frac{(k_a P_{{\rm cat}1})(k_{\rm esc} + 2k_a Q_{\rm cat 1})}{(k_{\rm esc} + 2k_a P_{\rm cat 1})(k_a Q_{{\rm cat}1})}, \vspace{2mm} \\
   Y= [K]/[P], \vspace{2mm} \\
   f_\alpha = P_1^\prime = \frac{k_a P_{{\rm cat}1}}{k_{\rm esc} + k_a P_{\rm cat 1}}, \vspace{2mm} \\
   f_\beta =  Q_1^\prime = \frac{k_a Q_{{\rm cat}1}}{k_{\rm esc} + k_a Q_{\rm cat 1}},
 \end{array}
  \label{solution_identical}
  \end{equation}
  are direct analogs of the quantities used in the sequential case. Equation (\ref{solution_identical}) is identical to Equation (5) of the main text, if the total intermediate concentration $[A_p] +[A^p]$ in Equation (\ref{solution_identical}) is mapped to the total intermediate concentration $[A_p]$ is Equation (5) of the main text. The only difference is that $\theta$ and $\phi$ are fundamentally constrained by the fact that the phosphorylation sites behave independently. Equation (\ref{solution_identical}) shows that $1 \leq \theta, \phi \leq 2$. For sequential phosphorylation, $\theta$ and $\phi$ are in principle unconstrained, except by biochemical details.  
  
 As a result, the behavior of a system with two independent and identical phosphorylation sites is the same as a sequential system with the same $\theta$, $\phi$, $\psi$, $Y$, $f_{\rm \alpha}$ and $f_{\rm \beta}$. However, in a sequential system it is in principle possible for $\theta \phi =\frac{k_{\rm eff} h_{\rm eff}}{k^\prime_{\rm eff}h^\prime_{\rm eff}} \rightarrow 0$, thereby limiting the undesired accumulation of phosphorylation intermediates through rapid conversion into fully-phosphorylated or unphosphorylated states. For a system with identical, independent sites $1 \leq \theta\phi \leq 4$, and it is impossible to limit finite concentrations of intermediates in this manner. From a more physical perspective, two intermediate states typically hold more proteins on aggregate than one, and it is difficult to keep their concentrations low as  if $A_p$ is drained rapidly, $A^p$ must be relatively long-lived.

The full solution for two independent phosphorylation sites is trivial to obtain but we have been unable to express it succinctly. Additionally, as discussed in the main text, we wish to make comparisons at fixed product yield, rather than fixed $Y=[K]/[P]$. In the case of independent, non-identical sites, solving for $Y$ at fixed $[A_p^p]/[A_0]$ corresponds to solving a cubic equation. Therefore it is difficult to draw conclusions from the analytic solutions - however, we can calculate specificity $X$ for specific systems as $k_{\rm esc}$ (and therefore binding probability) is varied whilst all other parameters are kept constant, analogously to the plots in Figures 5 and 6 in the main text.

\begin{figure*}
        \centering
        \subfloat[]{
                \includegraphics[width=15pc, angle=-90]{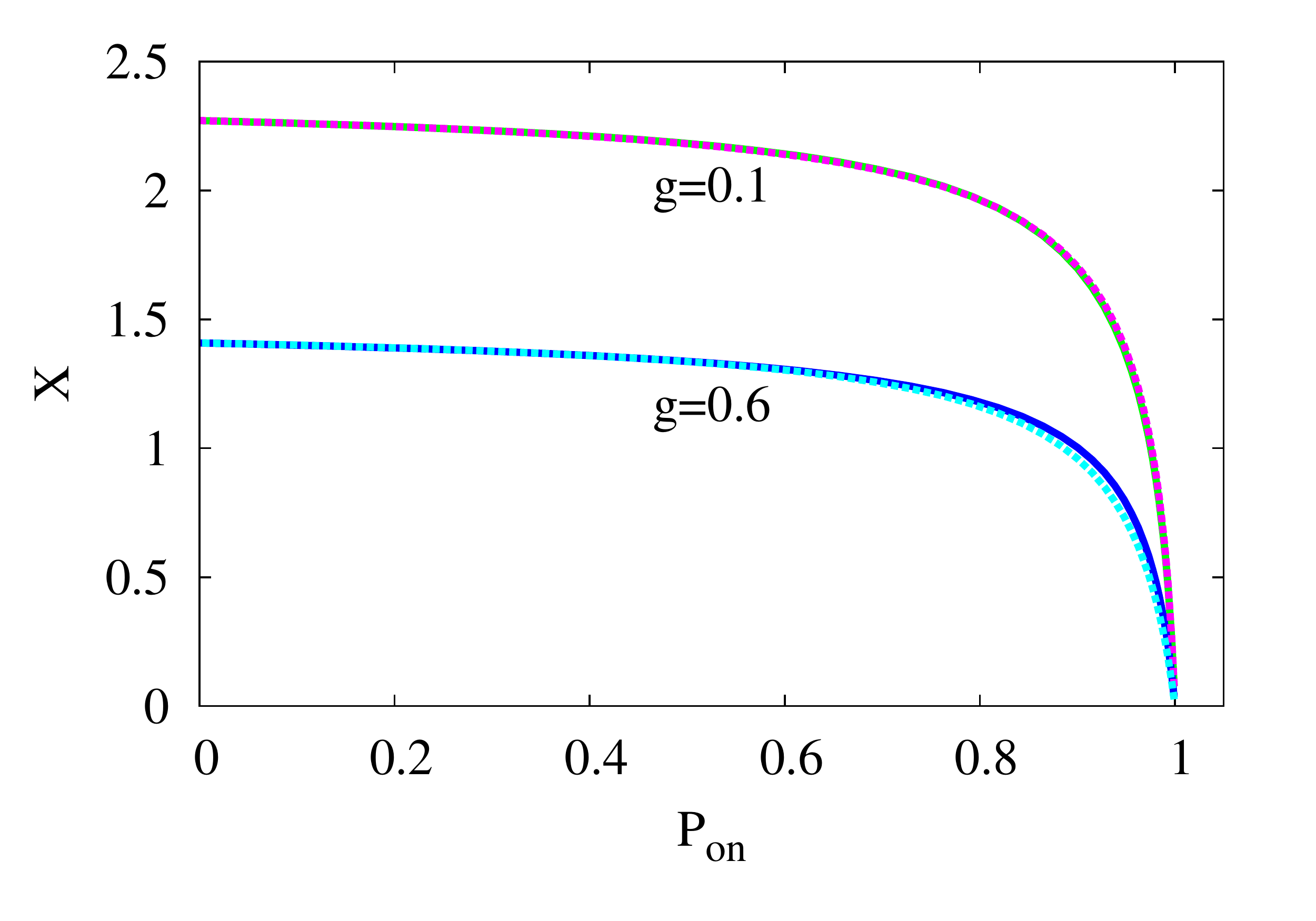}        
        }%
        ~ 
        \subfloat[]{
                \includegraphics[width=15pc, angle=-90]{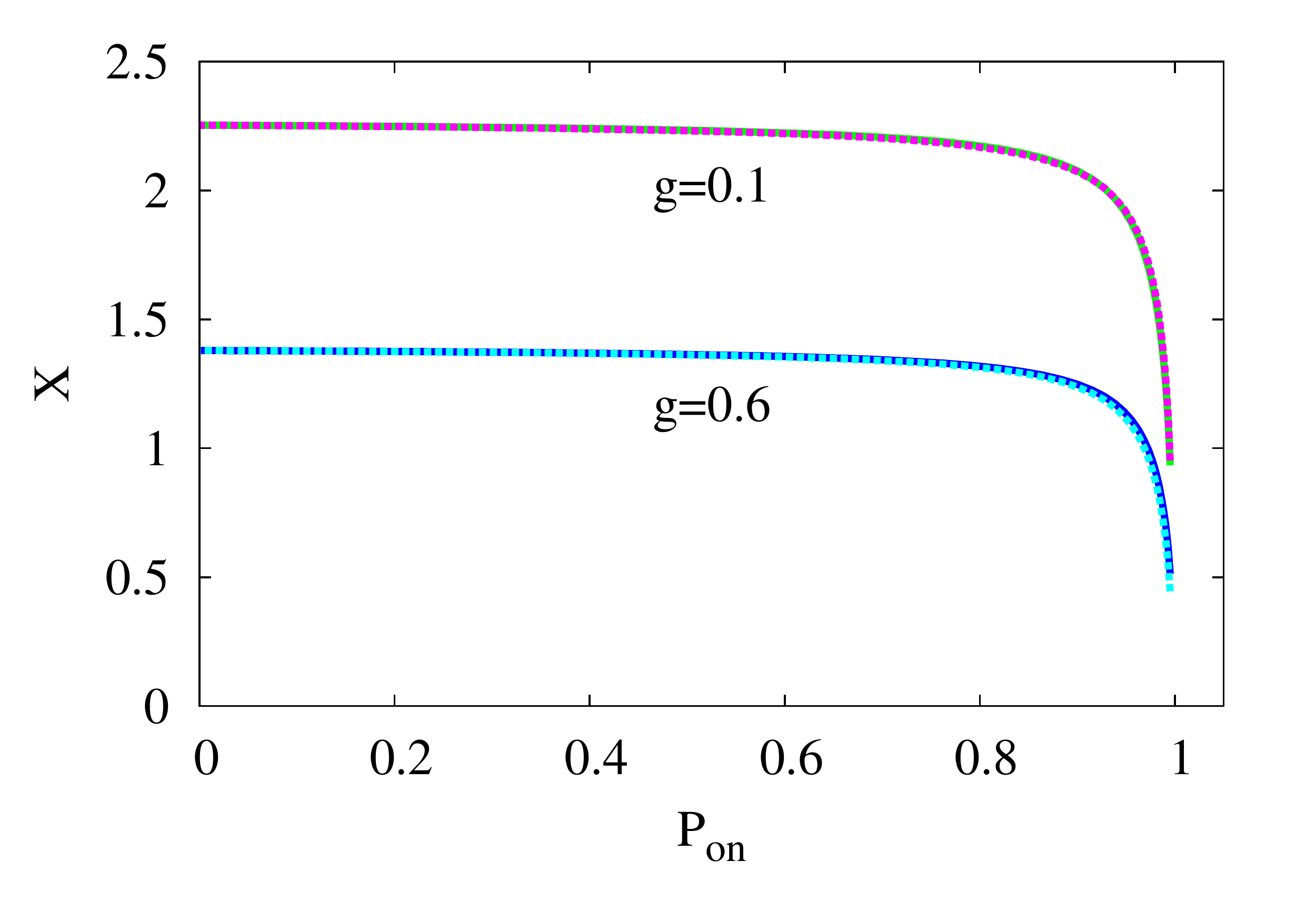}
        }
        ~ 
          
                  \subfloat[]{
               \includegraphics[width=15pc, angle=-90]{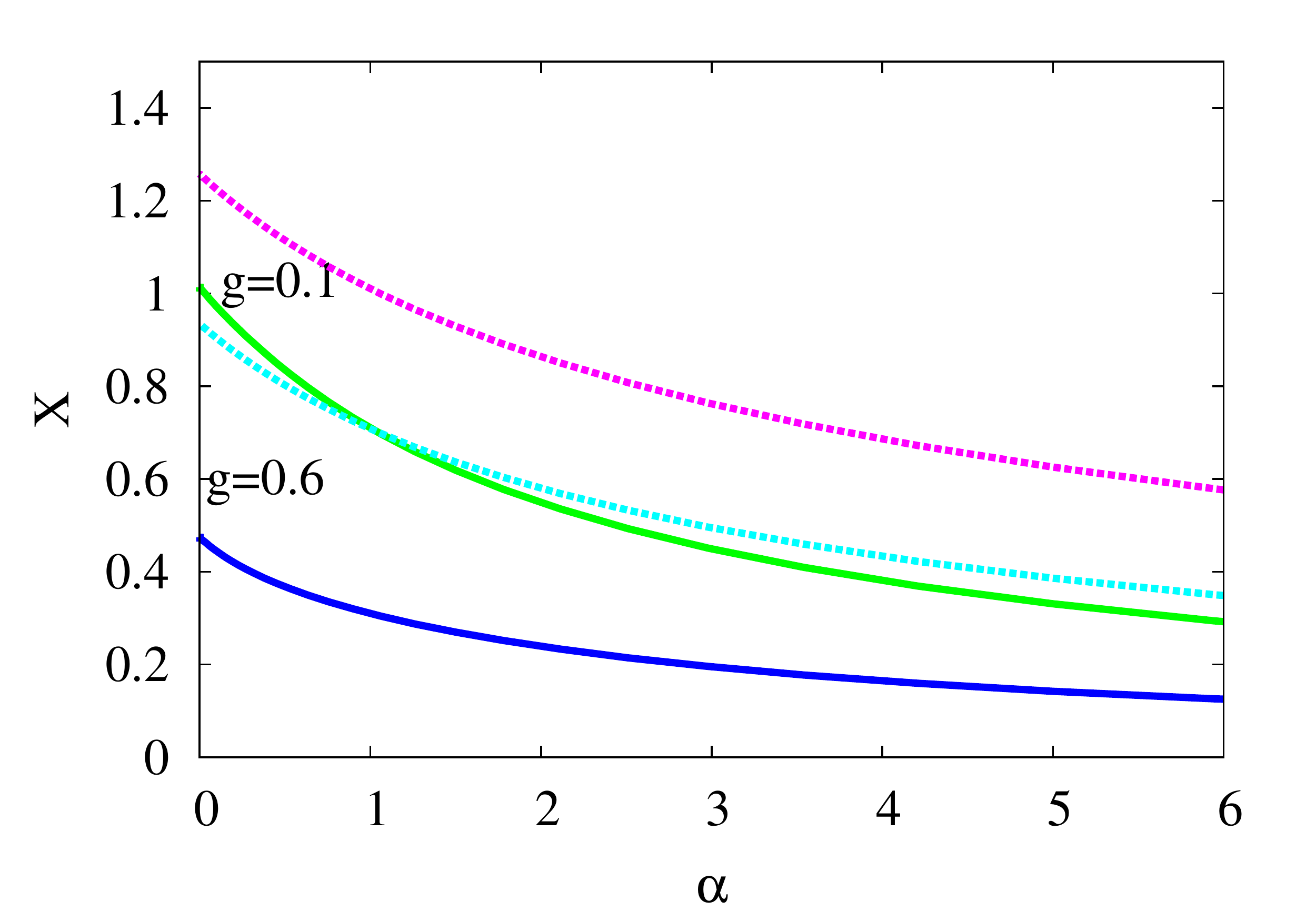}      
        }%
        ~ 
        \subfloat[]{
               \includegraphics[width=15pc, angle=-90]{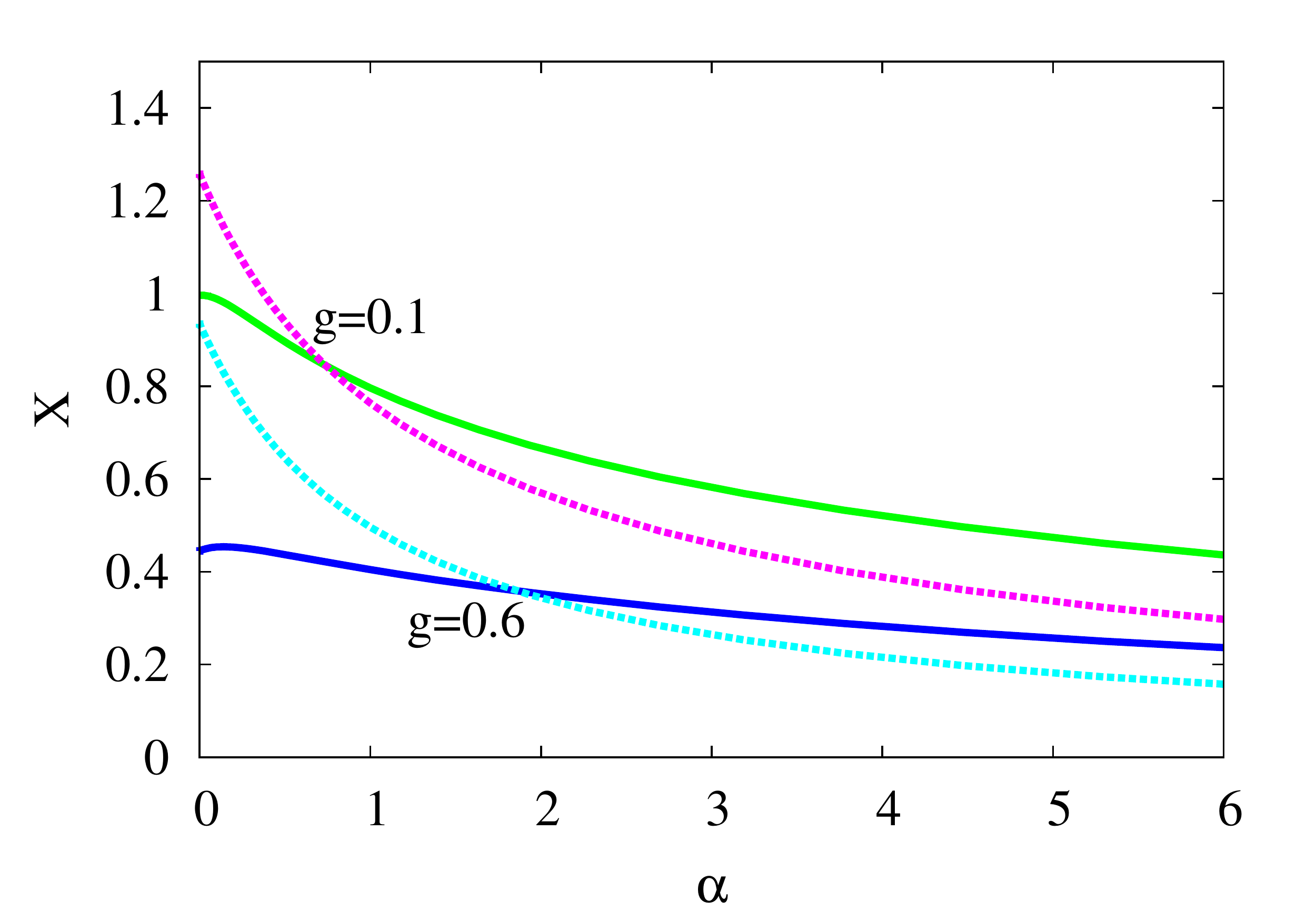}
        }
        ~ 
        \caption{\footnotesize{Example results for systems with two independent sites.  In (a) and (c), we plot specificity $X_{\rm independent}$ as a function of $P^{\rm tot}_{\rm on}$ at fixed $g=[A_{pp}]/[A_0]$ for two systems, along with comparison curves of equivalent systems in which separate kinases and phosphatases are needed for the two steps. We consider two product yields, $g= 0.1$ and $g = 0.6$. The results are so close that we do not separately label the single-enzyme and two-enzyme cases.In (c) and (d) we plot the specificity $X_{\rm ss}$ given a system with only the first phosphorylation site,  and the additional specificity provided by the second site $X_{\rm independent} -X_{\rm SS}$, against the processivity ratio $\alpha$ as defined in the main text. Only the potentially pseudo-processive systems with a single kinase and phosphatase species are considered. For (a) and (c) we take $P^{A}_{\rm cat}=0.2$, $P^{A \prime}_{\rm cat}=0.3$, 
$P^{B}_{\rm cat}=0.01$, $P^{B \prime}_{\rm cat}=0.015$, $Q_{\rm cat}=0.3$, $Q^{\prime}_{\rm cat}=0.2$. For (b) and (d) we take $P^{A}_{\rm cat}=0.1$, $P^{A \prime}_{\rm cat}=0.03$, 
$P^{B}_{\rm cat}=0.005$, $P^{B \prime}_{\rm cat}=0.0015$, $Q_{\rm cat}=0.2$, $Q^{\prime}_{\rm cat}=0.02$.} \label{two-sites1}}
\end{figure*}

We plot  $X_{\rm independent}$ as a function of $P_{\rm on}^{\rm tot} =  \frac{2 k_a}{2 k_a + k_{\rm esc}}$ (the probability that upstream and unphosphorylated downstream kinase bind given close proximity) for typical systems in Figure \ref{two-sites1}, with the specificity for the same system but with two distinct kinases and two distinct phosphatases shown for comparison. $X_{\rm independent}$ 
and $X_{\rm independent}-X_{\rm ss}$ are also plotted against the the pseudo-processivity factor 
\begin{equation}
\alpha =  \frac{P_1 P_2^{\prime} + P_2 P_1^{ \prime}}{P_1 (1-P_2^{\prime}) + P_2 (1-P_1^{\prime})} \\
\end{equation}
in Figure  \ref{two-sites1}. This factor is equivalent to that  for the ordered process, but averaged over the two pathways. 

The results are consistent with the general picture obtained with identical, independent sites. The findings of the main text still apply, but stronger saturation effects associated with the intermediates mean that adding a second interaction site is generally somewhat less favourable than if the
phosphorylation mechanism is ordered. In particular, Figures \ref{two-sites1}\,(c) and (d) show that the additional contribution to specificity of the second site is relatively low at higher yields ($g=0.6$) for these systems. This is nothing to do with pseudo-processivity, however: the additional specificity is already suppressed at small $\alpha$, and in purely distributive systems. It is also clear from Figures \ref{two-sites1}\,(a) and (b) that at least for these parameters the loss of specificity with increasing $P_{\rm on}$ is due to a loss of an ability to differentiate substrates due to rebinding, rather than 
pseudo-processivity {\it per se}, as generally found in the ordered case. Furthermore, specificity and proofreading can still be robust to moderate processivity, and would be more so if the intrinsic specifities $S$ and $S^\prime$ used for illustrative purposes were higher.

It is worth noting that ultrasensitivity is also adversely affected by switching from ordered phosphorylation to independent sites.\cite{Salazar2009} In the single-intermediate system, 
ultrasensitity was favoured by $\theta \phi <1$ (fast draining of intermediates), and disrupted by $\theta \phi >1$, because intermediate states with a significant population preclude a 
sharp transition. The existence of two intermediates that cannot both drain rapidly magnifies this issue, strongly limiting ultrasensitivity (and also compromising the mechanism
by which ultrasensitivty can be robust to increases in $\alpha$ that was discussed in Section \ref{sec-ultrasensitivity}).\\
\\

\section{Finite concentrations of bound enzymes and substrates}
This paper has been devoted to systems in which the concentrations of bound enzyme/substrate pairs, for instance $[KA_p]$, are negligible. This has made the analysis simpler, resulting in linear equations for the steady state that can be solved analytically, and greatly reducing the number of parameters that are relevant, allowing a systematic study. The additional complexities of the full non-linear system make a rigorous analysis of all regions of parameter space exceptionally challenging, and we do not attempt that here. Indeed, in some regimes the assumption of a single steady state will break down. \cite{Salazar2006,Ortega2006} Here, we merely demonstrate that moderate concentrations of enzyme/substrate pairs do not {\it necessarily} significantly compromise our earlier results on proofreading, by exploring a few examples.

We start from the model presented in Figure 2 of the main text, with the same simplification of rate constants summarized in Equation (6) of the main text, but do not make the assumption that the bound states such as $KA$ are swiftly resolved into either $K \circ A$ or $K \circ A_p$ (we still assume that these `close-proximity' states are short-lived). As kinase and phosphatase concentrations are not fixed but are influenced by both $A$ and $B$, we cannot solve separately for $A$ and $B$, so we must consider all species at once.
The resultant set of equations that must be solved for the steady state are
  \begin{widetext}
\begin{equation}
\begin{array}{c}
0 =[\dot{A}]= -k_D P_{\rm on} [K] [A] + k_{\rm cat} [P A_{p}]  + k^A_{d} (1-P_{\rm on}) [K A], \\ 
\\
0 = [\dot{A}_{pp}]=  - k_D P_{\rm on} [P] [A_{pp}]+k_{\rm cat} [K A_p]  +h^A_d (1- P_{\rm on})[PA_{\rm pp}],\\
\\
0=[\dot{B}]= -k_D P_{\rm on} [K] [B] + k_{\rm cat} [P B_{p}]  + k^B_{d} (1-P_{\rm on}) [K B], \\ 
\\
0 =[\dot{B}_{pp}]=  - k_D P_{\rm on} [P] [B_{pp}] + k_{\rm cat} [K B_p]+h^B_d (1- P_{\rm on})[PB_{\rm pp}],\\
\\
0= [\dot{K}]=-k_D P_{\rm on}([A] +[A_p] + [B] +[B_p])[K] +[KA]  (k^A_d(1-P_{\rm on}) +k _{\rm cat}(1 - P_{\rm on})) + [KA_p]  (k^{\prime A}_d(1-P_{\rm on}) +k_{\rm cat})  \vspace{2mm} \\+ [KB]  (k^B_d(1-P_{\rm on}) +k _{\rm cat}(1 - P_{\rm on}))  + [KB_p]  (k^{\prime B}_d(1-P_{\rm on}) +k_{\rm cat})  \\
\\
0= [\dot{P}]=-k_D P_{\rm on}([A_p] +[A_{pp}] + [B_p] +[B_{pp}])[P] +[PA_{pp}]  (h^A_d(1-P_{\rm on}) +k _{\rm cat}(1 - P_{\rm on})) + [PA_p]  (h^{\prime A}_d(1-P_{\rm on}) +k_{\rm cat})  \vspace{2mm} \\+ [PB]  (h^B_d(1-P_{\rm on}) +k _{\rm cat}(1 - P_{\rm on}))   + [PB_p]  (h^{\prime B}_d(1-P_{\rm on}) +k_{\rm cat}),  \\
\\
0 =[\dot{KA}]=  - (k^A_d(1-P_{\rm on}) +k_{\rm cat})[KA]+ k_D P_{\rm on} [K] [A], \\
\\
0 = [\dot{PA_{pp}}] = - (h^A_d(1-P_{\rm on}) +k_{\rm cat})[PA_{pp}] + k_D P_{\rm on} [P] [A_{pp}] \\
\\
0=[\dot{KB}] = - (k^B_d(1-P_{\rm on}) +k_{\rm cat})[KB]+k_D P_{\rm on} [K] [B] , \\
\\
0=[\dot{PB_{pp}}] = - (h^B_d(1-P_{\rm on}) +k_{\rm cat})[PB_{pp}]+k_D P_{\rm on} [P] [B_{pp}] ,\\
\\
0=[\dot{KB_p}] = -(k^{B\prime}_d (1-P_{\rm on}) -k_{\rm cat}) [KB_p]  + k_D P_{\rm on} [K][B_p] + k_{\rm cat}P_{\rm on}[KB],\\
\\
0=[\dot{PB_p}] = -(h^{B\prime}_d (1-P_{\rm on}) -k_{\rm cat}) [PB_p]  + k_D P_{\rm on} [P][B_p] + k_{\rm cat}P_{\rm on}[P B_{pp}].
\end{array}
\label{eqns-sat}
\end{equation}
  \end{widetext}
We solve this system using the `NSolve' routine in Mathematica. 

To control the degree of non-linearity, we allow $k_D$ to vary whilst keeping other rate constants and $[A_0]$, $[B_0]$ and $[K_0]$ (total kinase concentration) fixed.
Only binding rates scale explicitly with $k_D$, and so increasing $k_D$ leads to faster binding with respect to enzymatic action and unbinding and hence the possibility of  bound states that are long-lived compared to the timescale of binding. As before, we will make comparisons at constant yield $[A_{pp}]/[A_0]$. For each set of parameters considered, we iteratively find the concentration of phosphatases $[P_0]$ that gives the desired overall yield of phosphorylated product $A_{pp}$, and use that value to calculate $[B_{pp}]$ and  hence the specificity $X$. For all systems reported here, only one solution to Equations (\ref{eqns-sat}) was found by Mathematica's Nsolve routine with real, positive values for the concentration of each species.

We use rate constants that would reproduce the system considered in the main text in the limit $k_D \rightarrow 0$. 
\begin{equation}
\begin{array}{c}
k_{\rm cat} = 0.1\,s^{-1}, \\
\\ 
k^A_{\rm d} = 0.4\,s^{-1}, \hspace{2mm} 
k^{A \prime}_{\rm d} = 0.4\,s^{-1}, \\
\\
h^{A}_{\rm d} = 0.4\,s^{-1}, \hspace{2mm}  
h^{A \prime}_{\rm d} = 0.4\,s^{-1},\\
\\ 
k^B_{\rm d} = 9.9\,s^{-1}, \hspace{2mm} 
k^{B \prime}_{\rm d} = 9.9\,s^{-1}, \\
\\
h^{B}_{\rm d} = 0.4\,s^{-1}, \hspace{2mm}  
h^{B \prime}_{\rm d} = 0.4\,s^{-1}.\\
\end{array}
\end{equation}
As the close proximity states are still assumed to resolve quickly, the absolute values of $k_{\rm a}$ and $k_{\rm esc}$ are not important, only the probability $P_{\rm on}$ (we shall consider values of $P_{\rm on}=0.1$  and 0.9, corresponding to $f^A_\alpha= f^A_\beta=0.022$  and 0.64 respectively). Note that the specific values of rate constants are not that important --  identical steady-state concentrations would be obtained by scaling all rate constants in the system by the same amount. The key question is whether the resolution of a bound state (its dissociation into distinct enzyme and substrate) is fast compared to its formation, and hence whether enzyme/substrate complexes have appreciable concentrations. This is quantified by the Michealis-Menten constant $K_m$, which is $3.6/k_D$\,s$^{-1}$ for catalysis of $A$ when $P_{\rm on}=0.1$, and $0.044/k_D$\,s$^{-1}$ for catalysis of $A$ when $P_{\rm on}=0.9$. Thus when $k_D = 3.6 \times 10^6$\,M$^{-1}$\,s$^{-1}$ (for $P_{\rm on} =0.1$) or $k_D = 4.4 \times 10^4$\,M$^{-1}$\,s$^{-1}$ (for $P_{\rm on} =0.9$), $K_m = 1$\,$\mu$M (the maximum concentration of substrates/kinases we use in this section).

\begin{figure*}
        \centering
        \subfloat[]{
                \includegraphics[width=15pc, angle=-90]{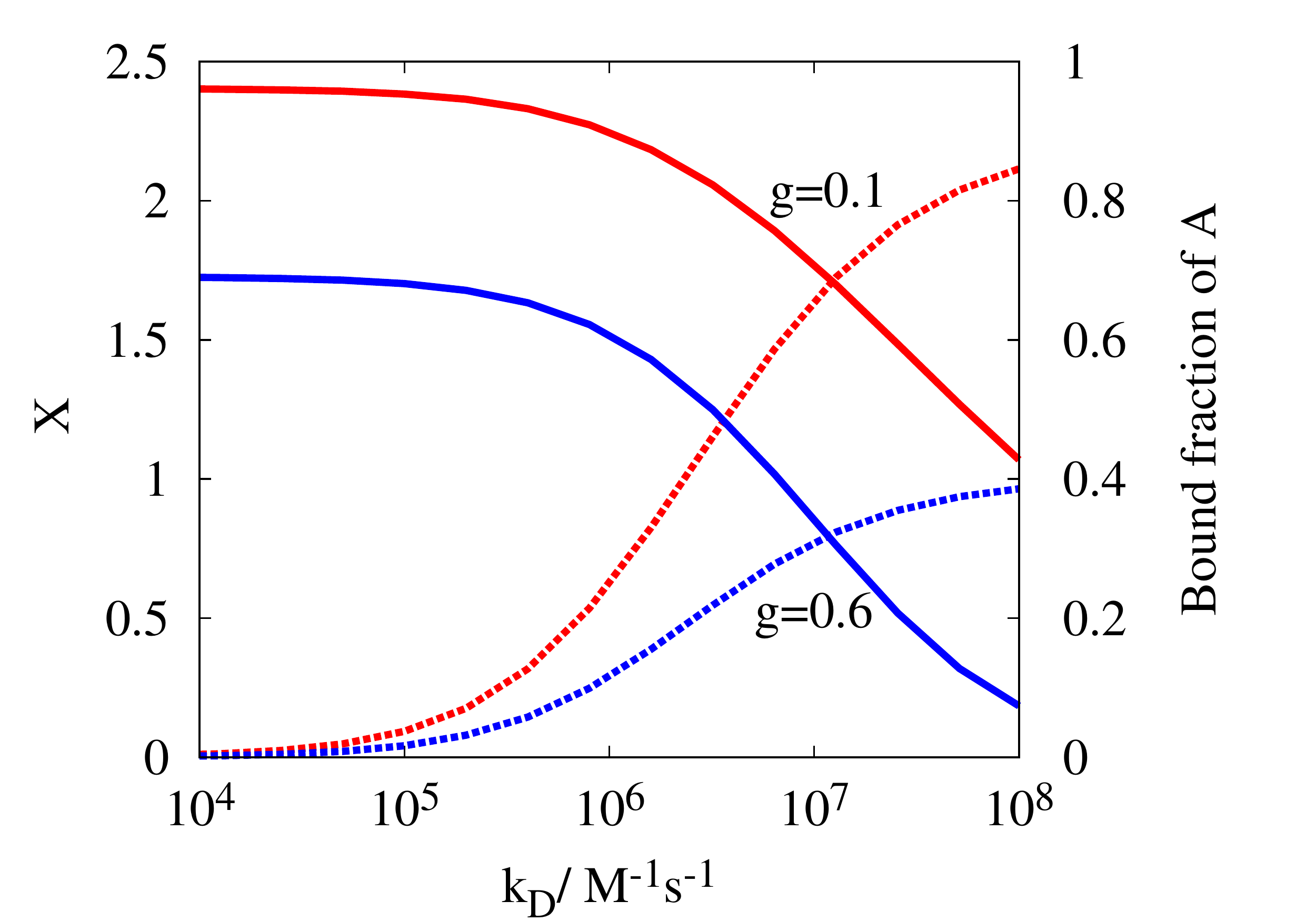}         
        }%
        ~ 
        \subfloat[]{
                \includegraphics[width=15pc, angle=-90]{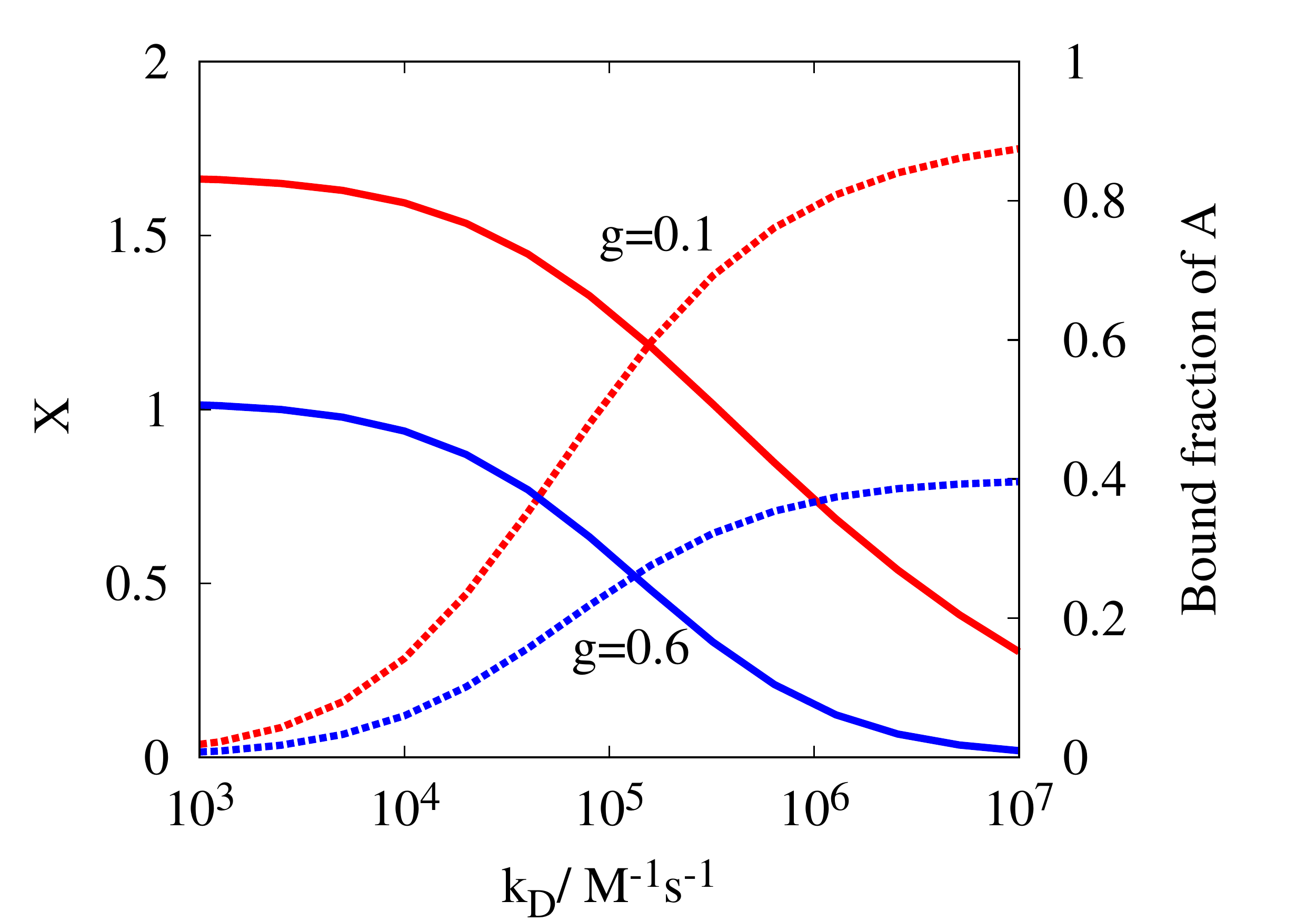}
        }
        
        \subfloat[]{
                \includegraphics[width=15pc, angle=-90]{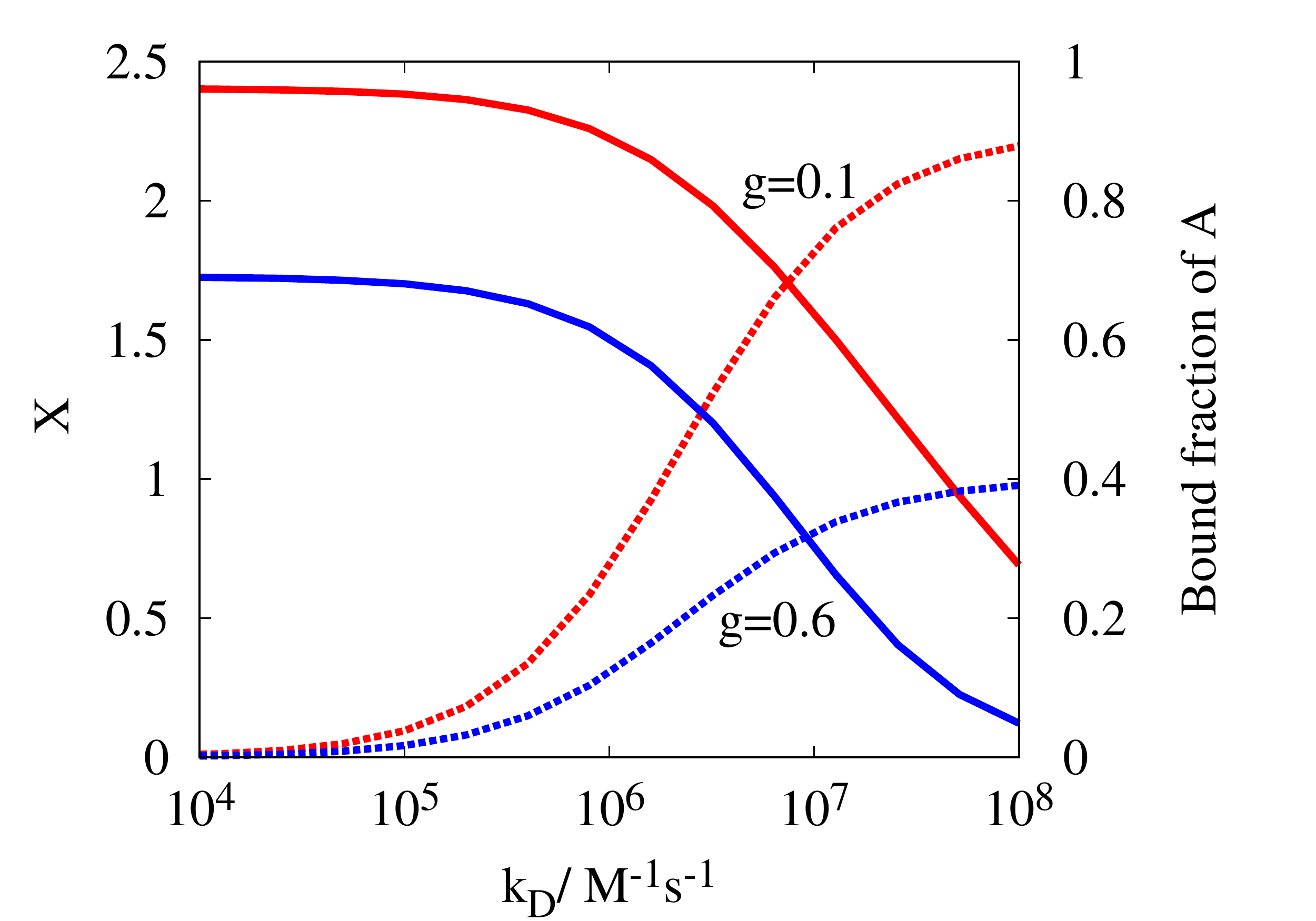}         
        }%
        ~ 
        \subfloat[]{
                \includegraphics[width=15pc, angle=-90]{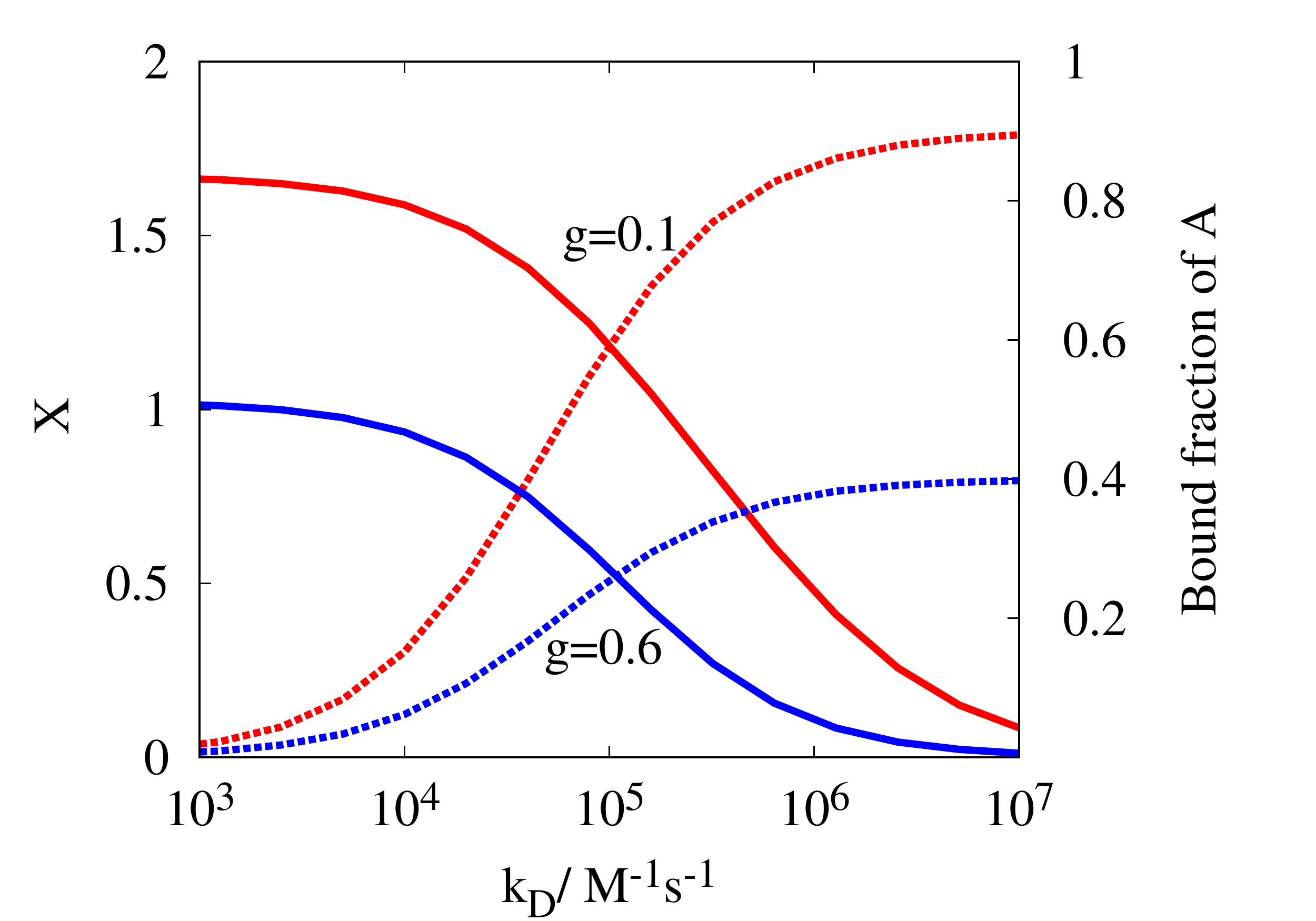}
        }    
             
           \subfloat[]{
                \includegraphics[width=15pc, angle=-90]{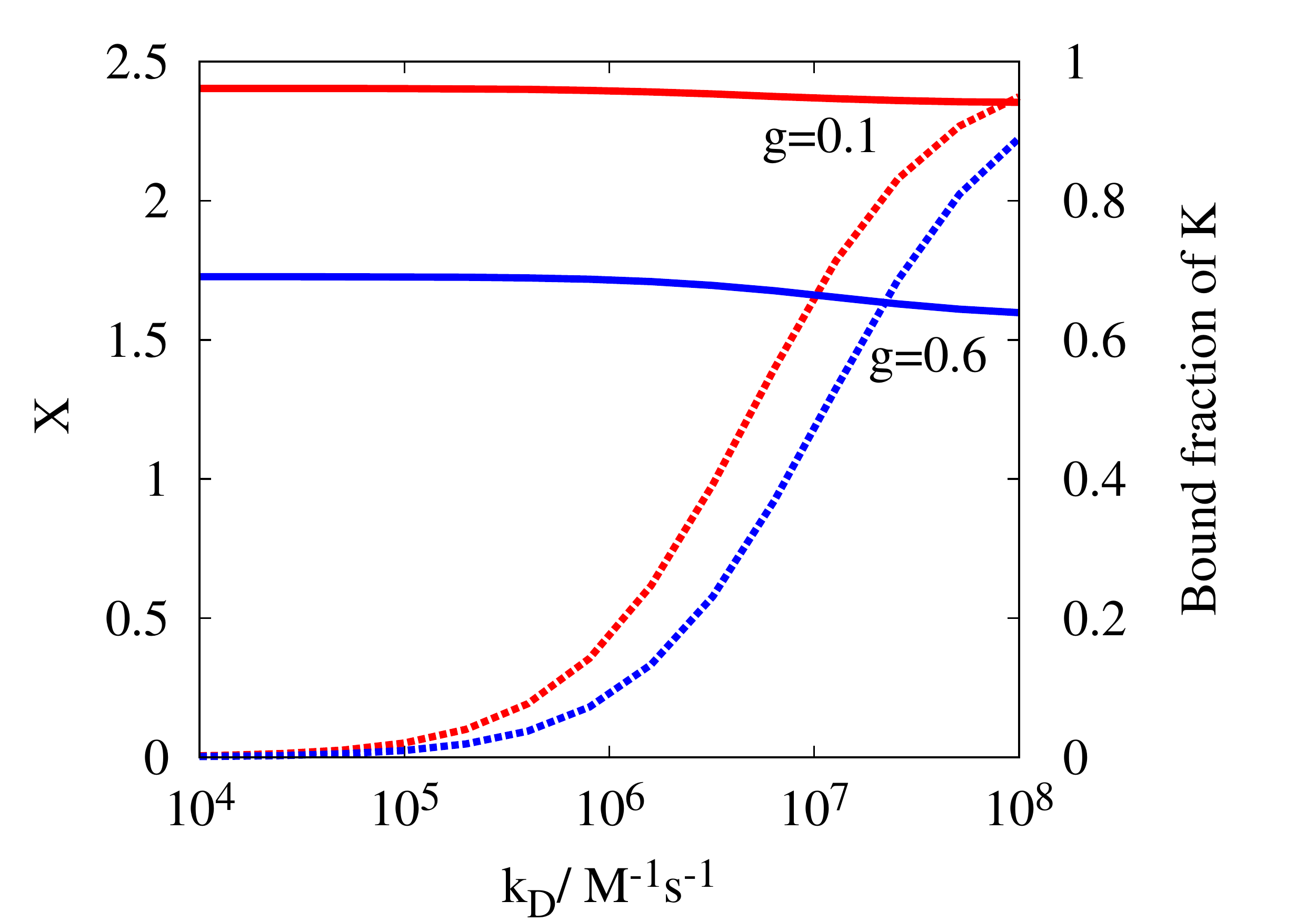}         
        }%
        ~ 
        \subfloat[]{
                \includegraphics[width=15pc, angle=-90]{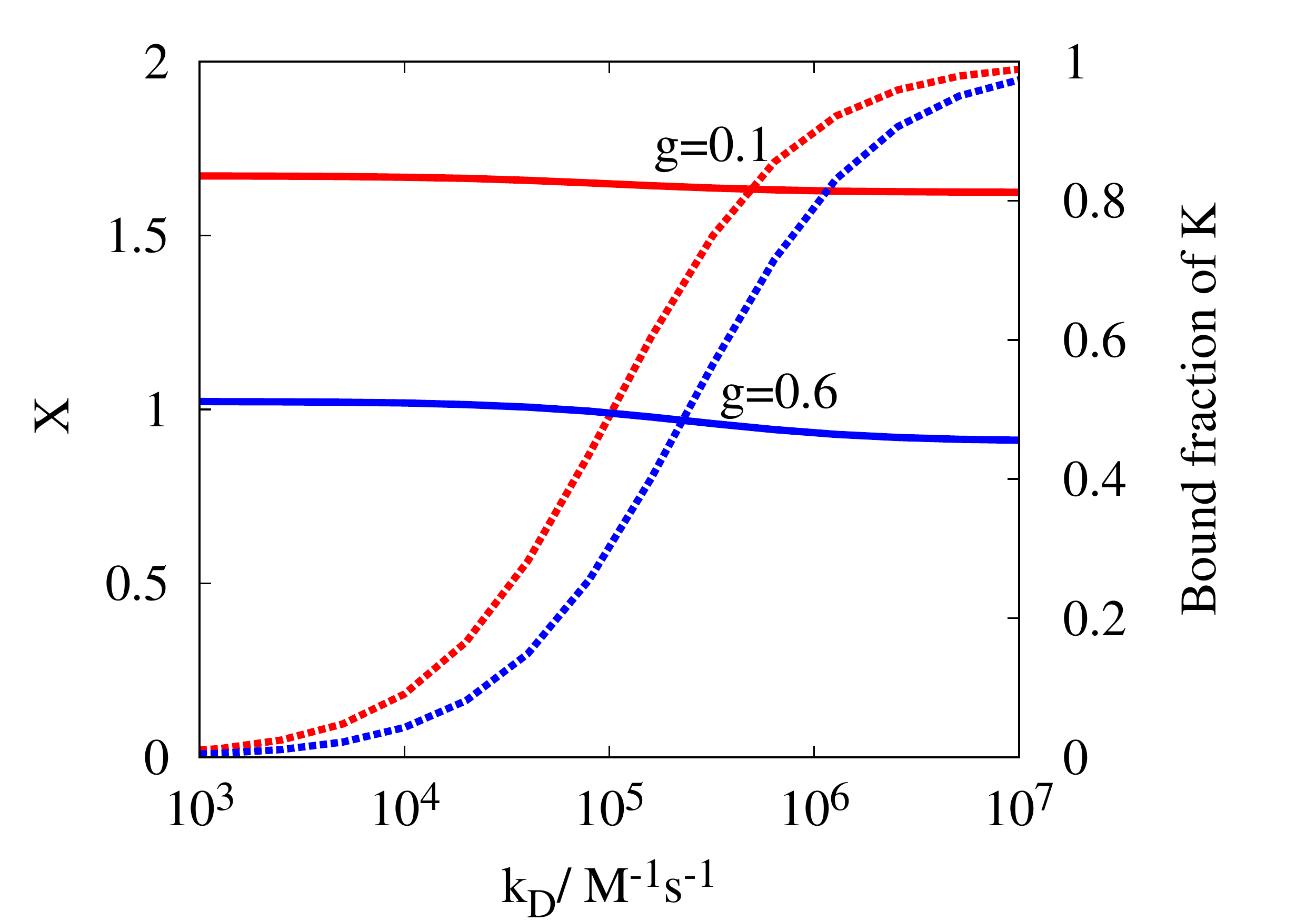}
        }
        ~ 
        \caption{\footnotesize{The variation in specificity with increasing concentrations of bound enzyme/substrate complexes, showing that moderate concentrations of complexes have small effects. We plot $X$ against $k_D$ at fixed yield $g$ (solid lines, left axes) for the system outlined in the text, with total concentrations (a,b) $[A_0] = [B_0] = [K_0] = 1\,\mu$M, (c,d) $[A_0] = [B_0] = 0.05\,\mu$M, $[K_0] = 1\,\mu$M, and (e,f) $[A_0] = [B_0] = 1\,\mu$M, $[K_0] = 0.05\,\mu$M. In each figure we consider two fixed yields, $g=0.1$ and $g=0.6$. In (a,c,e) we take $P_{\rm on}=0.1$, and in (b,d,f) we use $P_{\rm on}=0.9$. Also plotted (dashed lines, right axes) are the fraction of substrate molecules $A$ in complexes (a-d), and the fraction of kinase molecules $K$ in complexes (e,f).} 
        \label{saturation}}
\end{figure*}

In Figure \ref{saturation}\,(a) we plot the specificity $X$ of a system with the parameters listed above for concentrations $[A_0] = [B_0] = [K_0] = 1\,\mu$M, $P_{\rm on}=0.1$ and yield $g=[A_{pp}]/[A_0] = 0.1,0.6$, as $k_D$ is varied. Also shown is the fraction of substrates $A$ that are bound in a complex with an enzyme in the steady state. The same system, but with  $P_{\rm on}=0.9$, is shown in Figure \ref{saturation}\,(b). In Figures \ref{saturation}\,(c,d) we consider the same system but with $[A_0] = [B_0]= 0.05\,\mu$M, in which case substrates can be saturated by abundant enzymes at high $k_D$. In all four cases, specificity $X$ drops from the unsaturated limit considered in the main text (at $k_D\rightarrow0$) as $k_D\rightarrow \infty$. This drop is unsurprising, and there are at least two physical contributions.
\begin{enumerate}
\item When a larger fraction of $[A_0]$ is sequestered in enzyme complexes, the system must push the $[A_{pp}]/[A]$ ratio higher than before to get the same yield $g = [A_{pp}]/[A_0]$. Thus finite product yield becomes more of a problem when a substantial fraction of substrates are sequestered.
\item The distinction between $A$ and $B$ (due to unbinding rates) can become unimportant if a given substrate typically rebinds to an enzyme faster than it unbinds, as the difference in time spent attached (and waiting for catalysis to occur) is suppressed.
\end{enumerate}
We note, however, that in all cases in Figure \ref{saturation}\,(a-d), the drop in $X$ is not  really noticeable until the concentration of sequestered substrates is  at least $([A_0] - [A_{pp}])/4$ (when a quarter of the $A$ substrates that aren't fully phosphorylated are sequestered in enzyme complexes). Similarly, changes in $X$ are still fairly small at the point when the Michaelis-Menten constant is equal to the concentration of enzymes, $K_m = 1$\,$\mu$M ($k_D = 3.6 \times 10^6$\,M$^{-1}$\,s$^{-1}$ for $P_{\rm on} =0.1$ and $k_D = 4.4 \times 10^4$\,M$^{-1}$\,s$^{-1}$ for $P_{\rm on} =0.9$). Thus, at least in these cases, specificity is not strongly affected for small but non-negligible concentrations of enzyme/substrate complexes. 

In Figures \ref{saturation}\,(e,f) we consider the same system but with $[K_0] = 0.05\,\mu$M. In this case, the enzyme can become saturated by its substrates, so we plot the fraction of $K$ that is bound in complexes in the steady state. In this case, small but non-negligible fractions of $K$ in complexes have almost no effect on specificity, and even when the enzyme is heavily saturated we see only a small change in $X$. In the limit $[A_0], [B_0] \gg [K_0], [P_0]$, the time spent in complexes of enzymes and substrates becomes negligible relative to the time spent free for substrates $A$ and $B$ in all their phosphorylation states, regardless of $k_D$. In this limit, the differential equations for $A$ and $B$ are identical to those used in the unsaturated limit discussed in the main text, but with $[K] \neq [K_0]$ and $[P] \neq [P_0]$ due to the saturation of enzymes. The fact that $[K] \neq [K_0]$ and $[P] \neq [P_0]$ is irrelevant for the specificity at fixed product yield, however -- it just means that the quantity $Y_{\rm sat} = [K]/[P] \neq [K_0]/[P_0]$ must be used in Equation (5) of the main text, and the value of $[P_0]$ required to achieve a certain yield is not given by $[P_0] = [K_0]/Y_{\rm sat}$, but requires a more detailed calculation. In more physical terms, $A$ and $B$ compete for free enzymes in exactly the same manner as in the unsaturated case; the only difference is that the concentration of free enzymes is reduced by sequestration.  The very weak dependence of $X$ on $k_D$ in  Figures \ref{saturation}\,(e,f) reflects this behaviour.

\begin{figure*}
        \centering
        \subfloat[]{
                \includegraphics[width=15pc, angle=-90]{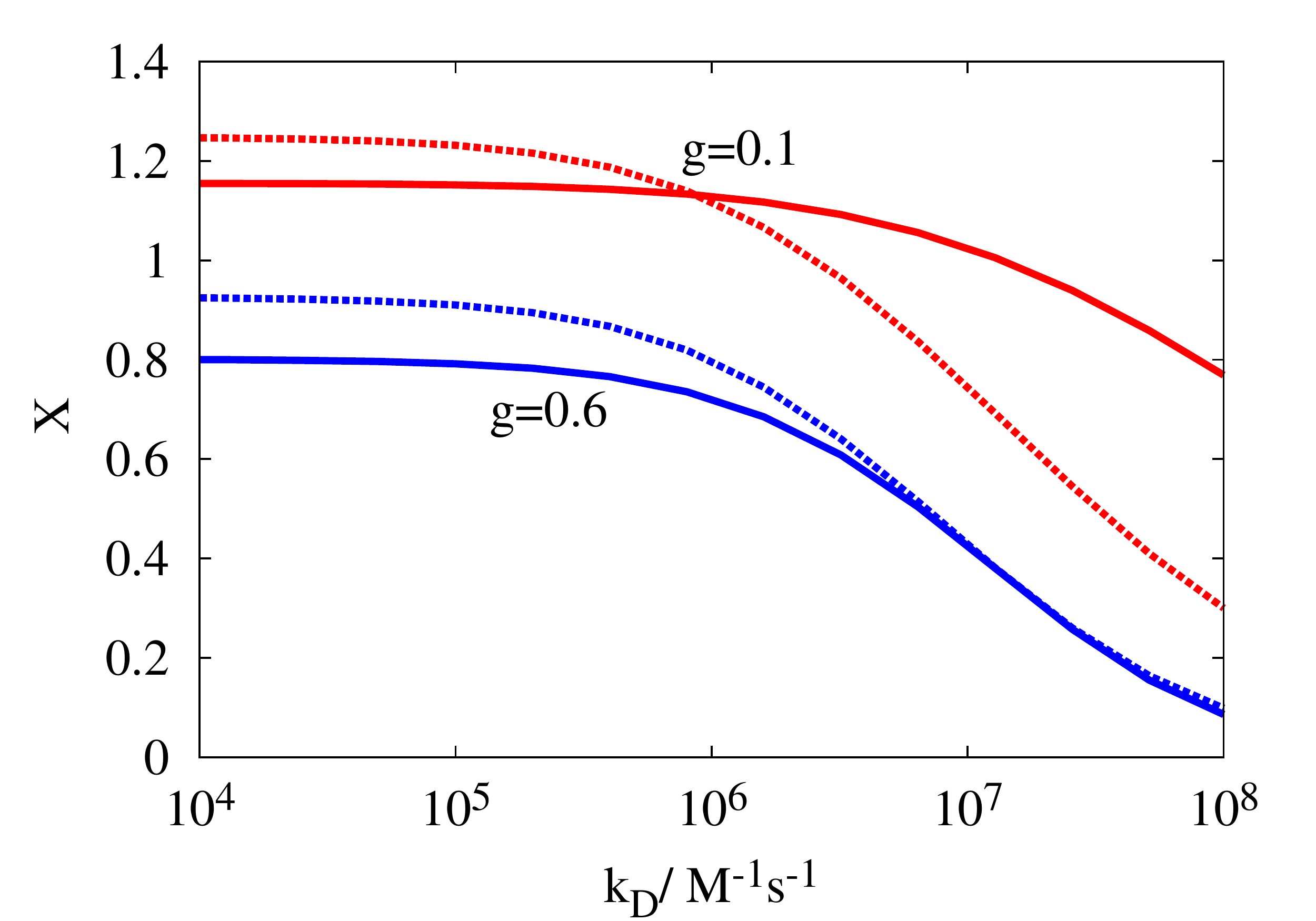}         
        }%
        ~ 
        \subfloat[]{
                \includegraphics[width=15pc, angle=-90]{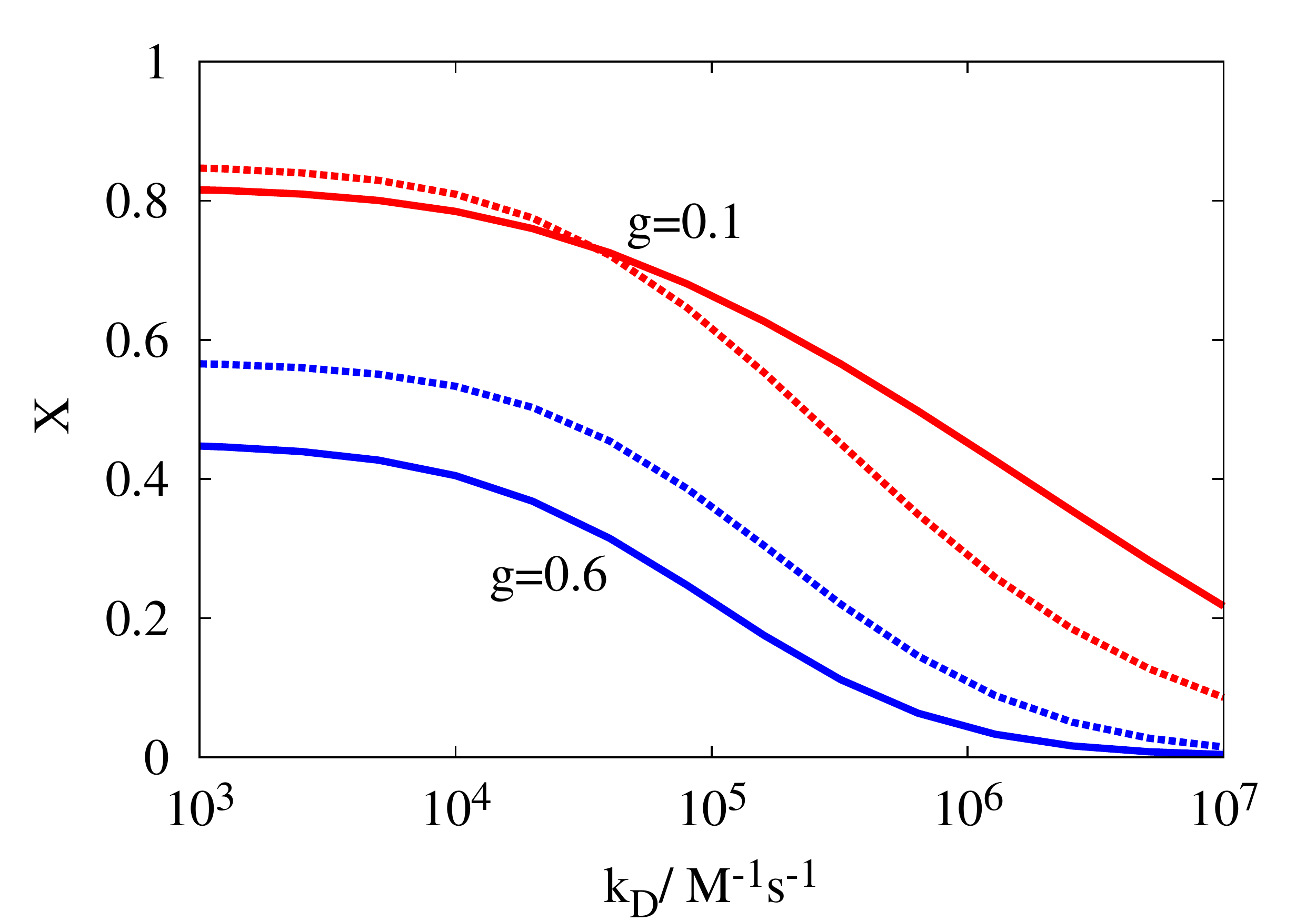}
        }
        
        \subfloat[]{
                \includegraphics[width=15pc, angle=-90]{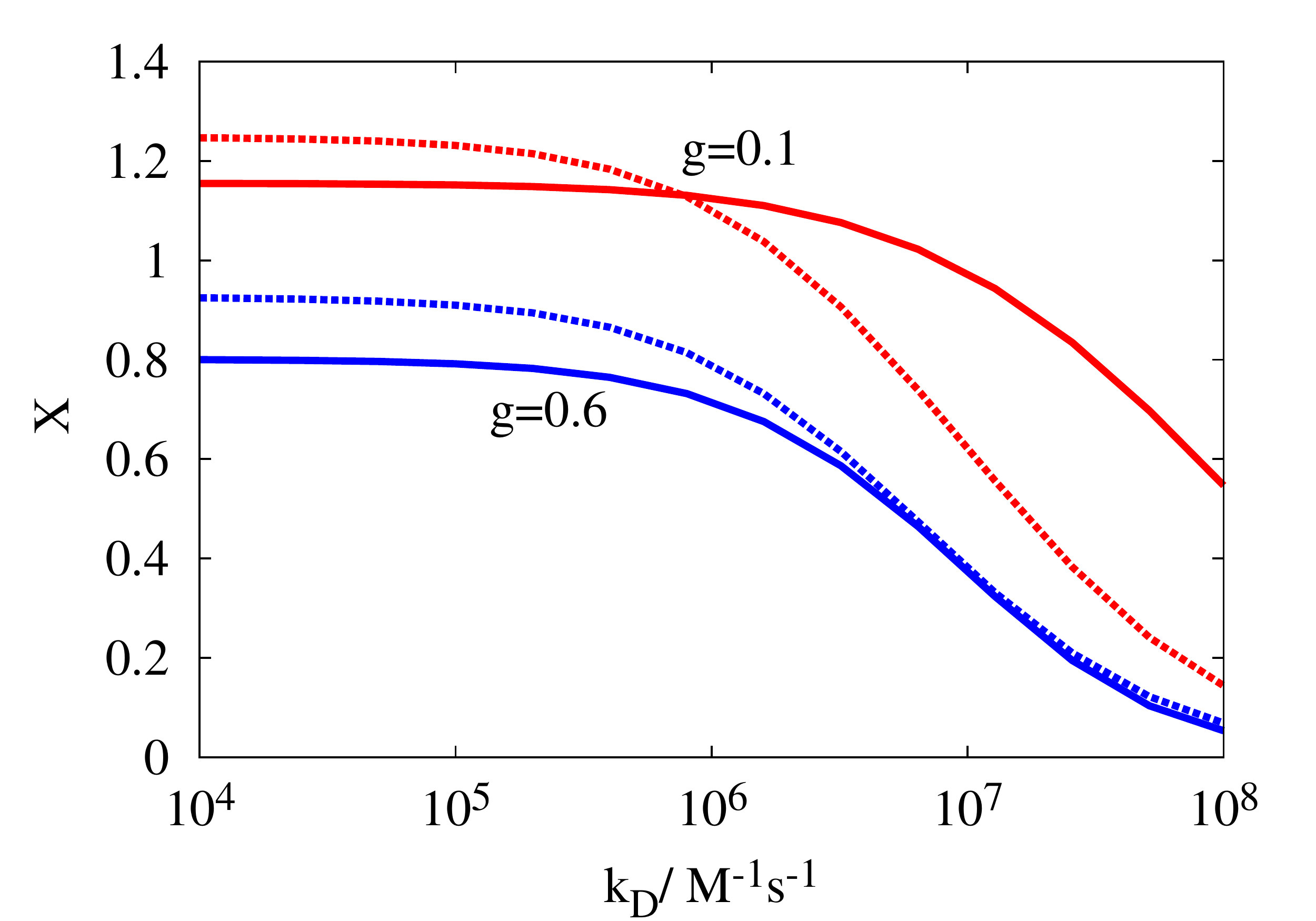}         
        }%
        ~ 
        \subfloat[]{
                \includegraphics[width=15pc, angle=-90]{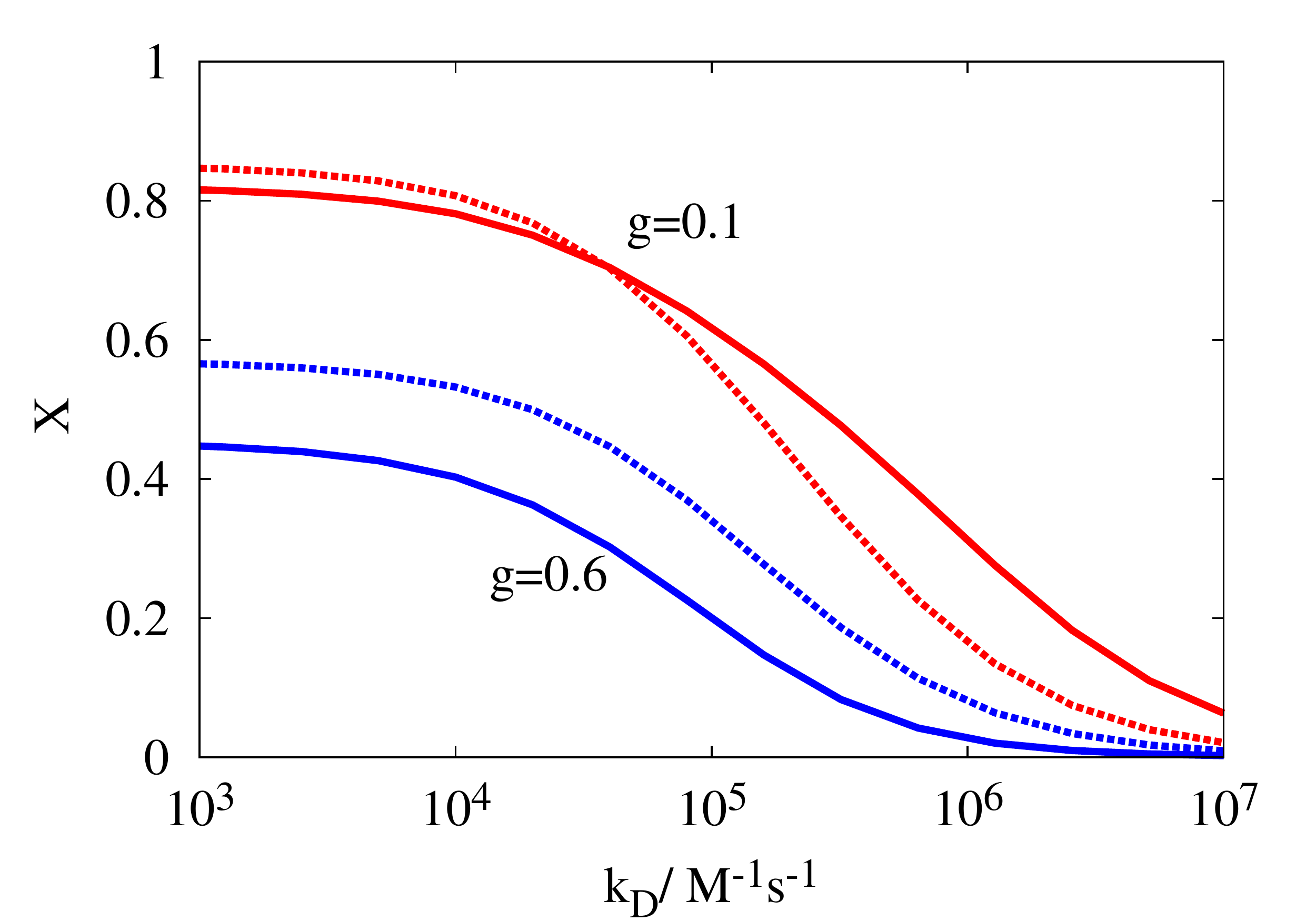}
        }    
             
           \subfloat[]{
                \includegraphics[width=15pc, angle=-90]{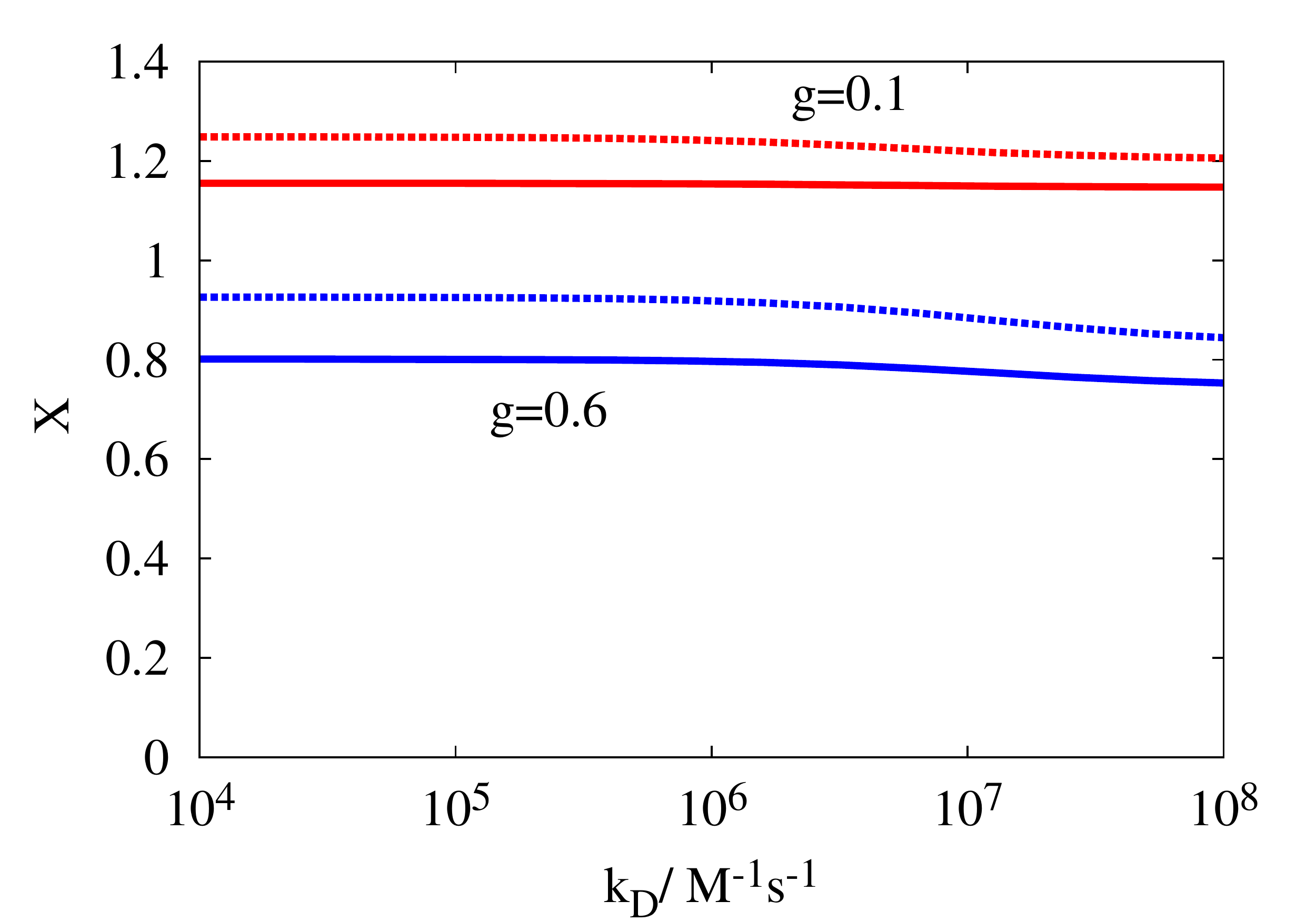}         
        }%
        ~ 
        \subfloat[]{
                \includegraphics[width=15pc, angle=-90]{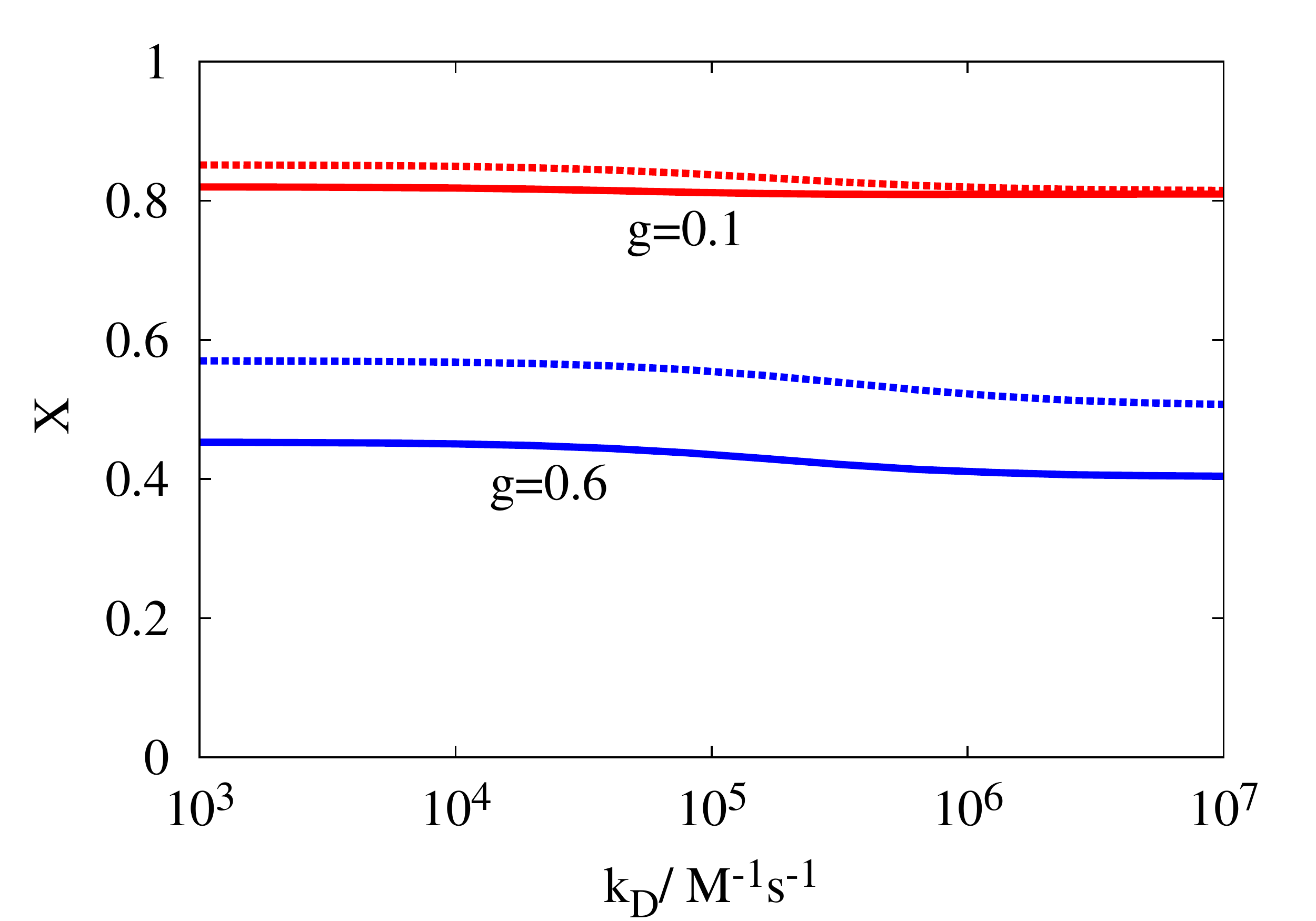}
        }
        ~ 
        \caption{\footnotesize{The variation in specificity $X_{\rm ss}$ for a single-site system with the first phosphorylation site (dashed lines), and the additional specificity gained from adding the second site $X-X_{\rm ss}$ (solid lines), with increasing concentrations of bound enzyme/substrate complexes. It is clear that the loss of specificity associated with saturation is not, in these cases at least, overwhelmingly due to a reduced benefit from the second site rather than the first. We consider total molecule concentrations of (a,b) $[A_0] = [B_0] = [K_0] = 1\,\mu$M, (c,d) $[A_0] = [B_0] = 0.05\,\mu$M, $[K_0] = 1\,\mu$M, and (e,f) $[A_0] = [B_0] = 1\,\mu$M, $[K_0] = 0.05\,\mu$M. In each figure we consider two fixed yields, $g=0.1$ and $g=0.6$. In (a,c,e) we take $P_{\rm on}=0.1$, and in (b,d,f) we use $P_{\rm on}=0.9$. All other parameters are stated in the text.} 
        \label{saturation2}}
\end{figure*}

In Figure \ref{saturation2}, we plot $X$ and $X-X_{\rm ss}$ for these systems, demonstrating that any loss in specificity is not generally related to a particular loss in efficacy of the second site relative to the first. In fact, in some cases (at low yield $g$ and when substrates are saturated by enzymes)  the specificity in a single-site system is more strongly compromised by the  finite lifetimes of complexes than the additional specificity of the second site. It is also worth noting that larger intrinsic selectivities $S$ and $S^\prime$ would mean that some  specificity due to both the first and second sites can be maintained at higher levels of saturation than in this case, just as larger intrinsic selectivities $S$ and $S^\prime$ allow specificity at higher yields $g$ in the unsaturated case. Thus our conclusions relating to the efficacy of proofreading are not necessarily compromised by non-negligible concentrations of enzyme/substrate complexes -- we defer a full investigation of these effects to later work. 


\end{appendix}

\end{document}